\pdfoutput=1
\documentclass[11pt]{article}
\usepackage[]{acl}
\usepackage{times,latexsym}
\usepackage[T1]{fontenc}
\usepackage[utf8]{inputenc}
\usepackage{microtype,inconsolata}
\usepackage{graphicx,booktabs,multirow,amsmath,amssymb,array,longtable}
\usepackage{xurl}
\usepackage[export]{adjustbox}
\usepackage{capt-of}
\usepackage{xcolor}
\usepackage{tikz}\usetikzlibrary{arrows.meta,calc,positioning}
\newcolumntype{P}[1]{>{\raggedright\arraybackslash}p{#1}}
\title{Protection Is Not One Event:\\
A Survey of RAG Protection Research\\
by Measured Endpoint}

\author{
\begin{tabular}{c}
Peiyang Liu$^{1,2}$, Xi Wang$^{2}$, Di Liang$^{3}$, Youru Li$^{4}$, Zhi Cai$^4$
\and Wei Ye$^{1,}$\thanks{\  \ Corresponding author}
\end{tabular}
\\
$^1$ National Engineering Research Center for Software Engineering, Peking University, Beijing, China, \\   
$^2$ School of Software and Microelectronics, Peking University, Beijing, China,\\
$^3$ Tencent Technology, Shenzhen, China,\\
$^4$ College of Computer Science, Beijing University of Technology, Beijing, China.\\
\href{liupeiyang@pku.edu.cn}{liupeiyang@pku.edu.cn}
}

\begin{document}
\maketitle
\begin{abstract}
Retrieval-augmented generation (RAG) systems face security, privacy, and copyright risks including content leakage, source reconstruction, unauthorized access, data poisoning, and prompt injection. Yet RAG protection studies use incompatible evaluation metrics, datasets, and threat models. We present a systematic survey and human-coded evidence map of 118 full-text-confirmed RAG defense papers, organized by measured endpoint rather than claimed goal or defense mechanism. Our seven-family taxonomy covers content disclosure, formal privacy proxies, access control and suppression, corpus-use detection, and pipeline integrity and attribution. Thirty-two annotators double-coded the papers ($\kappa=0.867$).

The survey identifies three gaps. First, retrieval reranking and context rewriting each span five or six endpoint families, so mechanisms do not identify protection outcomes. Second, only 17.8\% of studies use content disclosure or reconstruction as their primary endpoint; under the frozen copyright/ownership relation label, 73.3\% of records have corpus-use detection as their primary endpoint, and none have an emitted-expression primary endpoint. Third, most evaluations use general-purpose question-answering datasets and non-adaptive attacks rather than sensitive data and adaptive, multi-turn extraction.

Our controlled RAG testbed shows that preventing verbatim copying does not prevent private-fact recovery: a normalized 8-gram guard reduces 20-token copying to 0.0\%, but a three-query adaptive attacker recovers 48.3\% of hidden facts on Qwen3-8B. We release the taxonomy, dataset, coding protocols, and testbed to enable comparable, claim-aligned RAG security evaluation.
\end{abstract}

\section{Introduction}
\label{sec:introduction}
Retrieval-augmented generation (RAG) connects language models to external document collections at inference time \citep{lewis2020knowledgeintensivenlp,gao2023survey}. Because this external datastore can be searched, updated, and quoted directly, it creates new risks around privacy, copyright, and system security. Even if a document never appeared in pre-training, a RAG system might still leak its private details, copy its wording verbatim, or fall victim to poisoned entries.

A rapidly growing literature aims to ``protect'' RAG systems against these risks, yet papers using this label rarely measure the same thing. Consider two studies that both claim to protect the copyright of a retrieval database. The first plants a watermark to prove after the fact that the database was queried, while the second blocks unauthorized users from retrieving restricted documents. A third study defends against poisoned passages to keep answers accurate, and a fourth rewrites retrieved text so the generator cannot copy it word for word. All four papers report successful ``protection,'' but they evaluate completely different technical events. Treating them as interchangeable, or pooling their scores to compare defenses, is like comparing a burglar alarm to a fireproof safe.

Existing surveys organize this literature by threat model, pipeline stage, or system architecture. However, knowing where a defense sits in the pipeline tells us very little about what it actually achieves. In this work, we take a different approach: we survey RAG protection research based on the exact empirical endpoint each study measures. We define a frozen seven-family taxonomy of measured endpoints (Table~\ref{tab:families} and Figure~\ref{fig:taxonomy_tree}) that separates actual content disclosure from access control, usage detection, and pipeline integrity. Using this taxonomy, 32 annotators double-coded 118 full-text-confirmed RAG protection papers, reaching high agreement on what each paper measures ($\kappa=0.867$).

Looking at the literature through its measured endpoints reveals three major blind spots. First, defense mechanisms only weakly predict what a paper tests (\S\ref{sec:mechanisms}). Popular techniques such as retrieval reranking and context rewriting are evaluated across five or six different endpoint families, meaning readers cannot infer what was protected just from how the system was built. Second, research attention is heavily skewed away from actual data leakage (\S\ref{sec:crosscutting}). Only 21 of the 118 papers (17.8\%) have a primary endpoint that measures whether protected content is emitted to the user or reconstructed from the system. Among studies with the frozen copyright/ownership relation label, 73.3\% have corpus-use detection as their primary endpoint, whereas none have an emitted-expression primary endpoint. Third, evaluation setups often fail to stress-test the defenses they propose (\S\ref{sec:eval_practice}). Most studies rely on standard question-answering datasets that contain little sensitive content to begin with, and very few test against adaptive attackers who can ask follow-up questions. Our own controlled experiments confirm why this distinction matters: defenses that drive word-for-word copying down to zero still allow attackers to recover the underlying private facts (Appendix~\ref{app:casestudy}).

To help future work build on clear empirical foundations, we make four contributions:
\begin{itemize}
    \item \textbf{An endpoint-based survey} that organizes RAG defense mechanisms, evaluation metrics, datasets, and threat models by the exact technical event they measure (\S\ref{sec:families}, \S\ref{sec:eval_practice}).
    \item \textbf{A validated seven-family event taxonomy} that makes cross-paper comparisons reproducible (\S\ref{sec:map}).
    \item \textbf{A human-coded evidence map} of 118 direct RAG protection papers, complete with item-level annotations, operating points, and open-source verification tools (\S\ref{sec:map}, Appendix~\ref{app:operating_points}).
    \item \textbf{A critical audit and controlled testbed} examining reporting quality, and the gap between stopping surface copying and preventing fact recovery (\S\ref{sec:crosscutting}, \S\ref{sec:future}, Appendix~\ref{app:casestudy}).
\end{itemize}

\section{Related Work}
\label{sec:related}
Prior surveys of RAG security and privacy organize the field around system architecture and attack surfaces. Broad reviews of language model privacy cover threats like membership inference and data poisoning \citep{das2024securityprivacychallenges} alongside cryptographic and differentially private defenses \citep{chen2025surveyprivacyrisks}. Dedicated surveys on copyright and ownership focus on text and model watermarking \citep{yang2025watermarkingsurvey} or fingerprinting techniques \citep{xu2025llmcopyright}. Within the RAG literature specifically, existing reviews group studies by threat model, pipeline stage, or defensive module \citep{mu2026securerag,palanisamy2026ragsecurity,xu2026securingrag}. While this architectural view shows where a defense intervenes, it obscures what the evaluation actually tests. Individual empirical studies already hint at this divergence: papers evaluating literal and non-literal text reproduction \citep{chen2024copybench}, copyright-sensitive activation steering \citep{zhang2026scope}, proprietary knowledge base defense \citep{li2026ragfort}, and knowledge extraction benchmarks \citep{qi2026knowledgeextraction} all define ``protection'' through different operational outcomes. Furthermore, recent adaptation methods train models to decide when to retrieve, critique context, or withhold answers \citep{asai2024selfrag,lin2024radit}, blurring the boundary between parametric memory and external datastores. Rather than grouping papers by where their mechanisms sit in the pipeline, we human-code the exact technical events their experiments measure.

Our approach builds on meta-scientific audits of machine learning research \citep{lipton2019troubling,kapoor2023leakage} and classical measurement theory, which requires clearly defining target concepts and validating how they are measured \citep{cronbach1955construct,campbell1959convergent,messick1995validity}. These principles have shaped recent critiques of benchmark design and fairness evaluation in natural language processing \citep{jacobs2021measurement,blodgett2021salmon,raji2021everything,bowman2021benchmarking}. However, while prior audits typically expose flaws inside an experimental setup, we examine the link between two separate components of a paper: its high-level protection claim and its measured empirical endpoint. This focus on claim-measurement alignment parallels recent findings in question answering, where standard lexical overlap metrics often fail to capture true semantic accuracy \citep{bulian2022tomayto,kamalloo2023evaluating,adlakha2024evaluating}.

\definecolor{taxRoot}{HTML}{2C3340}
\definecolor{taxLeafBg}{HTML}{F5F2EA}
\definecolor{taxLeafBd}{HTML}{C9C3B5}
\definecolor{taxChipBg}{HTML}{FCFBF8}
\definecolor{taxCon}{HTML}{5B7393}
\definecolor{taxAcc}{HTML}{A06058}
\definecolor{taxUse}{HTML}{5E8466}
\definecolor{taxInt}{HTML}{836B91}

\begin{figure*}[t]
\vspace*{-0.55cm}
\setlength{\abovecaptionskip}{2pt}
\centering
\makebox[\textwidth][c]{\scalebox{0.80}{
\begin{tikzpicture}[
  x=1cm,y=1cm,
  root/.style={
    draw=taxRoot, fill=white, text=taxRoot,
    rounded corners=2.5pt, line width=0.8pt,
    font=\small\bfseries, align=center,
    inner xsep=8pt, inner ysep=5pt, text width=8.0cm
  },
  grp/.style={
    draw=#1, fill=#1, text=white,
    rounded corners=2.5pt, line width=0pt,
    font=\footnotesize\bfseries, align=center,
    inner xsep=2.4pt, inner ysep=4pt, minimum height=0.95cm
  },
  fam/.style={
    draw=#1!70, fill=taxLeafBg, text=black!85,
    rounded corners=2.5pt, line width=0.5pt,
    font=\fontsize{6.0}{6.5}\selectfont, align=left,
    inner xsep=2.2pt, inner ysep=3.2pt, text width=2.30cm,
    minimum height=1.42cm, anchor=north
  },
  cell/.style={
    draw=#1!45, fill=taxChipBg, text=black!82,
    rounded corners=2pt, line width=0.4pt,
    font=\fontsize{5.9}{6.3}\selectfont, align=left,
    inner xsep=2.2pt, inner ysep=2.8pt, text width=2.30cm,
    minimum height=1.44cm, anchor=north
  },
  sparse/.style={
    draw=#1!45, fill=taxChipBg, text=black!82,
    rounded corners=2pt, line width=0.4pt,
    font=\fontsize{5.9}{6.3}\selectfont, align=left,
    inner xsep=2.2pt, inner ysep=2.8pt, text width=2.30cm,
    minimum height=0.80cm, anchor=north
  },
  link/.style={-{Latex[length=2mm,width=1.5mm]}, line width=0.5pt, color=black!50},
  twig/.style={line width=0.5pt, color=black!50, rounded corners=3pt},
  stem/.style={line width=0.5pt, color=black!38}
]

\node[root, minimum height=45.833039pt] (root) at (0.00,0)
  {Protection \emph{event} a record measures\\[2pt]
   \fontsize{7.0}{7.6}\selectfont\mdseries
   \textbf{118} human double-coded
   ($\kappa=0.867$)};

\node[grp=taxCon, text width=6.35cm] (gCon) at (-3.83,-1.62)
  {Content disclosure\\\textnormal{\fontsize{6.6}{7}\selectfont 27 records}};
\node[grp=taxAcc, text width=2.30cm] (gAcc) at (2.55,-1.62)
  {Access\\suppression};
\node[grp=taxUse, text width=2.30cm] (gUse) at (5.11,-1.62)
  {Corpus-use\\detection};
\node[grp=taxInt, text width=2.30cm] (gInt) at (7.66,-1.62)
  {Integrity \&\\attribution};

\coordinate (rootFork) at (0.00,-1.05);
\draw[twig] (root.south) -- (rootFork);
\draw[twig] (rootFork -| gCon.north) -- (rootFork -| gInt.north);
\foreach \g in {gCon,gAcc,gUse,gInt} {\draw[link] (rootFork -| \g.north) -- (\g.north);}

\node[fam=taxCon] (fA) at (-7.66,-2.55)
  {\textcolor{taxCon}{\fontsize{7.2}{7.7}\selectfont\bfseries A}\\ emitted expression copying\\[1pt]
   \textcolor{taxCon}{\fontsize{6.2}{6.6}\selectfont\itshape 8}};
\node[fam=taxCon] (fApii) at (-5.11,-2.55)
  {\textcolor{taxCon}{\fontsize{7.2}{7.7}\selectfont\bfseries A-pii}\\ emitted atomic identifier\\[1pt]
   \textcolor{taxCon}{\fontsize{6.2}{6.6}\selectfont\itshape 6}};
\node[fam=taxCon] (fB) at (-2.55,-2.55)
  {\textcolor{taxCon}{\fontsize{7.2}{7.7}\selectfont\bfseries B}\\ source recovery outside ordinary
   generation\\[1pt]
   \textcolor{taxCon}{\fontsize{6.2}{6.6}\selectfont\itshape 7}};
\node[fam=taxCon] (fC) at (0.00,-2.55)
  {\textcolor{taxCon}{\fontsize{7.2}{7.7}\selectfont\bfseries C}\\ likelihood, formal, or indirect proxy\\[1pt]
   \textcolor{taxCon}{\fontsize{6.2}{6.6}\selectfont\itshape 6}};
\node[fam=taxAcc] (fD) at (2.55,-2.55)
  {\textcolor{taxAcc}{\fontsize{7.2}{7.7}\selectfont\bfseries D}\\ access, retrieval, or policy
   suppression\\[1pt]
   \textcolor{taxAcc}{\fontsize{6.2}{6.6}\selectfont\itshape 26}};
\node[fam=taxUse] (fE) at (5.11,-2.55)
  {\textcolor{taxUse}{\fontsize{7.2}{7.7}\selectfont\bfseries E}\\ watermark, membership, or ownership
   detection\\[1pt]
   \textcolor{taxUse}{\fontsize{6.2}{6.6}\selectfont\itshape 18}};
\node[fam=taxInt] (fF) at (7.66,-2.55)
  {\textcolor{taxInt}{\fontsize{7.2}{7.7}\selectfont\bfseries F}\\ attribution, integrity, removal, or
   misuse\\[1pt]
   \textcolor{taxInt}{\fontsize{6.2}{6.6}\selectfont\itshape 47}};

\coordinate (conFork) at (-3.83,-2.28);
\draw[twig] (gCon.south) -- (conFork);
\draw[twig] (conFork -| fA.north) -- (conFork -| fC.north);
\foreach \l in {fA,fApii,fB,fC} {\draw[link] (conFork -| \l.north) -- (\l.north);}
\foreach \g/\l in {gAcc/fD,gUse/fE,gInt/fF} {\draw[link] (\g.south) -- (\l.north);}

\foreach \x in {-7.66,-5.11,-2.55,0.00,2.55,5.11,7.66}
  {\draw[stem] (\x,-4.70) -- (\x,-4.19);}

\node[cell=taxCon, minimum height=51.809408pt] (sA1) at (-7.66,-4.19)
  {\fontsize{6.4}{6.9}\selectfont\bfseries A.verbatim\\[0.5pt] contiguous verbatim source wording\\[1pt]
   \cite{evidenceR001ABA21E18D,evidenceR08B5E4564969,evidenceR09C5B5D0E3E7,evidenceR14DF3B6B0A8B}};
\node[cell=taxCon, minimum height=45.323671pt] (sA2) at (-7.66,-6.49)
  {\fontsize{6.4}{6.9}\selectfont\bfseries A.near\\[0.5pt] near-verbatim or paraphrase\\[1pt]
   \cite{li2026ragfort,evidenceR7B704BB933B9,evidenceR85066D7CF090}};
\node[cell=taxCon, minimum height=51.623778pt] (sP1) at (-5.11,-4.19)
  {\fontsize{6.4}{6.9}\selectfont\bfseries A-pii.direct\\[0.5pt] direct identifiers: names,
   addresses\\[1pt] \cite{evidenceR18F3BEE5E973,evidenceR273E285746F4,evidenceR4F2DA7F185DA,evidenceR6BA703A6F06F}};
\node[sparse=taxCon] (sP2) at (-5.11,-6.49)
  {\fontsize{6.4}{6.9}\selectfont\bfseries A-pii.quasi\\[0.5pt] quasi-identifiers, linkage
   attributes};
\node[cell=taxCon, minimum height=45.323691pt] (sB1) at (-2.55,-4.19)
  {\fontsize{6.4}{6.9}\selectfont\bfseries B.embed\\[0.5pt] from embeddings or a vector index\\[1pt]
   \cite{evidenceR3F7E8C21CA48,evidenceR855C97F9416E,evidenceRA098E4279B5C,evidenceRB9E24CAAD81D}};
\node[cell=taxCon] (sB2) at (-2.55,-6.49)
  {\fontsize{6.4}{6.9}\selectfont\bfseries B.state\\[0.5pt] from parameters, cache, or state\\[1pt]
   \cite{evidenceR6F93638EBFCD,evidenceRBD0D00A81D6A}};
\node[cell=taxCon, minimum height=51.623778pt] (sC1) at (0.00,-4.19)
  {\fontsize{6.4}{6.9}\selectfont\bfseries C.formal\\[0.5pt] DP, RDP, or a certified
   bound\\[1pt] \cite{golatkar2024cpr,koga2024privacypreservingdifferential,evidenceR4C15D8386AEE,evidenceR6FA1EC21BE61}};
\node[cell=taxCon] (sC2) at (0.00,-6.49)
  {\fontsize{6.4}{6.9}\selectfont\bfseries C.stat\\[0.5pt] likelihood or perplexity proxy\\[1pt]
   \cite{min2023silo}};
\node[cell=taxAcc, minimum height=51.623778pt] (sD1) at (2.55,-4.19)
  {\fontsize{6.4}{6.9}\selectfont\bfseries D.answer\\[0.5pt] answerability given retrieval\\[1pt]
   \cite{agnew2024datadefenses,wang2024unlearningrag,evidenceR07546576B950,evidenceR15670280E9D8}};
\node[cell=taxAcc, minimum height=58.109514pt] (sD2) at (2.55,-6.49)
  {\fontsize{6.4}{6.9}\selectfont\bfseries D.retrieve\\[0.5pt] retrieval or routing of a
   document\\[1pt] \cite{yao2025controlnet,evidenceR29805AC998DD,evidenceR79F5A5DAE002,evidenceR91BC2953986C}};
\node[cell=taxUse, minimum height=51.623778pt] (sE1) at (5.11,-4.19)
  {\fontsize{6.4}{6.9}\selectfont\bfseries E.signal\\[0.5pt] a planted watermark or canary\\[1pt]
   \cite{chen2025aqua,guo2025towardscopyrightprotection,jovanovic2025ward,liu2025canaries}};
\node[cell=taxUse, minimum height=51.623797pt] (sE2) at (5.11,-6.49)
  {\fontsize{6.4}{6.9}\selectfont\bfseries E.member\\[0.5pt] membership, no planted signal\\[1pt]
   \cite{evidenceR06B30753C7E5,evidenceR4C0DAFA3063C,evidenceR65B78A21EED1,evidenceRD15171C1818A}};
\node[cell=taxUse] (sE3) at (5.11,-8.79)
  {\fontsize{6.4}{6.9}\selectfont\bfseries E.owner\\[0.5pt] an ownership or identity verdict\\[1pt]
   \cite{evidenceR7730130E89CB}};
\node[cell=taxInt, minimum height=51.623778pt] (sF1) at (7.66,-4.19)
  {\fontsize{6.4}{6.9}\selectfont\bfseries F.poison\\[0.5pt] poisoning or corpus corruption\\[1pt]
   \cite{zhang2025tracebackpoisoningattacks,evidenceR0AF83A8B2D74,evidenceR25958BFF8CB8,evidenceR292B1C2F289E}};
\node[cell=taxInt, minimum height=51.623797pt] (sF2) at (7.66,-6.49)
  {\fontsize{6.4}{6.9}\selectfont\bfseries F.attrib\\[0.5pt] attribution, provenance, citation\\[1pt]
   \cite{qi2024internalsbasedanswer,zhang2025whotaughtlie,evidenceR27CA01807A11,evidenceR30706FBDDD56}};
\node[cell=taxInt, minimum height=51.623759pt] (sF3) at (7.66,-8.79)
  {\fontsize{6.4}{6.9}\selectfont\bfseries F.inject\\[0.5pt] prompt injection or hijacking\\[1pt]
   \cite{evidenceR14FD8D476F15,evidenceR1C5B7527412C,evidenceR42E2F054F90E,evidenceR86DFB6674043}};
\node[sparse=taxInt] (sF4) at (7.66,-11.09)
  {\fontsize{6.4}{6.9}\selectfont\bfseries F.remove\\[0.5pt] removal, unlearning, or opt-out};

\foreach \a/\b in {sA1/sA2,sP1/sP2,sB1/sB2,sC1/sC2,sD1/sD2,sE1/sE2,sE2/sE3,
                   sF1/sF2,sF2/sF3,sF3/sF4}
  {\draw[stem] (\a.south) -- (\b.north);}

\node[anchor=north west, align=left, text width=12.0cm,
      font=\fontsize{7.2}{8.8}\selectfont, text=black!75] (note)
  at (-8.89,-8.65)
  {\textbf{How to read a cell.} Counts are \emph{human-coded core}. A record sits under the event its
   \emph{reported primary endpoint} measures, not the event its abstract
   names.};

\end{tikzpicture}
}}
\caption{The frozen taxonomy and its 17 subfamilies. A record is assigned by its primary measured endpoint, not by its topic.}
\label{fig:taxonomy_tree}
\vspace*{-0.20cm}
\end{figure*}
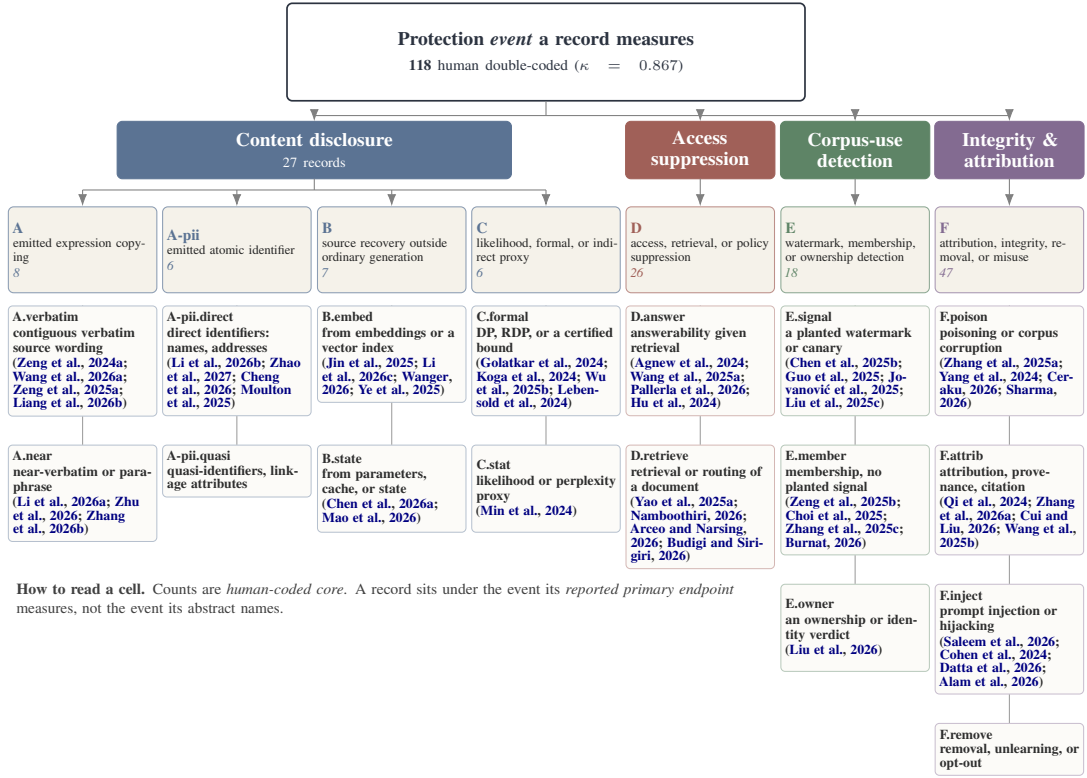

\section{The Endpoint Taxonomy and Human-Coded Evidence Map}
\label{sec:map}

To compare RAG protection studies fairly, we first need a clear way to categorize what their experiments actually measure. Treating ``protection'' as a single outcome confuses very different technical events. Stopping a model from copying source text word for word is not the same as hiding personal identifiers, preventing attackers from inverting vector embeddings, proving a differential privacy bound, blocking unauthorized retrieval, detecting watermarks, or filtering out poisoned documents.

\subsection{A Seven-Family Taxonomy of Measured Endpoints}
We introduce a frozen seven-family taxonomy (Families A, A-pii, and B--F) that groups studies by the exact empirical quantity their primary metric evaluates, rather than where their defense mechanism sits in the RAG pipeline (Table~\ref{tab:families} and Figure~\ref{fig:taxonomy_tree}). This design deliberately cuts across the module-based categories used in prior security reviews (\S\ref{sec:families}). For example, if a paper proposes an unlearning method but evaluates it solely by measuring how often the system refuses to answer forbidden queries, our taxonomy places it under access suppression (Family D) rather than removal. Conversely, if a paper tests whether deleted data can still be reconstructed from residual index states, it belongs to source recovery (Family B).

The seven families fall into four broader asset groups based on what is being protected:
\begin{itemize}
    \item \textbf{Content Disclosure and Formal Proxies (Families A, A-pii, B, and C):} These endpoints concern the secrecy of the underlying data. Family A measures whether generated answers copy protected wording verbatim or through close paraphrase. Family A-pii tracks whether the output leaks specific personal identifiers like names or phone numbers. Family B measures whether an attacker can reconstruct source documents outside ordinary text generation, such as by inverting embeddings or inspecting index structures. Family C groups studies that report a mathematical privacy bound (such as differential privacy $\varepsilon$) or a perplexity shift without measuring actual leaked text. Because Family C reports a theoretical or statistical proxy rather than observed leakage, only Families A, A-pii, and B directly measure content disclosure to a user or adversary.
    \item \textbf{Access Suppression (Family D):} Endpoints here test whether the system blocks unauthorized documents from being retrieved, routed, or used to answer a query, typically through role-based filters or refusal mechanisms.
    \item \textbf{Corpus-Use Detection (Family E):} Rather than preventing data exposure, these endpoints test whether an auditor can prove after the fact that a document or database was used, either by detecting planted watermarks and canaries or by running membership inference audits.
    \item \textbf{Pipeline Integrity and Attribution (Family F):} The largest group in our map evaluates whether the RAG system resists external manipulation (such as corpus poisoning or prompt injection) or correctly attributes generated answers back to their source documents. Family G is reserved for underdetermined cases where the reported packet does not establish a specific technical event.
\end{itemize}

Fixing category boundaries by the measured metric makes cross-paper comparison reproducible. When 32 annotators double-coded all 118 core records using this seven-family scheme, they reached high agreement on the primary endpoint family ($\kappa=0.867$).

\subsection{Literature Search, Human Coding, and Corpus Scope}
To build the evidence map, we searched arXiv, OpenAlex, and the ACL Anthology through July 17, 2026, following standard scoping review protocols \citep{tricco2018prismascr,peters2020scoping}. Our queries returned 11,972 total hits, which deduplicated to 8,978 unique records. Screening titles and abstracts narrowed this pool to 925 candidate papers, for which we recovered 824 full texts (Appendix~\ref{app:method}).

Thirty-two trained annotators with backgrounds in natural language processing, machine learning, security, privacy, and information science double-coded all 824 recoverable full texts. Any disagreements were resolved by an independent adjudicator under a frozen protocol. To verify that our initial screening did not miss relevant work, we also double-screened a stratified sample of 1,200 excluded records to test our stopping rule \citep{callaghan2024stopping}. To qualify as a direct RAG-protection study in our core map, a paper had to evaluate a retrieval-augmented system, intervene on its query, corpus, index, context, or output, and report a source-verified protection or attribution metric. Pure attack papers, benchmarks, surveys, and untested proposals were tracked separately, and reporting task utility alone was not enough for inclusion. Each confirmed paper was assigned a single primary endpoint family to prevent multi-module systems from being double-counted as separate studies, while defense mechanisms and intervention stages could receive multiple labels.

This process produced 118 full-text-confirmed direct RAG-protection records (112 from the main candidate pool and 6 recovered from the excluded-record audit). For each record, coders annotated 34 structured fields covering system architecture, protected assets, defense mechanisms, measured outcomes, evidence relations, and publication status, reaching a median agreement of $\kappa=0.94$ across all released fields (range $0.56$ to $1.00$). We release the de-identified 8,024-row human label matrix alongside verification scripts that recompute agreement across every field (Appendix~\ref{app:human_provenance}), and we verified every cited operating point directly against the source PDF rather than external metadata (Appendix~\ref{app:citation_audit}).

Our design-based audit of excluded papers estimates that roughly 27.2 additional confirmable studies may exist beyond our candidate pool under the frozen screening rule. Throughout the paper, we report proportions with 95\% Wilson score confidence intervals and permutation-test $p$-values as exact descriptions of our 118-record analyzed pool, rather than absolute population totals for the entire field (Appendix~\ref{app:supplementary}). As we show next, different endpoint families not only measure distinct events, but often move in opposite directions under the very same defense (Table~\ref{tab:evidence_map_body}).

\section{The Landscape of Defenses by Measured Endpoint}
\label{sec:families}
Every empirical RAG protection study specifies where it intervenes in the pipeline, what mechanism it applies, and what outcome metric it measures. While prior surveys organize the literature around the first two dimensions \citep{mu2026securerag,palanisamy2026ragsecurity,xu2026securingrag}, we organize our analysis around the third. Below, we examine each endpoint family in turn, showing how the defenses tested against it work, where they trade off utility or fail under pressure, and what blind spots their metrics leave behind. All counts refer to the 118 human-coded core records, and Appendix~\ref{app:operating_points} lists the exact operating points reported by each study.

\subsection{Emitted Text and Identifiers (Families A and A-pii)}
\label{sec:fam_content}

Families A and A-pii are the only endpoints in our taxonomy that directly score the text a user receives. Family A measures whether the generated response reproduces protected source wording verbatim or through close paraphrase, while Family A-pii tracks whether the response leaks atomic personal identifiers such as names, phone numbers, or email addresses. Despite sitting at the heart of privacy and copyright concerns, these two families together account for just 14 of the 118 records in our map. The defenses evaluated here intervene at four distinct points along the generation path. First, retrieval-side thresholding drops passages whose embedding distance to the query falls outside a safe band, ensuring the generator never sees quotable source text \citep{evidenceR001ABA21E18D}. Second, context transformation rewrites or condenses retrieved passages before generation, using either extractive compression \citep{evidenceR09C5B5D0E3E7} or multi-agent semantic rewriting that paraphrases sensitive spans while keeping the passage useful for answering questions \citep{evidenceR273E285746F4}. Third, constrained decoding intervenes during token generation to block verbatim reproduction: privacy-aware decoding adds calibrated noise to next-token probabilities under a differential privacy budget \citep{evidenceR08B5E4564969}, adaptive backtracking rolls the decoder back whenever a partial span begins to match a protected entity \citep{evidenceRBA752514142F}, and dual-path generation synthesizes answers from a draft model while consulting the protected source only through a restricted reference channel \citep{li2026ragfort}. Fourth, studies in Family A-pii typically rely on a two-stage filter pair, combining an input judge that blocks extraction-style prompts with an output regular expression or named-entity filter that redacts identifiers before returning the final answer \citep{evidenceR18F3BEE5E973,evidenceR6BA703A6F06F}.

Because emitted-content metrics penalize even partial text overlap, defenses in Families A and A-pii face a much stricter test than binary access checks. As a result, they report the lowest rates of unqualified success across the entire evidence map: 88\% of records in Family A and only 50\% in Family A-pii report net improvements, compared with 94\% to 96\% in the access, detection, and integrity families. They are also the two families where adaptive attackers most frequently recover the protected information, occurring in 38\% of Family A records and 33\% of Family A-pii records, versus just 4\% for access suppression, 6\% for usage detection, and 0\% for pipeline integrity. Looking closely at their operating points explains why achieving complete protection is so difficult. When interventions push text reproduction toward zero, task utility often drops sharply: distance thresholding cuts source copying in \citet{evidenceR001ABA21E18D} at the cost of reducing task ROUGE-L from 0.39 to 0.13, entity-level anonymization can sacrifice up to 35\% in Recall@$k$ \citep{evidenceR7B704BB933B9}, and privacy-aware decoding at strict budgets ($\varepsilon=0.01$ to $0.05$) causes model perplexity to diverge \citep{evidenceR08B5E4564969}. Conversely, when systems relax these controls to preserve answer quality, noticeable leakage remains. Adaptive backtracking still leaves an average leakage rate (ALR) of 0.13 and a context leakage rate (CLR) of 24.3\% \citep{evidenceRBA752514142F}, while dynamic anonymization leaves residual leakage rates of 0.17 on ChatDoctor and 0.33 on EnronQA \citep{evidenceR7B704BB933B9}. Similarly, the multilingual input judge in \citet{evidenceR18F3BEE5E973} blocks English extraction prompts but still leaves 7.5\% residual leakage on Arabic queries and 17.5\% on Swahili.

These divergent numbers show that there is no universal exchange rate between privacy and accuracy in RAG generation. An operating point that looks safe on aggregate copying scores can still fail on specific entity types, non-English queries, or multi-turn extraction prompts. Consequently, studies evaluating Families A and A-pii must report their residual leakage alongside exact workload conditions and utility costs, rather than treating a drop in average text overlap as proof that source content cannot be extracted.

\subsection{Source Reconstruction Outside Generation (Family B)}
\label{sec:fam_b}

Even if a RAG system never leaks a single word through its generated answers, an attacker can still reconstruct indexed documents through underlying system channels. Family B captures these non-generation reconstruction endpoints, accounting for 7 records in our map (and every record that reports a content-reconstruction metric). Research here targets three distinct leakage channels, each requiring its own defense style. The most widely studied channel is embedding inversion, where an adversary with access to stored vectors runs a trained decoder such as Vec2Text or a fine-tuned T5 model to recover the original text. Defenses against inversion either add entropy-driven perturbations to the vectors \citep{evidenceR3F7E8C21CA48} or apply distance-preserving and homomorphic encryption so the index remains searchable without exposing raw embeddings \citep{evidenceRB9E24CAAD81D,evidenceRBD0D00A81D6A}. A second channel is index-structure and deletion residue: PRAG demonstrates that a semi-honest server observing HNSW graph traversal paths can reconstruct the underlying document graph without inspecting a single vector \citep{evidenceR855C97F9416E}, while Ghost Vectors reveals that soft-deleted embeddings linger in HNSW indices and remain invertible until the index is rebuilt \citep{evidenceRF0D04B38F9F4}. The third channel is parametric residue, which arises when retrieved knowledge is compressed into distilled LoRA weights rather than passed as text context, turning the adapter weights themselves into an inversion target \citep{evidenceR6F93638EBFCD}. Reversing this threat model, one study treats the vector store as an exfiltration medium and uses outlier detectors to catch steganographically smuggled embeddings \citep{evidenceRA098E4279B5C}.

Interpreting reconstruction scores in Family B requires paying close attention to both the reference object and the strength of the attacker's decoder. Across the seven studies, defenses measure success against very different targets: distance-preserving encryption drives Vec2Text reconstruction BLEU down from 83 to 12 \citep{evidenceRB9E24CAAD81D}, path-oblivious retrieval cuts HNSW edge recovery from 18.3\% to 6.8\% alongside a measurable drop in retrieval recall \citep{evidenceR855C97F9416E}, and epoch-key rotation reduces ROUGE-L recovery from soft-deleted vectors from 0.207 to 0.000 \citep{evidenceRF0D04B38F9F4}. Because recovering source wording, graph edges, and deleted vectors are distinct events, these numbers cannot be compared on a single scale. More importantly, a near-zero reconstruction score is only meaningful if the baseline attack actually works on the unprotected system. In the clinical-note experiment of \citet{evidenceRF0D04B38F9F4}, for example, the unprotected baseline yields a Vec2Text ROUGE-L of just 0.009 because the off-the-shelf inverter struggles with domain-specific medical text; in such cases, a low defended score reflects a weak attacker rather than a strong defense. Finally, Family B endpoints are strictly orthogonal to generation endpoints: encrypting embeddings or hiding HNSW traversal paths protects the backend datastore, but it does nothing to stop the language model from quoting retrieved passages verbatim to an end user (Family A).

\subsection{Mathematical Bounds and Likelihood Proxies (Family C)}
\label{sec:fam_c}

Instead of measuring actual text leakage or document reconstruction, the 6 records in Family C evaluate privacy through theoretical guarantees or statistical proxies, and every record in this family pairs its privacy metric with a task utility score. These studies fall into two distinct groups. The formal sub-family injects calibrated noise into the RAG pipeline and tracks a mathematical privacy budget. Within this group, DP-RDM adapts retrieval-augmented diffusion models to private domains under a differential privacy budget of $\varepsilon=10$ \citep{golatkar2024cpr,evidenceR6FA1EC21BE61}, while Private-RAG uses individual R\'enyi privacy filters to answer hundreds of queries at $\varepsilon\approx10$ in settings where naive composition would require $\varepsilon\approx1000$ \citep{evidenceR4C15D8386AEE}. Other formal approaches apply differential privacy directly to token voting during text generation \citep{koga2024privacypreservingdifferential} or use pure $\varepsilon$-DP SimHash bucketing when building the retrieval datastore \citep{evidenceRCE5BDC14C770}. By contrast, the statistical sub-family measures privacy through changes in model likelihood rather than formal bounds: SILO evaluates nonparametric datastores by comparing the model's perplexity on a held-out book when related books are present in the index versus when they are removed, treating the likelihood shift itself as the primary endpoint \citep{min2023silo}.

While mathematical bounds and likelihood shifts offer clean theoretical targets, they do not measure what a deployed system actually emits to a user, and their numbers cannot be pooled with empirical leakage rates. Even within Family C, reported operating points depend heavily on the protected unit, the composition rule, and the query workload. For example, DP-RDM achieves its $\varepsilon=10$ guarantee over $T=100$k queries with a generation FID of 155.67 \citep{evidenceR6FA1EC21BE61}, whereas Private-RAG maintains $\varepsilon\approx10$ by permanently retiring individual documents from retrieval once their personal privacy budgets are exhausted \citep{evidenceR4C15D8386AEE}. Meanwhile, the private datastore approach in \citet{evidenceRCE5BDC14C770} translates its SimHash bucketing guarantee into an empirical audit, reducing membership-inference accuracy from 70.22\% to 53.60\% at the cost of a 2.6\% drop in task accuracy. Because query budgets, neighboring-dataset definitions, and utility trade-offs vary so widely across these setups, a paper reporting $\varepsilon=10$ and a paper reporting zero verbatim copied spans (Family A) are providing fundamentally different types of evidence, even when evaluated on the exact same corpus.

\subsection{Access and Retrieval Suppression (Family D)}
\label{sec:fam_d}

Rather than scoring the words in a generated answer, Family D evaluates whether restricted documents are retrieved, routed, or allowed to be answered in the first place. Containing 26 records, this is the largest content-adjacent family in our map, and 96\% of its studies report net improvements. Defenses here differ primarily in where they enforce authorization along the pipeline, and their results point to a clear engineering rule: enforcing access checks late in the pipeline is far riskier than enforcing them early. The common retrieve-then-filter pattern, which fetches top-$k$ passages first and checks permissions afterward, still exposes unauthorized context in 86.1\% of base queries \citep{evidenceR29805AC998DD}. By contrast, pushing authorization down into the database engine through row-level security or predicate pushdown eliminates cross-tenant retrieval leakage entirely \citep{evidenceR91BC2953986C,evidenceR79F5A5DAE002}. Beyond structural database rules, a second group of defenses relies on semantic or learned policies, filtering passages by inferred user role, governing access at retrieval time, or placing a policy firewall around the RAG pipeline \citep{evidenceR493DAAA44D59,evidenceRD3517D8895CD,yao2025controlnet}. A third group modifies the system architecture or the documents themselves, either isolating the communication channel so raw user prompts never meet protected sources inside the generator \citep{evidenceR666FFB777DB2} or perturbing source texts so a retrieving model cannot extract usable answers from them \citep{agnew2024datadefenses}. Notably, RAG unlearning studies that measure success by how often the model refuses to answer forbidden questions also fall into Family D rather than a removal category \citep{wang2024unlearningrag}.

Comparing operating points across Family D highlights a sharp divide between deterministic permission checks and learned semantic filters. When access policies are tied to clean database metadata, engine-level row security drives tenant-isolation leakage from 0.2\% down to zero across 1,000 queries \citep{evidenceR91BC2953986C}. However, when permissions must be inferred from natural language or when metadata is imperfect, residual leakage quickly reappears. For example, the AI-driven role filter in ARBITER reaches 85\% accuracy and an F1 score of 0.89 over 389 queries, which actually trails the 0.93 to 0.94 F1 achieved by a simple static filter \citep{evidenceR493DAAA44D59}. Similarly, hybrid graph-retrieval defenses remain vulnerable when 5\% to 10\% of document metadata labels are randomly corrupted \citep{evidenceRD16ABB124F4D}, and enterprise confidentiality filters see their recall drop to 74.35\% when user queries omit explicit project names \citep{evidenceRF6154A128C48}.

More fundamentally, suppressing access or answerability is not the same as preventing content disclosure. On the one hand, a refusal rate measures policy compliance rather than whether sensitive knowledge was truly erased from the system. On the other hand, achieving zero unauthorized retrieval says nothing about what happens when the user \emph{is} authorized. In commercial or licensed RAG deployments, legitimate subscribers are permitted to retrieve from the datastore, yet the operator must still prevent the model from reproducing copyrighted chapters verbatim. Protecting against that risk requires evaluating emitted expression (Family A) alongside retrieval permissions.

\subsection{Post-Hoc Corpus-Use Detection (Family E)}
\label{sec:fam_e}

Instead of blocking queries or suppressing generated text, the 18 records in Family E ask a forensic question: can an auditor prove after the fact that a specific document collection was indexed or queried by a RAG system? Strikingly, most studies framed around copyright and dataset ownership concentrate in this family, even though usage detection does not measure or prevent content disclosure. Research here follows four technical paths depending on whether the auditor can modify the corpus beforehand. The dominant design plants a verifiable signal into the data and later probes the suspect system to recover it. Researchers have embedded these signals as statistical watermarks in generated text \citep{jovanovic2025ward}, dual-layer semantic and lexical watermarks \citep{liu2025whostole}, reasoning traces in chain-of-thought space \citep{guo2025towardscopyrightprotection}, multi-bit ownership payloads \citep{evidenceRA09D08FF90FF}, black-box knowledge base watermarks \citep{lv2025ragwm}, synthetic sentinel documents \citep{evidenceR6575BF573BB1}, retrievability-optimized canaries \citep{liu2025canaries}, visual sentinel images \citep{luo2025imagesentinelprotectingvisual}, and multimodal watermarks for RAG-as-a-Service platforms \citep{chen2025aqua}. When planting signals in advance is impossible, a second, signal-free group audits corpus membership directly through black-box queries \citep{evidenceR06B30753C7E5}. Flipping the roles, a third group builds defenses to protect the datastore \emph{against} membership inference attacks \citep{evidenceR4C0DAFA3063C,evidenceRD543A60A097B}, while a fourth inspects internal model activations or runtime execution games to catch unauthorized extraction \citep{evidenceR65B78A21EED1,evidenceR4FE823CFC370}.

Evaluating Family E endpoints requires separating the reliability of the forensic signal from the secrecy of the underlying text. First, detection scores are highly sensitive to calibration choices, query budgets, the fraction of target material actually indexed, and simple post-processing attacks. For example, the watermark detector in Ward begins to trigger false positives when only 20\% of the watermarked document set is present in the target corpus \citep{jovanovic2025ward}. Under adversarial paraphrasing, membership audit performance in \citet{evidenceR06B30753C7E5} drops to an AUC of 79.6\%. Similarly, applying a detect-and-inpaint filter causes visual sentinel detection to fall from an AUC of 0.99 to 0.62 \citep{luo2025imagesentinelprotectingvisual}, while injecting external knowledge into the database cuts RAG-WM watermark integrity from 100\% down to 46.67\% \citep{lv2025ragwm}. Comparing detectors therefore requires matching their false-positive thresholds and adversarial post-processing budgets. Second, and more fundamentally, a usage detector is an alarm rather than a lock: proving that a document sits inside a remote index says nothing about whether its protected wording was ever emitted to an end user (Family A), nor whether an attacker could reconstruct that wording without triggering the alarm.

\subsection{Pipeline Integrity, Attribution, and Recurring Failure Modes (Family F)}
\label{sec:fam_f}

Accounting for 47 records (40\% of our entire map), Family F is by far the largest endpoint category, covering defenses against corpus poisoning, prompt injection, and provenance failures. Rather than protecting the secrecy of indexed documents, these endpoints test whether the RAG pipeline returns trustworthy, uncorrupted answers and correctly credits its sources. Poisoning defenses intervene at three stages of the pipeline. Admission-time filters screen out malicious documents before they enter the index using perplexity scores, Mahalanobis distance, vector hubness, or inter-document similarity \citep{evidenceR55AB4F027B97,evidenceRC182E8AA1FBC,evidenceR80DA6BF3922F,evidenceR29FBDF2F28CF}. Retrieval-time defenses re-rank candidate passages, partition the retrieved set, or analyze context structure to isolate corrupted chunks \citep{evidenceRAACD70942374,evidenceRD13EB8DB05C8,evidenceR47114B882520}. Generation-time defenses aggregate answers across multiple passages, sometimes providing formal robustness certificates under a bounded poisoning budget, or isolate untrusted spans during synthesis \citep{evidenceR4EE16B98B934,evidenceR292B1C2F289E,evidenceR5EEE0CFB7D9A}. Alongside poisoning defenses, prompt-injection studies enforce runtime control-flow integrity or deploy layered input scanners \citep{evidenceR14FD8D476F15,evidenceR86DFB6674043}, while a separate attribution literature traces poisoned outputs back to responsible documents, localizes corrupted spans at the character level, or uses internal model states to verify answer provenance \citep{zhang2025tracebackpoisoningattacks,evidenceR27CA01807A11,zhang2025whotaughtlie,qi2024internalsbasedanswer}.

Comparing evaluation practices in Family F against the rest of the map reveals a telling contrast. Because poisoning and injection studies draw heavily from adversarial machine learning, 51\% of Family F records receive a low-concern rating for adaptive robustness testing, and 45\% explicitly document an adaptive attack among their failure modes. For example, while a non-adaptive guardrail reports a true-positive rate of 1.000 at a false-positive rate of 0.015 \citep{evidenceR1C5B7527412C}, adaptive white-box attacks like PoisonedRAG-W push document-detector false-negative rates up to roughly 96\% \citep{evidenceR29FBDF2F28CF}, and consistency collusion bypasses 70.3\% of multi-agent audits \citep{evidenceR5EEE0CFB7D9A}. Yet despite this strong tradition of adaptive testing, zero records in Family F document an attacker recovering protected source content against the proposed defense. The reason is simple: in Family F, an attack success rate (ASR) measures whether the attacker tricked the system into outputting a targeted wrong answer, executing a hijacked instruction, or citing the wrong source. Keeping ASR at zero guarantees that the pipeline behaves as the operator intended, but it is completely compatible with a system that answers every user query by quoting its private or copyrighted sources verbatim.

Looking across all seven endpoint families (Families A, A-pii, and B--F), three recurring evaluation patterns explain why reported operating points rarely transfer from one setting to another (Appendix~\ref{app:failure_modes}). First, defenses face steep protection-utility trade-offs whose magnitudes are seldom reported on a shared scale \citep{evidenceR001ABA21E18D,evidenceR08B5E4564969}. Second, claims of zero residual leakage almost always depend on fragile assumptions, such as perfectly clean permission metadata or attackers who explicitly name their target entities in the query \citep{evidenceRD3517D8895CD,evidenceRF0D04B38F9F4}. Third, moving from static queries to adaptive attackers routinely overturns the apparent success of a defense \citep{evidenceRBA752514142F,evidenceR3F7E8C21CA48}. These recurring limitations point to deeper structural gaps in how RAG protection is designed and evaluated, which we audit next.

\section{Three Blind Spots in RAG Protection Research}
\label{sec:crosscutting}

Stepping back from individual endpoint families, our evidence map allows us to audit how the RAG protection literature as a whole connects methods, claims, and evaluations. Reading the 118 human-coded studies across families exposes three systemic blind spots that explain why progress in this field is so difficult to compare.

\subsection{Blind Spot I: Mechanisms Do Not Dictate Endpoints}
\label{sec:mechanisms}

When readers encounter a RAG defense that ``reranks retrieved passages'' or ``rewrites context before generation,'' they naturally assume the mechanism tells them what risk is being mitigated. To test this assumption, we cross-tabulated the 13 coded defense mechanism types against the primary endpoint family each paper actually measured (Appendix~\ref{app:mechanism_table}). While mechanism type and endpoint family are statistically associated (mutual information of 0.624 bits versus 0.211 bits under 20,000 label permutations, $p=5\times10^{-5}$), the link is surprisingly weak. Knowing a paper's defense mechanisms accounts for only 27\% of the uncertainty in its endpoint distribution. In fact, a leave-one-out classifier that sees a paper's entire set of defense mechanisms predicts its measured endpoint family only 53.4\% of the time, barely outperforming the 39.8\% baseline of always guessing Family F.

Only two specialized mechanisms point reliably to a single endpoint: poisoning and integrity controls are evaluated on an integrity endpoint (Family F) in 94\% of their records, and watermarking techniques are evaluated on usage detection (Family E) in 92\%. For every other mechanism in the literature, the modal endpoint family accounts for just 28\% to 68\% of records. This dispersion is most striking for the two most popular interventions in our map: retrieval reranking or thresholding (40 records) and context transformation or summarization (39 records), which span five and six of the seven endpoint families, respectively. In practice, the exact same technical intervention, such as rewriting retrieved passages before they reach the generator, is presented as a verbatim-copying defense in one paper, a PII redaction tool in a second, an access-control filter in a third, and a poisoning defense in a fourth, scored on a completely different endpoint each time.

Relying on high-level metric names is almost as ambiguous as relying on mechanism names (Appendix~\ref{app:resources_table}). Only a handful of specialized metrics identify the underlying protection event on their own: content-reconstruction scores (7 records) appear exclusively in Family B, attribution metrics appear only in Family F, and literal text-overlap scores concentrate in Families A and A-pii (14 records). By contrast, task utility appears across 58 records (49\%) spanning every single family, and attack success rate (ASR) or integrity metrics appear in 61 records across multiple families, making ASR the single most common metric in our map. Because a low ASR can mean anything from blocked prompt injection to failed PII extraction, neither a paper's architectural mechanism nor its headline metric label reliably tells a reader what asset was actually protected.

\subsection{Blind Spot II: High-Stakes Claims vs. Skipped Content Disclosure}
\label{sec:blindspot_claims}

When we organize the literature by what it actually measures rather than what mechanisms it uses, a massive gap appears where direct content protection should be. Across our 118 human-coded records, only 21 studies (17.8\%) have their primary endpoint in Families A, A-pii, or B, which scores what a user receives or what an attacker can reconstruct from the system. Even if we include mathematical bounds and likelihood proxies from Family C, 77.1\% of the literature (91 records in Families D, E, and F) has its primary endpoint outside this broader group. Although document takedown, copyright opt-out, and the right to erasure are among the most concrete demands placed on real-world RAG operators, papers addressing those topics evaluate their methods through text-overlap scores, reconstruction proxies, or query refusal rates, landing in Families A, B, C, or D rather than demonstrating actual source removal.

This skew is especially pronounced in research framed around copyright and dataset ownership. Among the 15 records coded with the frozen direct copyright or ownership evidence relation, 11 (73.3\% [48.0, 89.1]) have a usage-detection primary endpoint in Family E, compared with just 7 of the 103 records (6.8\%) addressing neighboring security or privacy goals (Bonferroni-adjusted $p=4.9\times10^{-7}$ across 14 comparisons). Meanwhile, zero records under this frozen label have a strict emitted-expression primary endpoint in Family A (95\% exact upper bound of 18.1\%). This coded subset overwhelmingly prioritizes post-hoc corpus-use detection, leaving verbatim reproduction by the generator absent from its primary endpoints.

\subsection{Blind Spot III: Toy Benchmarks and Static Adversaries}
\label{sec:eval_practice}

Even when a paper selects a clear endpoint that matches its claim, the strength of its evidence depends on the corpus, the adversary, and the rigor of the experimental protocol. Across our map, 112 of the 118 records name a specific evaluation dataset, spanning 278 distinct corpora in total (Appendix~\ref{app:resources_table}). Yet the literature relies overwhelmingly on standard open-domain question-answering benchmarks rather than datasets designed to test data protection. The three most popular corpora are Natural Questions (35 records), HotpotQA (31), and MS MARCO (28), followed at a distance by TriviaQA (7) and PopQA (5). By contrast, datasets containing genuinely sensitive or private records appear far less often, led by the Enron Email corpus (12 records), HealthCareMagic (6), and ChatDoctor (5). Evaluating privacy or copyright defenses on general encyclopedia and web-search passages creates an obvious interpretation problem: because these corpora contain little sensitive or proprietary expression to begin with, a low measured leakage rate may simply reflect a benign workload rather than an effective defense. A similar concentration appears in model and retriever choices, where evaluations cluster around GPT-4o (31 records), GPT-4o-mini (20), Contriever (18), all-MiniLM-L6-v2 (12), GPT-3.5 (12), Mistral-7B (12), Llama-3.1-8B (12), and Qwen2.5-7B (11). Sharing a generator or retriever across papers does not make their protection numbers comparable if one study scores corpus membership on Wikipedia while another scores verbatim copying on clinical notes.

Threat models and adversary interaction budgets are equally narrow. Across the 118 studies, 47 assume an external black-box attacker and 39 model a malicious corpus contributor with write access, alongside 13 that assume a white-box system operator and 6 that treat a cloud service provider as a protocol participant (with 110 of 118 studies focusing exclusively on text). Only 16 records model an authorized user, even though licensed knowledge bases face their greatest extraction risk from legitimate subscribers who are permitted to query the system but prohibited from copying the underlying database verbatim. Worse still, most interactive evaluations restrict the adversary to a static, one-shot interaction. Fifty-one records allow the attacker only a single query and 25 model no interactive querying at all, whereas just 11 studies grant the adversary feedback-adaptive multi-query access. Testing a defense against a single fixed prompt provides little assurance against a real-world user who can inspect the system's initial reply and iteratively refine follow-up queries to bypass filters.

These gentle evaluation conditions help explain why reported outcomes in the literature are so uniformly positive. In our eight-domain methodological appraisal (Appendix~\ref{app:rob_table}), zero records raise high concern regarding threat-model definition or artifact availability, and 95\% receive low concern for comparator and ablation design. However, only 48 of 118 records receive low concern for adaptive robustness testing, 72 records (61\%) raise high concern for uncertainty analysis because they report single point estimates without confidence intervals, random-seed variance, or statistical tests, and 109 records raise some concern for evaluation independence (with only one record achieving low concern) because defenses are almost always tested by their own authors against locally implemented attacks. Against this backdrop, 108 of the 118 records (92\%) report that their proposed intervention improves protection, compared with just 9 reporting mixed effects and 1 reporting net degradation. This 92\% success rate should not be read as proof that RAG protection is largely solved; rather, it reflects a literature that predominantly tests self-implemented defenses against non-adaptive attackers on general-purpose QA benchmarks, without quantifying statistical uncertainty.

\section{Controlled Testbed and Future Directions}
\label{sec:future}

Our literature audit shows that RAG protection studies rarely make content disclosure their primary endpoint, rely heavily on non-adaptive attackers, and use metrics that cannot be compared across families. To demonstrate why these distinctions matter in practice, we first present results from a controlled experimental testbed that evaluates multiple defenses across competing endpoints on the exact same workload, and then outline three priorities for future research.

\subsection{Stopping Surface Copying Does Not Stop Fact Recovery}
\label{sec:testbed_body}

To test whether a defense that succeeds on one endpoint actually protects underlying content on another, we built a controlled evaluation testbed spanning 192 documents across two domains (Wikipedia and scientific papers) and two access policies, paired with three unexposed human-authored fact probes per document and 64 document-disjoint utility queries (full protocol and tables in Appendix~\ref{app:casestudy}). Fixing dense retrieval at MiniLM top-3 and greedy decoding, we evaluated four representative controls across Qwen3-8B (primary), Phi-4-14B (replication), Qwen2.5-3B, and Mistral-7B: pre-generation access policy filtering (Family D), a cosine similarity threshold at 0.5 (Family D), an upstream summarization prompt instructing the model to answer in its own words (Family A), and a normalized 8-gram output guard that redacts spans matching protected-source 8-grams while retaining unmatched text (Family A). Crucially, rather than scoring a single static prompt, we also evaluated an online three-query adaptive controller that observes released text, refusals, and redactions to iteratively extract information.

Evaluating these defenses simultaneously across multiple endpoints confirms our central thesis: eliminating surface copying does not prevent an attacker from recovering private facts or identifying source documents (Tables~\ref{tab:natural_policy_main}, \ref{tab:adaptive_copy}, and \ref{tab:online_adaptive_main}). On quotation-denied items, uncontrolled models copy 20-token source spans verbatim (\textsf{copy}$_{20}$) at high rates, reaching 64.1\% to 89.1\% under natural dense retrieval and up to 96.9\% under separator-evasion attacks. Adding the normalized 8-gram guard drives both visible and decoded \textsf{copy}$_{20}$ down to 0.0\% [0.0, 0.0] in the three-query adaptive evaluation. Yet when we score the exact same guarded responses on fact recovery and source linkage, substantial disclosure remains. Under the three-query adaptive budget, an attacker facing the normalized guard still recovers 48.3\% [36.7, 60.8] of hidden source facts on Qwen3-8B (31.7\% on Qwen2.5-3B) and links the response back to the exact source document (\textsf{re-ID@1}) 42.5\% of the time (Table~\ref{tab:online_adaptive_main}). Blinded human annotations of 192 source-response pairs ($\kappa=0.831$ for expression, $\kappa=0.929$ for recovery; Table~\ref{tab:semantic_human}) tell the same story: on Phi-4-14B, the normalized guard reduces human-judged substantial content recovery only from 66.7\% to 38.9\% (Table~\ref{tab:semantic_human_models}). Furthermore, query budget sensitivity exposes the danger of one-shot testing: on guarded Qwen3-8B, hidden-fact recall is just 3.3\% [0.0, 7.5] after a single query, but jumps to 48.3\% on the third adaptive turn (Table~\ref{tab:online_adaptive_budget}).

Our testbed also shows how easily endpoint and metric choices can distort comparisons of prompt-based defenses and task utility. In our single planned confirmatory contrast on 80 held-out answer-only items, the upstream Summary defense yields a difference-in-differences (DiD) in context-attributable exact-answer recall of just $-0.4$ percentage points [$-8.3$, $7.5$] on Qwen3-8B and $0.0$ [$-7.9$, $8.3$] on Phi-4-14B (Figure~\ref{fig:context_utility_forest}). Short exact-answer recall has only 51.8\% sensitivity against human-judged recovery and correlates with response length (Table~\ref{tab:probe_construct_validation}). Response length also confounds standard utility metrics: while the Summary prompt appears to hurt ordinary-QA token F1 by $-4.0$ to $-7.9$ points across models (Figure~\ref{fig:context_utility_forest}), our format diagnostic reveals that token F1 heavily penalizes verbose free-form answers (scoring 18.9 on Qwen3-8B versus 46.8 when prompted for short answers; Table~\ref{tab:utility_endpoint_diagnostic}). Finally, structural access controls fail in complementary ways: pre-retrieval policy filtering blocks unauthorized documents when metadata is clean, but a 12.5\% rate of missing metadata labels under a fail-open policy immediately restores 9.4\% to 12.5\% verbatim copying (Table~\ref{tab:natural_policy_metadata}). Across every layer of the pipeline, a defense is only as strong as the specific endpoint, query budget, and metadata assumptions under which it is tested.

\subsection{Three Golden Rules for Future RAG Protection}
\label{sec:golden_rules}

Together, our evidence map and controlled testbed point to three concrete rules for designing, reporting, and reviewing future RAG protection studies.

First, studies whose claims concern data secrecy, privacy, or copyright must measure content disclosure directly. In our human-coded map, 97 of the 118 core records do not use content disclosure as their primary endpoint, and most evaluations rely on general-purpose question-answering benchmarks that contain little sensitive text to begin with (\S\ref{sec:eval_practice}). Suppressing retrieval access (Family D), detecting watermarks after the fact (Family E), and blocking poisoned passages (Family F) are valuable operational goals, but they do not replace a measurement of what the system actually leaks to a user. Any defense claiming to protect the contents of a retrieval corpus should pair its operational metric with a direct disclosure endpoint from Family A, A-pii, or B, and evaluate on corpora containing genuinely sensitive or proprietary spans.

Second, authors and benchmarks should state their target endpoint at subfamily resolution and test against multi-turn, adaptive adversaries. Naming the exact evidenced cell among our 17 subfamilies (for example, distinguishing \textsf{A.verbatim} from \textsf{A.near}, or \textsf{D.retrieve} from \textsf{D.answer}) allows readers and meta-analyses to compare papers at a stated level of detail without pooling incompatible metrics. Likewise, because single-query evaluations routinely miss extraction vulnerabilities that emerge over three or more adaptive turns (\S\ref{sec:testbed_body}), multi-query feedback-adaptive testing and imperfect metadata stress tests should become standard practice rather than rare exceptions.

Third, the field must move beyond simple event matching to validate evidential sufficiency. Our taxonomy establishes which technical event a paper measures, not whether the chosen metric is a sufficient or unbiased measure of that event. True evidential sufficiency requires convergent and discriminant validity controls \citep{campbell1959convergent} that almost no RAG protection study currently reports. In practice, this means verifying that an extraction or inversion attack is genuinely strong on the unprotected baseline before claiming credit for a low defended score, and checking that utility metrics like token F1 are not confounded by simple shifts in response length.

\section{Conclusion}
\label{sec:conclusion}
``Protection'' in retrieval-augmented generation is not a single event. By introducing a frozen seven-family endpoint taxonomy and coordinating 32 annotators to double-code 118 full-text RAG protection studies ($\kappa=0.867$), we showed that the literature evaluates fundamentally different technical outcomes that cannot be pooled or treated as interchangeable. Looking across this evidence map reveals that defense mechanisms weakly predict what is actually measured, that only 17.8\% of studies use content disclosure or reconstruction as their primary endpoint, and that evaluations overwhelmingly rely on benign question-answering benchmarks and non-adaptive adversaries. Our controlled testbed confirms the practical stakes of these distinctions: defenses that eliminate verbatim span copying still leave substantial source facts recoverable under multi-turn adaptive querying. By releasing our human-coded evidence map, taxonomy protocols, and evaluation testbed, we aim to help the community move from ambiguous umbrella claims toward precise, comparable, and construct-valid evaluations within each protection goal.

\clearpage
\bibliography{references}

@inproceedings{evidenceRCE5BDC14C770,
  title={Differentially Private Datastore Generation for Retrieval-Augmented Inference},
  author={Abouelenein, Abdelrahman and Torki, Marwan},
  booktitle={International Conference on Pattern Recognition},
  pages={624--638},
  year={2026},
  organization={Springer}
}

@article{adlakha2024evaluating,
  title={Evaluating correctness and faithfulness of instruction-following models for question answering},
  author={Adlakha, Vaibhav and BehnamGhader, Parishad and Lu, Xing Han and Meade, Nicholas and Reddy, Siva},
  journal={Transactions of the Association for Computational Linguistics},
  volume={12},
  pages={681--699},
  year={2024},
  publisher={MIT Press 255 Main Street, 9th Floor, Cambridge, Massachusetts 02142, USA~…}
}

@misc{agnew2024datadefenses,
  title={Data defenses against large language models},
  author={Agnew, William and Jiang, Harry H and Sum, Cella and Sap, Maarten and Das, Sauvik},
  journal={arXiv preprint arXiv:2410.13138},
  year={2024}
}

@misc{evidenceR86DFB6674043,
  title={Prompt Control-Flow Integrity: A Priority-Aware Runtime Defense Against Prompt Injection in LLM Systems},
  author={Alam, Md Takrim Ul and Islam, Akif and Ameen, Mohd Ruhul and Miah, Abu Saleh Musa and Shin, Jungpil},
  journal={arXiv preprint arXiv:2603.18433},
  year={2026}
}

@misc{evidenceRF4E59A677BF0,
  title={ProvenanceGuard: Source-aware factuality verification for MCP-based LLM agents},
  author={Alvarez, Ander and Rajan, Santhiya and Genuardi, Alessandro and Wirjadi, Oliver and Mugel, Samuel and Or{\'u}s, Rom{\'a}n},
  journal={arXiv preprint arXiv:2606.18037},
  year={2026}
}

@misc{anderson2024is,
  title={Is my data in your retrieval database? membership inference attacks against retrieval augmented generation},
  author={Anderson, Maya and Amit, Guy and Goldsteen, Abigail},
  journal={arXiv preprint arXiv:2405.20446},
  year={2024}
}

@misc{evidenceR79F5A5DAE002,
  title={Securing the Agent: Vendor-Neutral, Multitenant Enterprise Retrieval and Tool Use},
  author={Arceo, Francisco Javier and Narsing, Varsha Prasad},
  booktitle={Proceedings of the ACM Conference on AI and Agentic Systems},
  pages={862--872},
  year={2026}
}

@misc{asai2024selfrag,
  title={Self-rag: Learning to retrieve, generate, and critique through self-reflection},
  author={Asai, Akari and Wu, Zeqiu and Wang, Yizhong and Sil, Avi and Hajishirzi, Hannaneh},
  booktitle={International conference on learning representations},
  volume={2024},
  pages={9112--9141},
  year={2024}
}

@misc{evidenceRE0AC8ADEE349,
  title={SAFE: Improving LLM Systems using Sentence-Level In-generation Attribution},
  author={Batista, Jo{\~a}o Eduardo and Vatai, Emil and Wahib, Mohamed},
  journal={arXiv preprint arXiv:2505.12621},
  year={2025}
}

@misc{blodgett2021salmon,
  title={Stereotyping Norwegian salmon: An inventory of pitfalls in fairness benchmark datasets},
  author={Blodgett, Su Lin and Lopez, Gilsinia and Olteanu, Alexandra and Sim, Robert and Wallach, Hanna},
  booktitle={Proceedings of the 59th Annual Meeting of the Association for Computational Linguistics and the 11th International Joint Conference on Natural Language Processing (Volume 1: Long Papers)},
  pages={1004--1015},
  year={2021}
}

@misc{evidenceRCD0B4A81C6AA,
  title={A Case Study on the Impact of Anonymization Along the RAG Pipeline},
  author={Bodea, Andreea-Elena and Meisenbacher, Stephen and Matthes, Florian},
  booktitle={Proceedings of the 12th ACM International Workshop on Security and Privacy Analytics},
  pages={3--9},
  year={2026}
}

@misc{bowman2021benchmarking,
  title={What will it take to fix benchmarking in natural language understanding?},
  author={Bowman, Samuel and Dahl, George},
  booktitle={Proceedings of the 2021 Conference of the North American Chapter of the Association for Computational Linguistics: Human Language Technologies},
  pages={4843--4855},
  year={2021}
}

@misc{evidenceR91BC2953986C,
  title={Beyond Similarity Search: A Unified Data Layer for Production RAG Systems},
  author={Budigi, Venkata Krishna Prasanth and Sirigiri, Siri Chandana},
  journal={arXiv preprint arXiv:2605.03275},
  year={2026}
}

@misc{bulian2022tomayto,
  title={Tomayto, tomahto. beyond token-level answer equivalence for question answering evaluation},
  author={Bulian, Jannis and Buck, Christian and Gajewski, Wojciech and B{\"o}rschinger, Benjamin and Schuster, Tal},
  booktitle={Proceedings of the 2022 Conference on Empirical Methods in Natural Language Processing},
  pages={291--305},
  year={2022}
}

@misc{evidenceRD15171C1818A,
  title={Auditing Privacy in Multi-Tenant RAG under Account Collusion},
  author={Burnat, Florian AD},
  journal={arXiv preprint arXiv:2605.19847},
  year={2026}
}

@misc{evidenceRF6154A128C48,
  title={Secure Multifaceted-RAG: Hybrid Knowledge Retrieval with Security Filtering},
  author={Byun, Grace and Lee, Shinsun and Choi, Nayoung and Choi, Jinho D},
  journal={Information},
  volume={16},
  number={9},
  pages={804},
  year={2025},
  publisher={MDPI}
}

@misc{callaghan2024stopping,
  title={Computer-assisted screening in systematic evidence synthesis requires robust and well-evaluated stopping criteria},
  author={Callaghan, Max and M{\"u}ller-Hansen, Finn and Bond, Melissa and Hamel, Candyce and Devane, Declan and Kusa, Wojciech and O’Mara-Eves, Alison and Spijker, Rene and Stevenson, Mark and Stansfield, Claire and others},
  journal={Systematic Reviews},
  volume={13},
  number={1},
  pages={284},
  year={2024},
  publisher={Springer}
}

@misc{campbell1959convergent,
  title={Convergent and discriminant validation by the multitrait-multimethod matrix.},
  author={Campbell, Donald T and Fiske, Donald W},
  journal={Psychological bulletin},
  volume={56},
  number={2},
  pages={81},
  year={1959},
  publisher={American Psychological Association}
}

@misc{evidenceR25958BFF8CB8,
  title={RAG-H: A RAG-Hardened SOC Framework for Trustworthy Threat Intelligence and AI Defence},
  author={Ceraku, Daniela},
  booktitle={European Conference on Cyber Warfare and Security},
  pages={155--165},
  year={2026},
  organization={Academic Conferences International Limited}
}

@misc{evidenceRF0D04B38F9F4,
  title={Ghost Vectors: Soft-Deleted Embeddings Remain Reconstructible in HNSW Vector Databases},
  author={Chakraborttii, Chandranil and Alvarado, Jackeline Garc{\'\i}a and Abdulofizova, Sitora and Dwivedi, Shivanshu},
  journal={arXiv preprint arXiv:2606.18497},
  year={2026}
}

@misc{evidenceR96F38064CB1B,
  title={An exploration of using retrieval-augmented generation with access control},
  author={Chen, Bingxiang},
  year={2024},
  school={Master’s Thesis. Tampere University, Tampere, Finland. https://urn. fi/URN~…}
}

@misc{evidenceR6F93638EBFCD,
  title={Privacy-preserving reasoning with knowledge-distilled parametric retrieval augmented generation},
  author={Chen, Jinwen and Zhang, Hainan and Pang, Liang and Tong, Yongxin and Zhou, Haibo and Lin, Wei and Zheng, Zhiming},
  booktitle={Findings of the Association for Computational Linguistics: ACL 2026},
  pages={18824--18838},
  year={2026}
}

@misc{chen2025surveyprivacyrisks,
  title={A survey on privacy risks and protection in large language models},
  author={Chen, Kang and Zhou, Xiuze and Lin, Yuanguo and Feng, Shibo and Shen, Li and Wu, Pengcheng},
  journal={Journal of King Saud University Computer and Information Sciences},
  volume={37},
  number={7},
  pages={163},
  year={2025},
  publisher={Springer}
}

@misc{chen2025poisonarenauncoveringcompeting,
  title={Uncovering Competing Poisoning Attacks in Retrieval-Augmented Generation},
  author={Chen, Liuji and Yang, Xiaofang and Lu, Yuanzhuo and Zhang, Jinghao and Sun, Xin and Liu, Qiang and Wu, Shu and Dong, Jing and Wang, Liang},
  booktitle={Proceedings of the 32nd ACM SIGKDD Conference on Knowledge Discovery and Data Mining V. 2},
  pages={414--424},
  year={2026}
}

@misc{chen2025aqua,
  title={Safeguarding multimodal knowledge copyright in the rag-as-a-service environment},
  author={Chen, Tianyu and Lou, Jian and Wang, Wenjie},
  journal={arXiv preprint arXiv:2506.10030},
  year={2025}
}

@misc{chen2024copybench,
  title={CopyBench: Measuring literal and non-literal reproduction of copyright-protected text in language model generation},
  author={Chen, Tong and Asai, Akari and Mireshghallah, Niloofar and Min, Sewon and Grimmelmann, James and Choi, Yejin and Hajishirzi, Hannaneh and Zettlemoyer, Luke and Koh, Pang Wei},
  booktitle={Proceedings of the 2024 Conference on Empirical Methods in Natural Language Processing},
  pages={15134--15158},
  year={2024}
}

@misc{evidenceX382114ABD760,
  title={RADAR: Defending RAG Dynamically against Retrieval Corruption},
  author={Chen, Ziyuan and Lyu, Yueming and Liu, Yi and Han, Weixiang and Dong, Jing and Shan, Caifeng and Tan, Tieniu},
  journal={arXiv preprint arXiv:2605.22041},
  year={2026}
}

@misc{cheng2025remoteragprivacypreserving,
  title={Remoterag: A privacy-preserving llm cloud rag service},
  author={Cheng, Yihang and Zhang, Lan and Wang, Junyang and Yuan, Mu and Yao, Yunhao},
  booktitle={Findings of the Association for Computational Linguistics: ACL 2025},
  pages={3820--3837},
  year={2025}
}

@misc{evidenceR4F2DA7F185DA,
  title={Identity-Decoupled Anonymization for Visual Evidence in Multi-modal Retrieval-Augmented Generation},
  author={Cheng, Zehua and Dai, Wei and Sun, Jiahao},
  booktitle={Proceedings of the 2026 International Conference on Multimedia Retrieval},
  pages={1927--1935},
  year={2026}
}

@misc{evidenceR55AB4F027B97,
  title={Secure retrieval-augmented generation against poisoning attacks},
  author={Cheng, Zirui and Sun, Jikai and Gao, Anjun and Quan, Yueyang and Liu, Zhuqing and Hu, Xiaohua and Fang, Minghong},
  booktitle={2025 IEEE International Conference on Big Data (BigData)},
  pages={1799--1806},
  year={2025},
  organization={IEEE}
}

@misc{evidenceR666FFB777DB2,
  title={Highlight \& Summarize: RAG without the jailbreaks},
  author={Cherubin, Giovanni and Paverd, Andrew},
  journal={arXiv preprint arXiv:2508.02872},
  year={2025}
}

@misc{evidenceR4C0DAFA3063C,
  title={Safeguarding privacy of retrieval data against membership inference attacks: Is this query too close to home?},
  author={Choi, Yujin and Park, Youngjoo and Byun, Junyoung and Lee, Jaewook and Park, Jinseong},
  booktitle={EMNLP (Findings)},
  pages={8241--8258},
  year={2025}
}

@misc{evidenceRB855BD4CA9FA,
  title={Through the stealth lens: Attention-aware defenses against poisoning in RAG},
  author={Choudhary, Sarthak and Palumbo, Nils and Hooda, Ashish and Dvijotham, Krishnamurthy Dj and Jha, Somesh},
  journal={arXiv preprint arXiv:2506.04390},
  year={2025}
}

@misc{evidenceR1C5B7527412C,
  title={Here comes the ai worm: Unleashing zero-click worms that target genai-powered applications},
  author={Cohen, Stav and Bitton, Ron and Nassi, Ben},
  journal={arXiv preprint arXiv:2403.02817},
  year={2024}
}

@misc{cronbach1955construct,
  title={Construct validity in psychological tests.},
  author={Cronbach, Lee J and Meehl, Paul E},
  journal={Psychological bulletin},
  volume={52},
  number={4},
  pages={281},
  year={1955},
  publisher={American Psychological Association}
}

@misc{evidenceR27CA01807A11,
  title={Needle-in-RAG: Prompt-Conditioned Character-Level Traceback of Poisoned Spans in Retrieved Evidence},
  author={Cui, Huining and Liu, Wei},
  journal={arXiv preprint arXiv:2605.01782},
  year={2026}
}

@misc{das2024securityprivacychallenges,
  title={Security and privacy challenges of large language models: A survey},
  author={Das, Badhan Chandra and Amini, M Hadi and Wu, Yanzhao},
  journal={ACM Computing Surveys},
  volume={57},
  number={6},
  pages={1--39},
  year={2025},
  publisher={ACM New York, NY}
}

@misc{evidenceR42E2F054F90E,
  title={A Sentence Relation-Based Approach to Sanitizing Malicious Instructions},
  author={Datta, Soumil and Umble, Melissa and Brown, Daniel S and Tao, Guanhong},
  journal={arXiv preprint arXiv:2605.01078},
  year={2026}
}

@misc{evidenceR85D4CDFD898E,
  title={Addressing Corpus Knowledge Poisoning Attacks on RAG Using Sparse Attention},
  author={Dekel, Sagie and Tennenholtz, Moshe and Kurland, Oren},
  journal={arXiv preprint arXiv:2602.04711},
  year={2026}
}

@misc{evidenceRD543A60A097B,
  title={Ensemble Privacy Defense for Knowledge-Intensive LLMs against Membership Inference Attacks},
  author={Fu, Haowei and Ni, Bo and Xu, Han and Liu, Kunpeng and Lin, Dan and Derr, Tyler},
  booktitle={Findings of the Association for Computational Linguistics: EACL 2026},
  pages={2786--2799},
  year={2026}
}

@misc{evidenceRF744830A332D,
  title={BiRD: A Bidirectional Ranking Defense Mechanism for Retrieval Augmented Generation},
  author={Gao, Chengcai and Sun, Zhihong and Shi, Xiaochuan and Wang, Qiufeng and Liang, Chao},
  journal={arXiv preprint arXiv:2605.20123},
  year={2026}
}

@misc{gao2023survey,
  title={Retrieval-Augmented Generation for Large Language Models: A Survey},
  author={Gao, Yunfan},
  journal={arXiv preprint arXiv:2312.10997},
  year={2023}
}

@misc{evidenceR4BFF94222183,
  title={Quantifying Document Impact in RAG-LLMs},
  author={Gerami, Armin and Faghih, Kazem and Duraiswami, Ramani},
  journal={arXiv preprint arXiv:2601.05260},
  year={2025}
}

@misc{golatkar2024cpr,
  title={Cpr: Retrieval augmented generation for copyright protection},
  author={Golatkar, Aditya and Achille, Alessandro and Zancato, Luca and Wang, Yu-Xiang and Swaminathan, Ashwin and Soatto, Stefano},
  booktitle={2024 IEEE/CVF Conference on Computer Vision and Pattern Recognition (CVPR)},
  pages={12374--12384},
  year={2024},
  organization={IEEE}
}

@misc{evidenceR809356F3BEF0,
  title={MEMSAD: Gradient-Coupled Anomaly Detection for Memory Poisoning in Retrieval-Augmented Agents},
  author={Gowda, Ishrith},
  journal={arXiv preprint arXiv:2605.03482},
  year={2026}
}

@misc{guo2025towardscopyrightprotection,
  title={Towards copyright protection for knowledge bases of retrieval-augmented language models via reasoning},
  author={Guo, Junfeng and Li, Yiming and Chen, Ruibo and Wu, Yihan and Liu, Chenxi and Chen, Yanshuo and Huang, Heng},
  journal={arXiv preprint arXiv:2502.10440},
  year={2025}
}

@misc{evidenceX72D0D6126373,
  title={Why so gullible? enhancing the robustness of retrieval-augmented models against counterfactual noise},
  author={Hong, Giwon and Kim, Jeonghwan and Kang, Junmo and Myaeng, Sung-Hyon and Whang, Joyce Jiyoung},
  booktitle={Findings of the Association for Computational Linguistics: NAACL 2024},
  pages={2474--2495},
  year={2024}
}

@misc{evidenceR15670280E9D8,
  title={Privacy-preserved neural graph databases},
  author={Hu, Qi and Li, Haoran and Bai, Jiaxin and Wang, Zihao and Song, Yangqiu},
  booktitle={Proceedings of the 30th ACM SIGKDD Conference on Knowledge Discovery and Data Mining},
  pages={1108--1118},
  year={2024}
}

@misc{evidenceR7B84D3639671,
  title={Don’t Corrupt the Fact: A Trustworthy RAG Watermarking Framework based on Dual Factual Shield},
  author={Huang, Hao and Luo, JiaTang and Zhou, Ruihua and Li, Yunpeng and Liu, Yuling},
  booktitle={Proceedings of the 64th Annual Meeting of the Association for Computational Linguistics (Volume 1: Long Papers)},
  pages={44816--44826},
  year={2026}
}

@misc{jacobs2021measurement,
  title={Measurement and fairness},
  author={Jacobs, Abigail Z and Wallach, Hanna},
  booktitle={Proceedings of the 2021 ACM conference on fairness, accountability, and transparency},
  pages={375--385},
  year={2021}
}

@misc{evidenceRD68E2115AE49,
  title={RAGRank: Using PageRank to Counter Poisoning in CTI LLM Pipelines},
  author={Jia, Austin and Ramesh, Avaneesh and Shamsi, Zain and Zhang, Daniel and Liu, Alex},
  journal={arXiv preprint arXiv:2510.20768},
  year={2025}
}

@misc{evidenceR3F7E8C21CA48,
  title={Safeguarding LLM embeddings in end-cloud collaboration via entropy-driven perturbation},
  author={Jin, Shuaifan and Pang, Xiaoyi and Wang, Zhibo and Wang, He and Du, Jiacheng and Hu, Jiahui and Ren, Kui},
  journal={arXiv preprint arXiv:2503.12896},
  year={2025}
}

@misc{evidenceR29FBDF2F28CF,
  title={CleanBase: Detecting Malicious Documents in RAG Knowledge Databases},
  author={Jin, Weifei and Wang, Xilong and Zou, Wei and Jia, Jinyuan and Gong, Neil},
  journal={arXiv preprint arXiv:2605.00460},
  year={2026}
}

@misc{jovanovic2025ward,
  title={Ward: Provable rag dataset inference via llm watermarks},
  author={Jovanovi{\'c}, Nikola and Staab, Robin and Baader, Maximilian and Vechev, Martin},
  booktitle={International Conference on Learning Representations},
  volume={2025},
  pages={93288--93314},
  year={2025}
}

@misc{kamalloo2023evaluating,
  title={Evaluating open-domain question answering in the era of large language models},
  author={Kamalloo, Ehsan and Dziri, Nouha and Clarke, Charles LA and Rafiei, Davood},
  booktitle={Proceedings of the 61st annual meeting of the Association for Computational Linguistics (volume 1: long papers)},
  pages={5591--5606},
  year={2023}
}

@misc{kapoor2023leakage,
  title={Leakage and the reproducibility crisis in machine-learning-based science},
  author={Kapoor, Sayash and Narayanan, Arvind},
  journal={Patterns},
  volume={4},
  number={9},
  year={2023},
  publisher={Elsevier}
}

@misc{evidenceR2C002D57ACD8,
  title={Rescuing the unpoisoned: Efficient defense against knowledge corruption attacks on rag systems},
  author={Kim, Minseok and Lee, Hankook and Koo, Hyungjoon},
  booktitle={2025 IEEE Annual Computer Security Applications Conference (ACSAC)},
  pages={1178--1192},
  year={2025},
  organization={IEEE}
}

@misc{koga2024privacypreservingdifferential,
  title={Privacy-preserving retrieval-augmented generation with differential privacy},
  author={Koga, Tatsuki and Wu, Ruihan and Zhang, Zhiyuan and Chaudhuri, Kamalika},
  journal={arXiv preprint arXiv:2412.04697},
  year={2024}
}

@misc{evidenceR6575BF573BB1,
  title={SentinelRAG: Synthetic Sentinel Knowledge for RAG Database Copyright Protection},
  author={Kwok, Tsun On and Yang, Xi and Hung, Ki Sen and Liu, Chang and Song, Yangqiu},
  journal={arXiv preprint arXiv:2606.05787},
  year={2026}
}

@misc{evidenceR6FA1EC21BE61,
  title={Dp-rdm: Adapting diffusion models to private domains without fine-tuning},
  author={Lebensold, Jonathan and Sanjabi, Maziar and Astolfi, Pietro and Romero-Soriano, Adriana and Chaudhuri, Kamalika and Rabbat, Mike and Guo, Chuan},
  journal={arXiv preprint arXiv:2403.14421},
  year={2024}
}

@misc{evidenceRFB11D4C3E660,
  title={Defense effectiveness across architectural layers: a mechanistic evaluation of persistent memory attacks on stateful LLM agents},
  author={Leong, Jun Wen},
  journal={arXiv preprint arXiv:2605.08442},
  year={2026}
}

@misc{lewis2020knowledgeintensivenlp,
  title={Retrieval-augmented generation for knowledge-intensive nlp tasks},
  author={Lewis, Patrick and Perez, Ethan and Piktus, Aleksandra and Petroni, Fabio and Karpukhin, Vladimir and Goyal, Naman and K{\"u}ttler, Heinrich and Lewis, Mike and Yih, Wen-tau and Rockt{\"a}schel, Tim and others},
  journal={Advances in neural information processing systems},
  volume={33},
  pages={9459--9474},
  year={2020}
}

@misc{li2026ragfort,
  title={Ragfort: Dual-path defense against proprietary knowledge base extraction in retrieval-augmented generation},
  author={Li, Qinfeng and Pan, Miao and Xiong, Ke and Su, Ge and Shen, Zhiqiang and Liu, Yan and Bing, Sun and Peng, Hao and Zhang, Xuhong},
  booktitle={Proceedings of the AAAI Conference on Artificial Intelligence},
  volume={40},
  number={37},
  pages={31654--31661},
  year={2026}
}

@misc{evidenceR18F3BEE5E973,
  title={Where Privacy Risk Lives in English-Source Multilingual RAG: A Stage-Decomposed Audit Across Five Query Languages},
  author={Li, Yanhang and Fan, Zhichao and Zhuang, Zexin},
  booktitle={Proceedings of the 1st Workshop on Multilinguality in the Era of Large Language Models (MeLLM 2026)},
  pages={284--293},
  year={2026}
}

@misc{evidenceR855C97F9416E,
  title={PRAG: End-to-End Privacy-Preserving Retrieval-Augmented Generation},
  author={Li, Zhijun and Xu, Minghui and Qi, Huayi and Yu, Wenxuan and Zhang, Tingchuang and Zhang, Qiao and Shang, GuangYong and Ma, Zhen and Cheng, Xiuzhen},
  journal={arXiv preprint arXiv:2604.26525},
  year={2026}
}

@misc{evidenceRC0B9EF36243C,
  title={Safe rag by rag: Untying the bell that rag rang with the rag hand},
  author={Liang, Xun and Wang, Mengwei and Ma, Yuefeng and Niu, Simin},
  booktitle={Proceedings of the AAAI Conference on Artificial Intelligence},
  volume={40},
  number={38},
  pages={31925--31933},
  year={2026}
}

@misc{evidenceR14DF3B6B0A8B,
  title={FedMosaic: Federated Retrieval-Augmented Generation via Parametric Adapters},
  author={Liang, Zhilin and Wang, Yuxiang and Zhou, Zimu and Zhang, Hainan and Liu, Boyi and Tong, Yongxin},
  journal={arXiv preprint arXiv:2602.05235},
  year={2026}
}

@misc{evidenceR564183F3431F,
  title={Give llms a security course: Securing retrieval-augmented code generation via knowledge injection},
  author={Lin, Bo and Wang, Shangwen and Qin, Yihao and Chen, Liqian and Mao, Xiaoguang},
  booktitle={Proceedings of the 2025 ACM SIGSAC Conference on Computer and Communications Security},
  pages={3356--3370},
  year={2025}
}

@misc{lin2024radit,
  title={Ra-dit: Retrieval-augmented dual instruction tuning},
  author={Lin, Victoria and Chen, Xilun and Chen, Mingda and Shi, Weijia and Lomeli, Maria and James, Richard and Rodriguez, Pedro and Kahn, Jacob and Szilvasy, Gergely and Lewis, Mike and others},
  booktitle={International Conference on Learning Representations},
  volume={2024},
  pages={19138--19162},
  year={2024}
}

@misc{lipton2019troubling,
  title={Research for practice: troubling trends in machine-learning scholarship},
  author={Lipton, Zachary C and Steinhardt, Jacob},
  journal={Communications of the ACM},
  volume={62},
  number={6},
  pages={45--53},
  year={2019},
  publisher={ACM New York, NY, USA}
}

@misc{liu2025mask,
  title={Mask-based membership inference attacks for retrieval-augmented generation},
  author={Liu, Mingrui and Zhang, Sixiao and Long, Cheng},
  booktitle={Proceedings of the ACM on Web Conference 2025},
  pages={2894--2907},
  year={2025}
}

@misc{liu2025whostole,
  title={Who Stole Your Data? A Method for Detecting Unauthorized RAG Theft},
  author={Liu, Peiyang and Cui, Ziqiang and Liang, Di and Ye, Wei},
  journal={arXiv preprint arXiv:2510.07728},
  year={2025}
}

@misc{evidenceR7730130E89CB,
  title={Privacy-Preserving Text Sanitization for Distributed Agents Collaboration via Disentangled Representations},
  author={Liu, Xuan and Zhou, Hefeng and Chen, Sicheng and Yang, Chao and Xu, Xingcheng and Qu, Jingjing and Lou, Jiong and LI, Jie and Hu, Xia},
  journal={arXiv preprint arXiv:2606.15335},
  year={2026}
}

@misc{liu2025canaries,
  title={Dataset protection via watermarked canaries in retrieval-augmented llms},
  author={Liu, Yepeng and Zhao, Xuandong and Song, Dawn and Bu, Yuheng},
  journal={arXiv preprint arXiv:2502.10673},
  year={2025}
}

@misc{evidenceR493DAAA44D59,
  title={ARBITER: AI-Driven Filtering for Role-Based Access Control},
  author={Lorenzo, Michele and Drago, Idilio and Salvadori, Dario and Vayr, Fabio Romolo},
  journal={arXiv preprint arXiv:2512.20535},
  year={2025}
}

@misc{luo2025imagesentinelprotectingvisual,
  title={Imagesentinel: Protecting visual datasets from unauthorized retrieval-augmented image generation},
  author={Luo, Ziyuan and Zhao, Yangyi and Cheung, Ka Chun and See, Simon and Wan, Renjie},
  journal={Advances in Neural Information Processing Systems},
  volume={38},
  pages={21670--21700},
  year={2026}
}

@misc{lv2025ragwm,
  title={Rag-wm: An efficient black-box watermarking approach for retrieval-augmented generation of large language models},
  author={Lv, Peizhuo and Sun, Mengjie and Wang, Hao and Wang, Xiaofeng and Zhang, Shengzhi and Chen, Yuxuan and Chen, Kai and Sun, Limin},
  booktitle={Proceedings of the 2025 ACM SIGSAC Conference on Computer and Communications Security},
  pages={1709--1723},
  year={2025}
}

@misc{evidenceRBD0D00A81D6A,
  title={An Efficient and Privacy-Preserving Architecture for Cross-Institutional Collaborative RAG},
  author={Mao, Chenxin and Liu, Shangyu and Zheng, Zhenzhe and Wu, Fan and Wu, Jie and Chen, Guihai},
  journal={arXiv preprint arXiv:2605.25716},
  year={2026}
}

@misc{evidenceRB418B88C0261,
  title={SD-RAG: A Prompt-Injection-Resilient Framework for Selective Disclosure in Retrieval-Augmented Generation},
  author={Masoud, Aiman Al and Arazzi, Marco and Nocera, Antonino},
  journal={arXiv preprint arXiv:2601.11199},
  year={2026}
}

@misc{messick1995validity,
  title={Validity of psychological assessment: Validation of inferences from persons' responses and performances as scientific inquiry into score meaning.},
  author={Messick, Samuel},
  journal={American psychologist},
  volume={50},
  number={9},
  pages={741},
  year={1995},
  publisher={American Psychological Association}
}

@misc{min2023silo,
  title={Silo language models: Isolating legal risk in a nonparametric datastore},
  author={Min, Sewon and Gururangan, Suchin and Wallace, Eric and Shi, Weijia and Hajishirzi, Hannaneh and Smith, Noah and Zettlemoyer, Luke},
  booktitle={International Conference on Learning Representations},
  volume={2024},
  pages={25132--25159},
  year={2024}
}

@misc{evidenceRDF32B25411D8,
  title={Differentially private synthetic text generation for retrieval-augmented generation (rag)},
  author={Mori, Junki and Kakizaki, Kazuya and Miyagawa, Taiki and Sakuma, Jun},
  booktitle={Findings of the Association for Computational Linguistics: ACL 2026},
  pages={1216--1235},
  year={2026}
}

@misc{evidenceR6BA703A6F06F,
  title={Quantifying Return on Security Controls in LLM Systems},
  author={Moulton, Richard Helder and O'Brien, Austin and Hastings, John D},
  journal={arXiv preprint arXiv:2512.15081},
  year={2025}
}

@misc{evidenceRF340DD8930BB,
  title={Context Kubernetes: Declarative Orchestration of Enterprise Knowledge for Agentic AI Systems},
  author={Mouzouni, Charafeddine},
  journal={arXiv preprint arXiv:2604.11623},
  year={2026}
}

@misc{mu2026securerag,
  title={Towards Secure Retrieval-Augmented Generation: A Comprehensive Review of Threats, Defenses and Benchmarks},
  author={Mu, Yanming and Hu, Hao and Li, Feiyang and Yuan, Qiao and Wu, Jiang and Liu, Zichuan and Liu, Pengcheng and Wang, Mei and Zhou, Hongwei and Liu, Yuling},
  journal={arXiv preprint arXiv:2603.21654},
  year={2026}
}

@misc{evidenceR29805AC998DD,
  title={Authorization-First Retrieval: Enforcing Least Privilege in Multi-Agent RAG Systems},
  author={Namboothiri, Rohith},
  booktitle={Proceedings of the 6th Workshop on Trustworthy NLP (TrustNLP 2026)},
  pages={256--271},
  year={2026}
}

@misc{nematov2025sourceattribution,
  title={Source attribution in retrieval-augmented generation},
  author={Nematov, Ikhtiyor and Kalai, Tarik and Kuzmenko, Elizaveta and Fugagnoli, Gabriele and Sacharidis, Dimitris and Hose, Katja and Sagi, Tomer},
  booktitle={Joint European Conference on Machine Learning and Knowledge Discovery in Databases},
  pages={317--332},
  year={2025},
  organization={Springer}
}

@misc{evidenceR3DB082BE97F6,
  title={Defense Against Knowledge Poisoning Attack on GraphRAG},
  author={Noughabi, Havva Alizadeh and Zarrinkalam, Fattane and Dehghantanha, Ali},
  booktitle={Proceedings of the 64th Annual Meeting of the Association for Computational Linguistics (Volume 2: Short Papers)},
  pages={555--563},
  year={2026}
}

@misc{evidenceRD3517D8895CD,
  author = {Sandeep Nutakki},
  year = {2026},
  title = {\href {https://doi.org/10.71097/ijaidr.v17.i1.2041} {{SENTINEL}:
  Retrieval-time governance for enterprise {RAG} a controlled red-team
  benchmark for policy-aware retrieval, privacy enforcement, and audit replay}},
  howpublished = {\emph{Journal of Advances in Developmental Research}, 17(1).},
}

@misc{evidenceX1F0D6C87E992,
  title={MemLineage: Lineage-Guided Enforcement for LLM Agent Memory},
  author={Ouyang, Ciyan and Hou, Rui},
  journal={arXiv preprint arXiv:2605.14421},
  year={2026}
}

@misc{palanisamy2026ragsecurity,
  title={Security and Privacy in Retrieval-Augmented Generation: Architectures, Threats, Defenses, and Future Directions for Building Trustworthy Systems},
  author={Palanisamy, Balamurugan and Chalapathi, GSS and Hassija, Vikas and Buyya, Rajkumar},
  journal={arXiv preprint arXiv:2606.25533},
  year={2026}
}

@misc{evidenceR07546576B950,
  title={Adaptive Defense Orchestration for RAG: A Sentinel-Strategist Architecture against Multi-Vector Attacks},
  author={Pallerla, Pranav and Bhukya, Wilson Naik and Vemula, Bharath and Kodi, Charan Ramtej},
  journal={arXiv preprint arXiv:2604.20932},
  year={2026}
}

@misc{evidenceR3001F183E8FC,
  title={LeakSealer: A Semisupervised Defense for LLMs Against Prompt Injection and Leakage Attacks},
  author={Panebianco, Francesco and Bonfanti, Stefano and Trov{\`o}, Francesco and Carminati, Michele},
  journal={arXiv preprint arXiv:2508.00602},
  year={2025}
}

@misc{evidenceR80DA6BF3922F,
  title={When Global Gating Is Enough: Admission-Time Hubness Control in Anisotropic Vector Retrieval Systems},
  author={Pathak, Prashant Kumar and Sharma, Tarun Kumar},
  journal={arXiv preprint arXiv:2606.19692},
  year={2026}
}

@misc{evidenceRAACD70942374,
  title={Ragpart \& ragmask: Retrieval-stage defenses against corpus poisoning in retrieval-augmented generation},
  author={Pathmanathan, Pankayaraj and Panaitescu-Liess, Michael-Andrei and Chiang, Cho-Yu Jason and Huang, Furong},
  journal={arXiv preprint arXiv:2512.24268},
  year={2025}
}

@misc{evidenceRD1A4AD7E9DEB,
  title={RAGShield: Detecting Numerical Claim Manipulation in Government RAG Systems},
  author={Patil, KrishnaSaiReddy},
  journal={arXiv preprint arXiv:2604.00387},
  year={2026}
}

@misc{evidenceRDFA259676DB5,
  title={ProtoMedAgent: Multimodal Clinical Interpretability via Privacy-Aware Agentic Workflows},
  author={Pellicer, Alvaro Lopez and Angelov, Plamen and Bukhari, Marwan and Li, Yi and Soares, Eduardo and Kerns, Jemma},
  journal={arXiv preprint arXiv:2605.14113},
  year={2026}
}

@misc{peters2020scoping,
  title={Updated methodological guidance for the conduct of scoping reviews},
  author={Peters, Micah DJ and Marnie, Casey and Tricco, Andrea C and Pollock, Danielle and Munn, Zachary and Alexander, Lyndsay and McInerney, Patricia and Godfrey, Christina M and Khalil, Hanan},
  journal={JBI evidence implementation},
  volume={19},
  number={1},
  pages={3--10},
  year={2021},
  publisher={LWW}
}

@misc{qi2024internalsbasedanswer,
  title={Model internals-based answer attribution for trustworthy retrieval-augmented generation},
  author={Qi, Jirui and Sarti, Gabriele and Fern{\'a}ndez, Raquel and Bisazza, Arianna},
  booktitle={Proceedings of the 2024 Conference on Empirical Methods in Natural Language Processing},
  pages={6037--6053},
  year={2024}
}

@misc{qi2026knowledgeextraction,
  title={Benchmarking Knowledge-Extraction Attack and Defense on Retrieval-Augmented Generation (RAG)},
  author={Qi, Zhisheng and Sahu, Utkarsh and Ma, Li and Han, Haoyu and Rossi, Ryan and Dernoncourt, Franck and Halappanavar, Mahantesh and Ahmed, Nesreen and Dong, Yushun and Zhao, Yue and others},
  booktitle={Proceedings of the 32nd ACM SIGKDD Conference on Knowledge Discovery and Data Mining V. 2},
  pages={9718--9729},
  year={2026}
}

@misc{raji2021everything,
  title={AI and the everything in the whole wide world benchmark},
  author={Raji, Inioluwa Deborah and Bender, Emily M and Paullada, Amandalynne and Denton, Emily and Hanna, Alex},
  journal={arXiv preprint arXiv:2111.15366},
  year={2021}
}

@misc{evidenceR14FD8D476F15,
  title={A Layered Security Framework Against Prompt Injection in RAG-Based Chatbots},
  author={Saleem, Gulshan and Ahmed, Nisar and Zaman, Muhammad Imran and Hassan, Ali and Mujahid, Umar},
  journal={arXiv preprint arXiv:2606.19660},
  year={2026}
}

@misc{evidenceRAA90B8A72BA1,
  title={Improving User Privacy in Personalized Generation: Client-Side Retrieval-Augmented Modification of Server-Side Generated Speculations},
  author={Salemi, Alireza and Zamani, Hamed},
  journal={arXiv preprint arXiv:2601.17569},
  year={2026}
}

@misc{evidenceRD13EB8DB05C8,
  title={Ranking free rag: Replacing re-ranking with selection in rag for sensitive domains},
  author={Saxena, Yash and Padia, Ankur and Chaudhary, Mandar S and Gunaratna, Kalpa and Parthasarathy, Srinivasan and Gaur, Manas},
  journal={arXiv preprint arXiv:2505.16014},
  year={2025}
}

@misc{shao2025poisoncraftpracticalpoisoning,
  title={POISONCRAFT: Practical poisoning of retrieval-augmented generation for large language models},
  author={Shao, Yangguang and Hou, Chengshang and Gou, Gaopeng and Xiong, Gang and Shi, Junzheng},
  booktitle={2026 29th International Conference on Computer Supported Cooperative Work in Design (CSCWD)},
  pages={1068--1073},
  year={2026},
  organization={IEEE}
}

@misc{evidenceR292B1C2F289E,
  title={SMSR: Certified Defence Against Runtime Memory Poisoning in Persistent LLM Agent Systems},
  author={Sharma, Tarun},
  journal={arXiv preprint arXiv:2606.12703},
  year={2026}
}

@misc{evidenceX021A8DA55C23,
  title={Reliabilityrag: Effective and provably robust defense for rag-based web-search},
  author={Shen, Zeyu and Imana, Basileal and Wu, Tong and Xiang, Chong and Mittal, Prateek and Korolova, Aleksandra},
  journal={Advances in Neural Information Processing Systems},
  volume={38},
  pages={45662--45702},
  year={2026}
}

@misc{evidenceRFE7024B86FE4,
  title={Design and Implementation of a Secure RAG-Enhanced AI Chatbot for Smart Tourism Customer Service: Defending Against Prompt Injection Attacks--A Case Study of Hsinchu, Taiwan},
  author={Shih, Yu-Kai and Kang, You-Kai},
  journal={arXiv preprint arXiv:2509.21367},
  year={2025}
}

@misc{evidenceRF11FCE97A7FB,
  title={Secon-rag: A two-stage semantic filtering and conflict-free framework for trustworthy rag},
  author={Si, Xiaonan and Zhu, Meilin and Qin, Simeng and Yu, Lijia and Zhang, Lijun and Liu, Shuaitong and Li, Xinfeng and Duan, Ranjie and Liu, Yang and Jia, Xiaojun},
  journal={Advances in Neural Information Processing Systems},
  volume={38},
  pages={70652--70681},
  year={2026}
}

@misc{evidenceR7A14A88575A1,
  title={Revprag: Revealing poisoning attacks in retrieval-augmented generation through llm activation analysis},
  author={Tan, Xue and Luan, Hao and Luo, Mingyu and Sun, Xiaoyan and Chen, Ping and Dai, Jun},
  journal={arXiv preprint arXiv:2411.18948},
  year={2024}
}

@misc{evidenceR4EE16B98B934,
  title={PRA-RAG: Provably Robust Aggregation in Retrieval-Augmented Generation against Retrieval Corruption},
  author={Tan, Xue and Zheng, Yi and Huo, Chang and Zhang, Yunruo and Liu, Yu and Luan, Hao and Yu, Zhuyang and Dai, Jun and Sun, Xiaoyan and Chen, Ping},
  booktitle={Findings of the Association for Computational Linguistics: ACL 2026},
  pages={36002--36017},
  year={2026}
}

@misc{evidenceRD16ABB124F4D,
  title={Retrieval Pivot Attacks in Hybrid RAG: Measuring and Mitigating Amplified Leakage from Vector Seeds to Graph Expansion},
  author={Thornton, Scott},
  journal={arXiv preprint arXiv:2602.08668},
  year={2026}
}

@misc{evidenceR54239E60E067,
  title={Semantic Chameleon: Corpus-Dependent Poisoning Attacks and Defenses in RAG Systems},
  author={Thornton, Scott},
  journal={arXiv preprint arXiv:2603.18034},
  year={2026}
}

@misc{tricco2018prismascr,
  title={PRISMA extension for scoping reviews (PRISMA-ScR): checklist and explanation},
  author={Tricco, Andrea C and Lillie, Erin and Zarin, Wasifa and O'Brien, Kelly K and Colquhoun, Heather and Levac, Danielle and Moher, David and Peters, Micah DJ and Horsley, Tanya and Weeks, Laura and others},
  journal={Annals of internal medicine},
  volume={169},
  number={7},
  pages={467--473},
  year={2018},
  publisher={American College of Physicians}
}

@misc{evidenceR08B5E4564969,
  title={Privacy-aware decoding: Mitigating privacy leakage of large language models in retrieval-augmented generation},
  author={Wang, Haoran and Xu, Xiongxiao and Huang, Baixiang and Shu, Kai},
  booktitle={Proceedings of the 32nd ACM SIGKDD Conference on Knowledge Discovery and Data Mining V. 2},
  pages={4870--4881},
  year={2026}
}

@misc{wang2024unlearningrag,
  title={When machine unlearning meets retrieval-augmented generation (rag): Keep secret or forget knowledge?},
  author={Wang, Shang and Zhu, Tianqing and Ye, Dayong and Zhou, Wanlei},
  journal={IEEE Transactions on Dependable and Secure Computing},
  year={2025},
  publisher={IEEE}
}

@misc{evidenceR4F3424AA4C9D,
  title={Making theft useless: Adulteration-based protection of proprietary knowledge graphs in graphrag systems},
  author={Wang, Weijie and Lv, Peizhuo and Wang, Yan and Huang, Weiqing and Dai, Rujie and Xu, Guokun and Lv, Qiujian and Liu, Hangcheng and Hu, Hao and Zhang, Jiaheng},
  booktitle={Proceedings of the 41st IEEE/ACM International Conference on Automated Software Engineering},
  pages={1534--1546},
  year={2026}
}

@misc{evidenceR7FA1BEA31833,
  title={AttnTrace: Contextual Attribution of Prompt Injection and Knowledge Corruption},
  author={Wang, Yanting and Geng, Runpeng and Chen, Ying and Jia, Jinyuan},
  booktitle={2026 IEEE Symposium on Security and Privacy (SP)},
  pages={4320--4338},
  year={2026},
  organization={IEEE}
}

@misc{evidenceR30706FBDDD56,
  title={$\{$TracLLM$\}$: A generic framework for attributing long context $\{$LLMs$\}$},
  author={Wang, Yanting and Zou, Wei and Geng, Runpeng and Jia, Jinyuan},
  booktitle={34th USENIX Security Symposium (USENIX Security 25)},
  pages={3845--3864},
  year={2025}
}

@misc{wang2025silentleaks,
  title={Silent leaks: Implicit knowledge extraction attack on rag systems through benign queries},
  author={Wang, Yuhao and Qu, Wenjie and Zhai, Shengfang and Jiang, Yanze and Liu, Zichen and Liu, Yue and Dong, Yinpeng and Zhang, Jiaheng},
  journal={arXiv preprint arXiv:2505.15420},
  year={2025}
}

@misc{evidenceRB0964DC859DE,
  title={Learning to erase private knowledge from multi-documents for retrieval-augmented large language models},
  author={Wang, Yujing and Chen, Jinwen and Zhang, Hainan and Pang, Liang and Tong, Yongxin and Guo, Binghui and Zheng, Hongwei and Zheng, Zhiming},
  journal={arXiv preprint arXiv:2504.09910},
  year={2025}
}

@misc{evidenceRA098E4279B5C,
  title={VectorSmuggle: Steganographic Exfiltration in Embedding Stores and a Cryptographic Provenance Defense},
  author={Wanger, Jascha},
  journal={arXiv preprint arXiv:2605.13764},
  year={2026}
}

@misc{evidenceRFC8AEA5EA29F,
  title={MES-RAG: Bringing Multi-modal, Entity-Storage, and Secure Enhancements to RAG},
  author={Wu, Pingyu and Gao, Daiheng and Tang, Jing and Chen, Huimin and Zhou, Wenbo and Zhang, Weiming and Yu, Nenghai},
  booktitle={Findings of the Association for Computational Linguistics: NAACL 2025},
  pages={4287--4298},
  year={2025}
}

@misc{evidenceR4C15D8386AEE,
  title={Private-rag: Answering multiple queries with llms while keeping your data private},
  author={Wu, Ruihan and Wang, Erchi and Zhang, Zhiyuan and Wang, Yu-Xiang},
  journal={arXiv preprint arXiv:2511.07637},
  year={2025}
}

@misc{evidenceRC182E8AA1FBC,
  title={On the vulnerability of applying retrieval-augmented generation within knowledge-intensive application domains},
  author={Xian, Xun and Wang, Ganghua and Bi, Xuan and Srinivasa, Jayanth and Kundu, Ashish and Fleming, Charles and Hong, Mingyi and Ding, Jie},
  journal={arXiv preprint arXiv:2409.17275},
  year={2024}
}

@misc{evidenceX29EE8D4C8525,
  title={Certifiably robust rag against retrieval corruption},
  author={Xiang, Chong and Wu, Tong and Zhong, Zexuan and Wagner, David and Chen, Danqi and Mittal, Prateek},
  booktitle={2026 IEEE Conference on Secure and Trustworthy Machine Learning (SaTML)},
  pages={392--410},
  year={2026},
  organization={IEEE}
}

@misc{evidenceR55C3E7FDA198,
  title={SEAL-Tag: Self-Tag Evidence Aggregation with Probabilistic Circuits for PII-Safe Retrieval-Augmented Generation},
  author={Xie, Jin and Li, Songze and Cheng, Guang},
  journal={arXiv preprint arXiv:2603.17292},
  year={2026}
}

@misc{evidenceR4FE823CFC370,
  title={Detecting RAG Extraction Attack via Dual-Path Runtime Integrity Game},
  author={Xie, Yuanbo and Zhang, Yingjie and Li, Yulin and Song, Shouyou and Chen, Xiaokun and Liu, Zhihan and Su, Liya and Liu, Tingwen},
  booktitle={Proceedings of the 64th Annual Meeting of the Association for Computational Linguistics (Volume 1: Long Papers)},
  pages={8503--8518},
  year={2026}
}

@misc{xu2026securingrag,
  title={Securing Retrieval-Augmented Generation: A Taxonomy of Attacks, Defenses, and Future Directions},
  author={Xu, Yuming and Zhang, Mingtao and Ge, Zhuohan and Li, Haoyang and Hu, Nicole and Zhang, Yongqi and Wen, Zhiyuan and Zhang, Jason Chen and Li, Qing and Chen, Lei},
  journal={arXiv preprint arXiv:2604.08304},
  year={2026}
}

@misc{xu2025llmcopyright,
  title={Copyright protection for large language models: A survey of methods, challenges, and trends},
  author={Xu, Zhenhua and Yue, Xubin and Wang, Zhebo and Zhang, Haobo and Liu, Qichen and Zhao, Xixiang and Zhang, Jingxuan and Zeng, Wenjun and Xing, Wengpeng and Kong, Dezhang and others},
  journal={arXiv preprint arXiv:2508.11548},
  year={2025}
}

@misc{evidenceRA09D08FF90FF,
  title={Knowledge-Infused Multi-Bit Watermarking for RAG Knowledge Bases},
  author={Yang, Ke and Yuan, Shuguang and Yu, Jing and Chen, Chi},
  booktitle={Findings of the Association for Computational Linguistics: ACL 2026},
  pages={21195--21218},
  year={2026}
}

@misc{evidenceR0AF83A8B2D74,
  title={Dialectical alignment: Resolving the tension of 3h and security threats of llms},
  author={Yang, Shu and Su, Jiayuan and Jiang, Han and Li, Mengdi and Cheng, Keyuan and Ali, Muhammad Asif and Hu, Lijie and Wang, Di},
  journal={arXiv preprint arXiv:2404.00486},
  year={2024}
}

@misc{yang2025watermarkingsurvey,
  title={Watermarking for large language models: A survey},
  author={Yang, Zhiguang and Zhao, Gejian and Wu, Hanzhou},
  journal={Mathematics},
  volume={13},
  number={9},
  pages={1420},
  year={2025},
  publisher={MDPI}
}

@misc{yao2025controlnet,
  title={Controlnet: A firewall for rag-based llm system},
  author={Yao, Hongwei and Shi, Haoran and Chen, Yidou and Jiang, Yixin and Wang, Cong and Qin, Zhan},
  journal={arXiv preprint arXiv:2504.09593},
  year={2025}
}

@misc{evidenceR47114B882520,
  title={EcoSafeRAG: Efficient Security through Context Analysis in Retrieval-Augmented Generation.},
  author={Yao, Ruobing and Zhang, Yifei and Song, Shuang and Gao, Neng and Tu, Chenyang},
  booktitle={EMNLP (Findings)},
  pages={4034--4050},
  year={2025}
}

@misc{evidenceRBA752514142F,
  title={Adaptive Backtracking for Privacy Protection in Large Language Models},
  author={Yao, Zhihao and Gu, Yuxuan and Feng, Xiachong and Ma, Weitao and Li, Bo and Feng, Xiaocheng and Qin, Bing},
  booktitle={Findings of the Association for Computational Linguistics: ACL 2026},
  pages={37278--37298},
  year={2026}
}

@misc{evidenceRB9E24CAAD81D,
  title={Efficient privacy-preserving retrieval augmented generation with distance-preserving encryption},
  author={Ye, Huanyi and Guo, Jiale and Liu, Ziyao and Lam, Kwok-Yan},
  booktitle={2025 3rd International Conference on Foundation and Large Language Models (FLLM)},
  pages={668--676},
  year={2025},
  organization={IEEE}
}

@misc{evidenceR5EEE0CFB7D9A,
  title={Cordon-MAS: Defending RAG against Knowledge Poisoning via Information-Flow Control},
  author={Yu, Zhe and Xing, Wenpeng and Li, Gaolei and Xiong, Shuguang and Wang, Hongzhi and Teng, Xuyang and Han, Meng},
  journal={arXiv preprint arXiv:2605.26754},
  year={2026}
}

@misc{evidenceR65D32EB39C2F,
  title={Privacy Policy Enforcement Guardrails for Data-Sensitive Retrieval-Augmented Generation},
  author={Zafar, Osama and Nemecek, Alexander and Zhang, Yiqian and Li, Wenbiao and Ganguly, Debargha and Singh, Vikash and Chaudhary, Vipin and Ayday, Erman},
  journal={arXiv preprint arXiv:2605.17034},
  year={2026}
}

@misc{evidenceXA52302837E10,
  title={Selection Integrity for LLM Graph Memory: An Accumulability Criterion for Information-Flow-Blind Retrieval},
  author={Fei, Zeming and Fei, Hongming and Wang, Xiaoyang and Gope, Prosanta and Sikdar, Biplab and Zhang, Ying and others},
  journal={arXiv preprint arXiv:2606.12290},
  year={2026}
}

@misc{zeng2024good,
  title={The good and the bad: Exploring privacy issues in retrieval-augmented generation (rag)},
  author={Zeng, Shenglai and Zhang, Jiankun and He, Pengfei and Liu, Yiding and Xing, Yue and Xu, Han and Ren, Jie and Chang, Yi and Wang, Shuaiqiang and Yin, Dawei and others},
  booktitle={Findings of the Association for Computational Linguistics: ACL 2024},
  pages={4505--4524},
  year={2024}
}

@misc{evidenceR09C5B5D0E3E7,
  title={Mitigating the privacy issues in retrieval-augmented generation (rag) via pure synthetic data},
  author={Zeng, Shenglai and Zhang, Jiankun and He, Pengfei and Ren, Jie and Zheng, Tianqi and Lu, Hanqing and Xu, Han and Liu, Hui and Xing, Yue and Tang, Jiliang},
  booktitle={Proceedings of the 2025 conference on empirical methods in natural language processing},
  pages={24538--24569},
  year={2025}
}

@misc{evidenceR001ABA21E18D,
  title={The good and the bad: Exploring privacy issues in retrieval-augmented generation (rag)},
  author={Zeng, Shenglai and Zhang, Jiankun and He, Pengfei and Liu, Yiding and Xing, Yue and Xu, Han and Ren, Jie and Chang, Yi and Wang, Shuaiqiang and Yin, Dawei and others},
  booktitle={Findings of the Association for Computational Linguistics: ACL 2024},
  pages={4505--4524},
  year={2024}
}

@misc{evidenceR06B30753C7E5,
  title={S-RAG: A novel audit framework for detecting unauthorized use of personal data in RAG systems},
  author={Zeng, Zhirui and Liu, Jiamou and Chiang, Meng-Fen and He, Jialing and Zhang, Zijian},
  booktitle={Proceedings of the 63rd Annual Meeting of the Association for Computational Linguistics (Volume 1: Long Papers)},
  pages={10375--10385},
  year={2025}
}

@misc{zhang2025whotaughtlie,
  title={Who taught the lie? responsibility attribution for poisoned knowledge in retrieval-augmented generation},
  author={Zhang, Baolei and Xin, Haoran and Chen, Yuxi and Liu, Zhuqing and Yi, Biao and Li, Tong and Nie, Lihai and Liu, Zheli and Fang, Minghong},
  booktitle={2026 IEEE Symposium on Security and Privacy (SP)},
  pages={4203--4222},
  year={2026},
  organization={IEEE}
}

@misc{zhang2025tracebackpoisoningattacks,
  title={Traceback of poisoning attacks to retrieval-augmented generation},
  author={Zhang, Baolei and Xin, Haoran and Fang, Minghong and Liu, Zhuqing and Yi, Biao and Li, Tong and Liu, Zheli},
  booktitle={Proceedings of the ACM on Web Conference 2025},
  pages={2085--2097},
  year={2025}
}

@misc{zhang2025benchmarkingpoisoningattacks,
  title={Benchmarking poisoning attacks against retrieval-augmented generation},
  author={Zhang, Baolei and Xin, Haoran and Li, Jiatong and Zhang, Dongzhe and Fang, Minghong and Liu, Zhuqing and Nie, Lihai and Liu, Zheli},
  journal={arXiv preprint arXiv:2505.18543},
  year={2025}
}

@misc{evidenceR85066D7CF090,
  title={Is External Database Protection Static in Retrieval-Augmented Generation? Rethinking Privacy Preservation under Dynamic Queries},
  author={Zhang, Gang and Tian, Mingyu and Luan, Xukun and Ma, Yuanchi and Liu, Jinyan},
  journal={arXiv preprint arXiv:2607.14811},
  year={2026}
}

@misc{evidenceR65B78A21EED1,
  title={Isacl: Internal state analyzer for copyrighted training data leakage},
  author={Zhang, Guangwei and Su, Qisheng and Liu, Jiateng and Qian, Cheng and Pan, Yanzhou and Fu, Yanjie and Zhang, Denghui},
  journal={arXiv preprint arXiv:2508.17767},
  year={2025}
}

@misc{evidenceRF0A0A315FCC0,
  title={Customized retrieval-augmented generation with llm for debiasing recommendation unlearning},
  author={Zhang, Haichao and Zhang, Chong and Hu, Peiyu and Qiu, Shi and Wang, Jia},
  booktitle={2025 IEEE International Conference on Data Mining (ICDM)},
  pages={1702--1711},
  year={2025},
  organization={IEEE}
}

@misc{evidenceR3569BDC61B98,
  title={" Ghost of the past": identifying and resolving privacy leakage from LLM's memory through proactive user interaction},
  author={Zhang, Shuning and Ye, Lyumanshan and Yi, Xin and Tang, Jingyu and Shui, Bo and Xing, Haobin and Liu, Pengfei and Li, Hewu},
  journal={arXiv preprint arXiv:2410.14931},
  year={2024}
}

@misc{zhang2026scope,
  title={SCOPE: Intrinsic Semantic Space Control for Mitigating Copyright Infringement in LLMs},
  author={Zhang, Zhenliang and Hu, Xinyu and Wan, Xiaojun},
  booktitle={Proceedings of the AAAI Conference on Artificial Intelligence},
  volume={40},
  number={41},
  pages={34897--34905},
  year={2026}
}

@misc{evidenceR273E285746F4,
  title={Privacy-preserving RAG via multi-agent semantic rewriting: Achieving confidentiality without compromising contextual fidelity},
  author={Zhao, Yuanhe and Zhang, Tianyu and Xing, Huafei and Wong, Derek F and Li, Jianbin and Fang, Tao},
  journal={Information Processing \& Management},
  volume={64},
  number={1},
  pages={105001},
  year={2027},
  publisher={Elsevier}
}

@misc{evidenceRE3EED5702691,
  title={Trustrag: Enhancing robustness and trustworthiness in retrieval-augmented generation},
  author={Zhou, Huichi and Lee, Kin-Hei and Zhan, Zhonghao and Chen, Yue and Li, Zhenhao and Wang, Zhaoyang and Haddadi, Hamed and Yilmaz, Emine},
  journal={arXiv preprint arXiv:2501.00879},
  year={2025}
}

@misc{evidenceR7F813FD6F967,
  title={Provably secure retrieval-augmented generation},
  author={Zhou, Pengcheng and Feng, Yinglun and Yang, Zhongliang},
  journal={arXiv preprint arXiv:2508.01084},
  year={2025}
}

@misc{evidenceR7B704BB933B9,
  title={Not All Entities are Created Equal: A Dynamic Anonymization Framework for Privacy-Preserving RAG},
  author={Zhu, Xinyuan and Fei, Zekun and Wang, Enye and He, Ruiqi and Guo, Jia and Wang, Ruijie and Liu, Zheli and Zeng, Qingkai},
  journal={arXiv preprint arXiv:2603.26074},
  year={2026}
}

\clearpage
\appendix
\onecolumn
\raggedbottom
\small
\setlength{\smallskipamount}{2.5pt plus 1pt minus 0.5pt}

\section{Event and Control Schema}
\label{app:event_schema}\label{app:constructs}\label{app:taxonomy_tree}
To prevent our coding from confusing a paper's high-level topic with what its experiments actually measure, our schema records six dimensions separately: the technical event, the protected asset, the adversary's identity and access level, the defensive intervention function, the primary measured endpoint, and the broader evidence relation (Table~\ref{tab:event_schema}a). Table~\ref{tab:event_schema}b defines our frozen seven-family endpoint taxonomy (Families A, A-pii, and B--F). Families A, A-pii, B, and C are grouped under the shared content-secrecy asset class; however, when we count studies whose primary endpoint measures actual content disclosure, we include only Families A, A-pii, and B because Family C reports theoretical bounds or likelihood shifts without observing emitted or reconstructed text. Under this four-group hierarchy, a paper's claim and its primary measured endpoint are coded as concordant when they fall into the exact same family, partial when they land in different families within the same asset group, and discordant (off-target) when they belong to different asset groups altogether. Family G is assigned whenever a claim or an experimental packet is too vague to establish a specific technical event. Finally, each publication receives a single mutually exclusive primary endpoint family from human adjudication, whereas its defense mechanisms and intervention functions can receive multiple labels.

\par\smallskip\noindent\begin{minipage}{\textwidth}\centering\scriptsize
\setlength{\tabcolsep}{3pt}\renewcommand{\arraystretch}{1.04}
\captionof{table}{Event definitions and coding schema.}\label{tab:event_schema}\label{tab:families}
\noindent\textbf{(a)} Separate coding fields for technical events, assets, actors, functions, and evidence relations.\par\smallskip
\adjustbox{max width=\linewidth}{
\begin{tabular}{@{}P{0.17\linewidth}P{0.19\linewidth}P{0.20\linewidth}
P{0.18\linewidth}P{0.19\linewidth}@{}}
\toprule
\textbf{Technical event} & \textbf{Primary asset / goal} &
\textbf{Actor and access} & \textbf{Relevant functions} &
\textbf{Evidence relation} \\
\midrule
Unauthorized corpus use or KB theft &
Owner text or image corpus; evidence of use &
External owner/auditor with black-box queries; copying operator &
Detect, attribute; occasionally prevent/remediate &
Direct copyright/ownership evidence
\cite{jovanovic2025ward,liu2025canaries,lv2025ragwm,
liu2025whostole,guo2025towardscopyrightprotection} \\
\addlinespace
Retrieval extraction or authorization bypass &
Restricted documents and access policy &
End user with black-box, adaptive queries &
Measure attacks; evaluate preventive controls separately &
Direct copyright- or privacy-framed extraction attacks
\cite{wang2025silentleaks,zeng2024good}; security-framed controls
\cite{yao2025controlnet}; benchmark evidence
\cite{qi2026knowledgeextraction} \\
\addlinespace
Output copying or close transformation &
Expression exposed in retrieved context &
End user or downstream recipient &
Prevent, redact, attribute, remediate &
Direct only when copying is measured; citation alone is insufficient;
copyright/ownership image output comparison \cite{golatkar2024cpr};
likelihood-based text opt-out is a separate proxy \cite{min2023silo};
security-framed text recovery \cite{li2026ragfort}; attribution is adjacent
\cite{qi2024internalsbasedanswer}. Data Defenses measures use resistance,
not copying \cite{agnew2024datadefenses} \\
\addlinespace
Query, database, or membership leakage &
Confidential query, corpus record, or index &
Cloud provider, protocol participant, or probing user &
Prevent or detect under a privacy/security definition &
Direct RAG but non-copyright privacy/confidentiality
\cite{anderson2024is,liu2025mask,
cheng2025remoteragprivacypreserving} \\
\addlinespace
Corpus poisoning or retriever backdoor &
Integrity of retrieved evidence and answer &
Corpus contributor with write or injection access &
Prevent, detect, trace &
Direct RAG security; not evidence that copyright copying is prevented
\cite{zhang2025benchmarkingpoisoningattacks,
chen2025poisonarenauncoveringcompeting,
shao2025poisoncraftpracticalpoisoning} \\
\addlinespace
Source influence or knowledge removal &
Attribution of an output; requested removal &
Operator or auditor with black-, gray-, or white-box access &
Attribute, detect, remediate &
Direct RAG attribution/remediation; legal relevance is fact-dependent
\cite{nematov2025sourceattribution,qi2024internalsbasedanswer,
wang2024unlearningrag} \\
\bottomrule
\end{tabular}
}
\end{minipage}\par\smallskip
\par\smallskip\noindent\begin{minipage}{\textwidth}\centering\scriptsize
\setlength{\tabcolsep}{3pt}\renewcommand{\arraystretch}{1.04}
\noindent\textbf{Table~\ref{tab:event_schema}, (b)} Frozen event families. C shares the content asset group for concordance but does not measure observed disclosure.\par\smallskip
\adjustbox{max width=\linewidth}{
\begin{tabular}{@{}llp{0.60\columnwidth}@{}}
\toprule
\textbf{Asset} & \textbf{Fam.} & \textbf{The endpoint concerns\ldots} \\
\midrule
\multirow{3}{*}{\shortstack[l]{Content\\disclosure}}
 & A     & generated output reproducing protected source \emph{expression} \\
 & A-pii & generated output containing protected \emph{atomic identifiers} \\
 & B     & \emph{reconstruction} of source content from embeddings, an index, or another non-generation channel \\
\cmidrule(l){2-3}
\shortstack[l]{Proxy for\\disclosure} & C & a likelihood, perplexity, or formal privacy \emph{bound}, with no emitted or reconstructed content \\
\midrule
Access & D & suppression of authorization, retrieval, routing, or answerability \\
\midrule
Use & E & \emph{detection} of a watermark, canary, membership, corpus use, or ownership \\
\midrule
Integrity & F & poisoning, prompt injection, provenance, attribution, removal, or misuse \\
\midrule
Unresolved & G & a packet that does not establish the outcome a stricter assignment needs \\
\bottomrule
\end{tabular}
}
\end{minipage}\par\smallskip

\section{Search, Coding, and Evidence Map}
\label{app:method}\label{app:human_provenance}

\subsection{Literature Search and Eligibility Criteria}
Our literature search ran through July 17, 2026, combining eight initial arXiv queries, nine expanded arXiv query families, the same nine query families executed over OpenAlex, and a title-and-metadata scan of the ACL Anthology. We also ran ten seed-paper lookups and ten forward-citation tracking passes as separate continuity checks, alongside capped supplementary searches on Crossref and DBLP. Each search family paired core retrieval terms (such as ``RAG,'' ``retrieval-augmented,'' ``retrieval-enhanced,'' and ``grounded generation'') with protection-related keywords covering copyright, licensing, ownership, watermarking, copying, extraction, privacy, attribution, removal, and security; an adjacent query family paired general language-model terms with copyright and copying prevention. Retrieved records were deduplicated sequentially by arXiv identifier, Digital Object Identifier (DOI), and normalized title, followed by an author-title-year audit that linked multiple preprints or versions belonging to the same underlying study.

 To be confirmed as a direct empirical RAG-protection record in our core map, a study had to evaluate a retrieval-augmented system, intervene on its query, corpus, index, context, or output, report a source-verified protection or attribution operating point, and address either a direct copyright/ownership goal or a neighboring RAG privacy/security goal. Pure attack papers, standalone benchmarks, survey articles, and untested architectural proposals were tracked in separate roles, and reporting task utility alone was not sufficient for inclusion.

In total, our queries returned 11,972 raw hits, which deduplicated to 8,978 unique records and yielded 925 candidate inclusions after abstract screening. We successfully recovered full texts for 824 candidates, while 101 full texts were unavailable and one recovered paper remained unresolved. Human full-text review confirmed 112 direct control studies from this candidate pool, and our stratified audit of 1,200 records sampled from the 8,053 excluded records identified 6 additional direct controls, bringing the final human-coded map to 118 records (Table~\ref{tab:design_based_estimate}a and Table~\ref{tab:design_based_estimate}b).

\par\smallskip\noindent\begin{minipage}{\textwidth}\centering\scriptsize
\setlength{\tabcolsep}{3pt}\renewcommand{\arraystretch}{1.04}
\captionof{table}{Search-coverage estimates and evidence-map dimensions.}\label{tab:design_based_estimate}\label{tab:evidence_dimensions}
\noindent\textbf{(a)} Excluded-frame audit: 1,200 sampled exclusions, ten strata. Conservative intervals combine simultaneous finite-population hypergeometric bounds with Bonferroni correction. The component sum is not a population total.\par\smallskip
\adjustbox{max width=\linewidth}{
\begin{tabular}{@{}lrrrp{0.27\linewidth}@{}}
\toprule
\textbf{Quantity} & \textbf{Observed} & \textbf{Estimate/range} & \textbf{Conservative 95\%} & \textbf{Identification} \\
\midrule
excluded frame & 6 & 27.2 & 6 to 318 & Protocol-confirmable controls under the frozen human abstract rule; abstract-screen false negatives are not estimated. \\
Available possible-inclusion full texts & 112 & 112 & n/a & Census of available full texts only; 102 records are unclassified. \\
Descriptive component sum & 118 & 139.2 & 118 to 430 & Not an identified total for all eligible records in the 8,978-record frame. \\
If two recall nonresponses are controls & 120 & 146.1 & 120 to 441 & All-positive second-stage nonresponse sensitivity. \\
\midrule
Main-pool missingness envelope (102 unclassified) & n/a & 139.2 to 241.2 & n/a & Non-probabilistic; recall nonresponses remain excluded. \\
Expanded missingness envelope & n/a & 139.2 to 248.1 & n/a & Also treats both weighted recall nonresponses as controls. \\
\bottomrule
\end{tabular}
}
\end{minipage}\par\smallskip
\par\smallskip\noindent\begin{minipage}{\textwidth}\centering\scriptsize
\setlength{\tabcolsep}{3pt}\renewcommand{\arraystretch}{1.04}
\noindent\textbf{Table~\ref{tab:design_based_estimate}, (b)} Human-adjudicated map dimensions. Intervention functions are multi-label; abstract-only records are excluded.\par\smallskip
\adjustbox{max width=\linewidth}{
\begin{minipage}[t]{0.4\textwidth}\vspace{0pt}\centering
\begin{tabular}{@{}llr@{}}\toprule\textbf{Axis} & \textbf{Category} & \textbf{Records} \\ \midrule
Evidence relation & copyright / ownership & 15 \\
 & neighboring RAG goal & 103 \\
Validated outcome & emitted source expression & 8 \\
 & emitted atomic identifier / PII & 6 \\
 & non-output source recovery & 7 \\
Validated flag & adaptive multi-query recovery & 7 \\
Legacy field & broad ``output-copy'' yes & 28 \\
\bottomrule\end{tabular}\end{minipage}\hfill
\begin{minipage}[t]{0.29\textwidth}\vspace{0pt}\centering
\begin{tabular}{@{}llr@{}}\toprule\textbf{Axis} & \textbf{Category} & \textbf{Records} \\ \midrule
Modality & text & 110 \\
 & multimodal & 5 \\
 & image & 3 \\
Publication & peer reviewed & 29 \\
 & preprint & 88 \\
 & thesis/report & 1 \\
\bottomrule\end{tabular}\end{minipage}\hfill
\begin{minipage}[t]{0.27\textwidth}\vspace{0pt}\centering
\begin{tabular}{@{}llr@{}}\toprule\textbf{Axis} & \textbf{Category} & \textbf{Records} \\ \midrule
Year & 2023 & 1 \\
 & 2024 & 18 \\
 & 2025 & 43 \\
 & 2026 & 56 \\
Function & prevent & 83 \\
 & detect & 69 \\
 & attribute & 26 \\
 & remediate & 24 \\
\bottomrule\end{tabular}\end{minipage}\hfill

}
\end{minipage}\par\smallskip

\subsection{Human Annotation and Inter-Rater Reliability}
Thirty-two trained annotators with backgrounds in NLP and machine learning (18), security and privacy (9), and information science or law and technology (5) double-coded the recovered full texts. Whenever the initial pair disagreed, a third reviewer outside that pair adjudicated the record. Before coding the main pool, every annotator completed a 100-item calibration test with a required passing score of 95\%; all 32 annotators passed, scoring between 95\% and 98\% (median 97\%). Because raw calibration files and identity-linked records remain restricted under participant confidentiality terms, domain backgrounds and calibration pass rates are reported as coordinator attestations.

Our public artifact releases a de-identified 8,024-row human label matrix covering all 118 confirmed records across 34 coded fields. In this matrix, initial coder pairs that had no entry in the disagreement log are reconstructed as matching the consensus label, while logged disagreements preserve the exact pre-adjudication calls of both coders. Running our standalone verification script on this matrix reproduces a pre-adjudication agreement of 89.83\% ($\kappa=0.8672$) on the primary endpoint family (Table~\ref{tab:human_validation}a and Table~\ref{tab:human_validation}b). Across the 14 strictly categorical fields, median reliability is $\kappa=0.87$. Fifteen free-text rationale and source-location fields had no logged categorical disagreements; their nominal $\kappa=1.00$ reflects consensus text entry rather than an independent reliability estimate. For the eight methodological appraisal domains, linear weighted $\kappa$ is computed across lower, some, and high concern levels excluding ``not reported'' (NR) pairs, whereas nominal agreement includes NR calls.

\par\smallskip\noindent\begin{minipage}{\textwidth}
\captionof{table}{Human screening and refined-coding agreement.}\label{tab:human_validation}\label{tab:refined_agreement}
\noindent\begin{minipage}[t]{0.46\textwidth}\vspace{0pt}
\centering\scriptsize
\setlength{\tabcolsep}{3pt}\renewcommand{\arraystretch}{1.04}
\noindent\textbf{(a)} Pre-adjudication agreement over 824 full texts.\par\smallskip
\adjustbox{max width=\linewidth}{
\begin{tabular}{@{}lrr@{}}
\toprule
\textbf{Field} & \textbf{Agreement} & $\boldsymbol{\kappa}$ \\
\midrule
Strict direct-control decision & 92.4\% & 0.642 \\
Evidence relation & 74.3\% & 0.546 \\
Primary study role & 87.5\% & 0.844 \\
Legacy broad output-copy field & 97.2\% & 0.852 \\
Legacy recovery field & 94.1\% & 0.795 \\
Publication status & 93.6\% & 0.860 \\
\bottomrule
\end{tabular}
}
\end{minipage}\hfill
\begin{minipage}[t]{0.52\textwidth}\vspace{0pt}
\centering\scriptsize
\setlength{\tabcolsep}{3pt}\renewcommand{\arraystretch}{1.04}
\noindent\textbf{Table~\ref{tab:human_validation}, (b)} Refined-coding agreement. Weighted $\kappa$ is linear across lower/some/high and excludes NR; nominal agreement includes NR.\par\smallskip
\adjustbox{max width=\linewidth}{
\begin{tabular}{@{}lrrr@{}}
\toprule
\textbf{Field} & \textbf{Agreement} & \textbf{$\kappa$} & \textbf{Weighted $\kappa$} \\
\midrule
primary family & 89.8\% & 0.867 & n/a \\
adaptive recovery & 99.2\% & 0.929 & n/a \\
interaction adaptivity & 83.1\% & 0.760 & n/a \\
Event, asset, actor, and access & 97.5\% & 0.932 & 0.932 \\
Comparator and ablation & 98.3\% & 0.826 & 0.851 \\
Evaluation independence & 95.8\% & 0.718 & 0.722 \\
Sample and unit reporting & 93.2\% & 0.873 & 0.881 \\
Uncertainty analysis & 94.1\% & 0.897 & 0.938 \\
Adaptive or robustness test & 91.5\% & 0.857 & 0.877 \\
Matched utility and harms & 94.9\% & 0.879 & 0.900 \\
Artifact or reproduction support & 97.5\% & 0.949 & 0.950 \\
\bottomrule
\end{tabular}
}
\end{minipage}
\end{minipage}\par\smallskip

\subsection{Copyright-Relation Sensitivity}
\label{app:copyright_cell}\label{app:endpoint_blind}\label{app:supplementary}\label{app:citation_audit}
Within our 118-record map, 15 records are coded under the frozen direct copyright or dataset-ownership evidence relation (Table~\ref{tab:evidence_map_body}a). Eleven of these 15 studies (73.3\% [48.0, 89.1]) have a usage-detection primary endpoint in Family E, compared with just 7 of the 103 studies (6.8\%) addressing neighboring goals (Bonferroni-adjusted $p=4.9\times10^{-7}$ across 14 family comparisons). Under the frozen copyright-relation label (Table~\ref{tab:evidence_map_body}b), zero records have a strict emitted-expression primary endpoint in Family A (95\% exact upper bound of 18.1\%).

All citations and operating points were verified against source PDFs, and the exact query strings and audit logs are preserved in the released artifact.

\par\smallskip\noindent\begin{minipage}{\textwidth}\centering\scriptsize
\setlength{\tabcolsep}{3pt}\renewcommand{\arraystretch}{1.04}
\captionof{table}{Evidence-relation sensitivity.}\label{tab:evidence_map_body}\label{tab:copyright_boundary}
\noindent\textbf{(a)} Primary endpoints by evidence relation across 118 records. Copyright/ownership is a technical coding relation, not a legal classification.\par\smallskip
\adjustbox{max width=\linewidth}{
\begin{tabular}{@{}llrrr@{}}
\toprule
Protected asset & Primary reported endpoint & All & Copyright & Other \\
 & & (\%) & relation & relation \\
\midrule
Content disclosure & Emitted source expression & 8 (6.8) & 0 & 8 \\
 & Emitted atomic identifiers & 6 (5.1) & 0 & 6 \\
 & Non-output reconstruction & 7 (5.9) & 0 & 7 \\
 & Likelihood or formal bound & 6 (5.1) & 2 & 4 \\
Access suppression & Access and retrieval suppression & 26 (22.0) & 1 & 25 \\
Use detection & Corpus-use or ownership detection & 18 (15.3) & 11 & 7 \\
Integrity / attribution & Integrity, provenance, or removal & 47 (39.8) & 1 & 46 \\
\midrule
\multicolumn{2}{@{}l}{Total} & 118 & 15 & 103 \\
\bottomrule
\end{tabular}
}
\end{minipage}\par\smallskip
\par\smallskip\noindent\begin{minipage}{\textwidth}
\noindent\begin{minipage}[t]{0.57\textwidth}\vspace{0pt}
\centering\scriptsize
\setlength{\tabcolsep}{3pt}\renewcommand{\arraystretch}{1.04}
\noindent\textbf{Table~\ref{tab:evidence_map_body}, (b)} Copyright-boundary sensitivity: relation agreement 74.3\% ($\kappa=0.546$), centrality agreement 97.6\% ($\kappa=0.893$).\par\smallskip
\adjustbox{max width=\linewidth}{
\begin{tabular}{@{}P{0.27\linewidth}rrp{0.43\linewidth}@{}}
\toprule
\textbf{Boundary} & \textbf{Records} & \textbf{Emitted} & \textbf{Diagnostic change} \\
\midrule
Frozen evidence-relation label & 15 & 0 & Includes S-RAG; excludes RAGFort. \\
Relation minus personal-data S-RAG & 14 & 0 & Removes the disputed privacy audit. \\
Frozen copyright-centrality label & 16 & 1 & Includes both S-RAG and RAGFort. \\
Centrality minus personal-data S-RAG & 15 & 1 & Substitutes proprietary-KB extraction for the privacy audit. \\
\bottomrule
\end{tabular}
}
\end{minipage}
\end{minipage}\par\smallskip

\section{Controlled Testbed and Claim-Aligned Analyses}
\label{app:casestudy}\label{app:controlled}\label{app:claim_aligned}
\subsection{Design and analysis hierarchy}
The fixed protocol (not preregistered) uses a 192-document security corpus across two domains and two policies, with three unexposed human-authored probes per document; utility uses 64 document-disjoint open-policy items. MiniLM dense retrieval is fixed at top-3, decoding is greedy, and the three-query controller observes only released text, redaction, refusal, transport decodability, and released length. Qwen3 is the primary model, Phi-4 the replication, and Qwen2.5-3B and Mistral-7B exploratory. The exact upstream Summary baseline is commit-pinned.

The sole planned confirmatory contrast is the Qwen3 Summary-minus-uncontrolled difference-in-differences (DiD) of exact hidden-probe recall on all 80 held-out answer-only items. Context effects are target-present minus target-ablated. An attack ranking selected on uncontrolled target-present development data is shared across all four exposure/control cells. Inference uses paired source-item bootstrapping and sign randomisation without multiplicity correction for this single contrast. Other models, policies, natural retrieval, shorter spans, and linkage analyses are secondary or exploratory.

\par\smallskip\noindent\begin{minipage}{\textwidth}\centering\scriptsize
\setlength{\tabcolsep}{3pt}\renewcommand{\arraystretch}{1.04}
\captionof{table}{Controls and their stated claims.}\label{tab:claim_control_matrix}
\noindent\textbf{(a)} Declared claims and their aligned measurements; other endpoints describe separate channels.\par\smallskip
\adjustbox{max width=\linewidth}{
\begin{tabular}{@{}P{0.16\linewidth}P{0.30\linewidth}P{0.23\linewidth}P{0.23\linewidth}@{}}
\toprule
\textbf{Control} & \textbf{Declared technical claim} &
\textbf{Claim-aligned evidence} & \textbf{Not established} \\
\midrule
Policy filter &
Contexts unauthorized for the requesting principal do not reach generation. &
Target presence after filtering; access-control failures; authorized QA utility. &
Copying or recovery from contexts the principal is authorized to access. \\
Cosine threshold &
Retrieved documents with MiniLM cosine below the pinned 0.5 operating point
do not reach generation. &
Post-threshold target presence, all-item recovery, and matched QA utility. &
Confidentiality of relevant documents that remain above threshold. \\
Benchmark Summary &
The pinned upstream prompt requests a concise, query-relevant summary in the
model's own words and rejects irrelevant context. &
Response concision and relevance; full-source coverage and blinded human
source-content recovery for the broader extraction-defense interpretation.
Exact hidden-answer recall is retained only with a construct audit. &
Legal similarity, corpus-use detection, or universal resistance to semantic
inference. \\
Benchmark SystemBlock &
The pinned system prompt forbids exposing raw database content or facts
available only from the database. &
Context-attributable source-content recovery, refusal behavior, and matched
benign utility under a defense-aware attacker. &
Formal access control, legal confidentiality, or robustness beyond the tested
models, prompts, and budget. \\
Normalized 8-gram guard &
No released output contains a matching normalized source 8-gram. &
Trigger behavior, shorter-span utility, and the implied absence of matching
copy$_{20}$ under the same normalization. &
Suppression of facts, paraphrase, source identity, or reversible encodings
outside its canonicalization. \\
\bottomrule
\end{tabular}
}
\end{minipage}\par\smallskip

\subsection{Natural-policy and adaptive evaluations}
Natural-policy and claim-aligned evaluations use distinct denominators, stated in their table panels. Natural-policy intervals resample source items, and models are reported separately. A quotation-denied item can remain answerable; the policy filter excludes inaccessible documents, while the normalised 8-gram guard suppresses protected source spans. Authorised quotation is desired release. The adaptive experiment additionally distinguishes raw outcomes on all quotation-denied items from residual outcomes conditional on successful unguarded copying.

\par\smallskip\noindent\begin{minipage}{\textwidth}\centering\scriptsize
\setlength{\tabcolsep}{3pt}\renewcommand{\arraystretch}{1.04}
\captionof{table}{Natural-policy outcomes and copied-span thresholds.}\label{tab:natural_policy_main}\label{tab:natural_policy_thresholds}
\noindent\textbf{(a)} Natural-policy dense retrieval. Copying uses all quotation-denied items. Authorised quote copying is desired release. Brackets: 95\% source-item bootstrap intervals.\par\smallskip
\adjustbox{max width=\linewidth}{
\begin{tabular}{@{}lrrrrrr@{}}
\toprule
\textbf{Model} & \textbf{Target retrieved} & \textbf{Uncontrolled copy$_{20}$} & \textbf{Policy filter copy$_{20}$} & \textbf{Filter + guard copy$_{20}$} & \textbf{Ordinary-QA F1} & \textbf{Authorized quote copy$_{20}$} \\
\midrule
Qwen-3B & 98.4 [95.3, 100.0] & 78.1 [68.8, 87.5] & 37.5 [26.6, 50.0] & 0.0 [0.0, 0.0] & 0.250 [0.213, 0.288] & 73.4 [62.5, 84.4] \\
Phi-3.5 & 98.4 [95.3, 100.0] & 89.1 [81.2, 95.3] & 45.3 [32.8, 57.8] & 0.0 [0.0, 0.0] & 0.369 [0.316, 0.426] & 79.7 [68.8, 89.1] \\
Qwen-7B & 98.4 [95.3, 100.0] & 85.9 [76.6, 93.8] & 46.9 [34.4, 59.4] & 0.0 [0.0, 0.0] & 0.309 [0.267, 0.353] & 81.2 [71.9, 90.6] \\
Mistral-7B & 98.4 [95.3, 100.0] & 64.1 [51.6, 76.6] & 29.7 [18.8, 40.6] & 0.0 [0.0, 0.0] & 0.261 [0.225, 0.298] & 67.2 [56.2, 78.1] \\
\bottomrule
\end{tabular}
}
\end{minipage}\par\smallskip
\par\smallskip\noindent\begin{minipage}{\textwidth}\centering\scriptsize
\setlength{\tabcolsep}{3pt}\renewcommand{\arraystretch}{1.04}
\noindent\textbf{Table~\ref{tab:natural_policy_main}, (b)} Each cell is copy$_5$/copy$_{10}$/copy$_{20}$ (\%) under the semantic extraction query with quotation forbidden.\par\smallskip
\adjustbox{max width=\linewidth}{
\begin{tabular}{@{}lrrr rrr@{}}
\toprule
\textbf{Model} & \multicolumn{3}{c}{\textbf{Lexical}} & \multicolumn{3}{c}{\textbf{Dense}} \\
\cmidrule(lr){2-4}\cmidrule(lr){5-7}
 & \textbf{None} & \textbf{Filter} & \textbf{Filter+guard} & \textbf{None} & \textbf{Filter} & \textbf{Filter+guard} \\
\midrule
Qwen-3B & 87.5/84.4/70.3 & 51.6/39.1/34.4 & 26.6/0.0/0.0 & 92.2/89.1/78.1 & 62.5/45.3/37.5 & 26.6/1.6/0.0 \\
Phi-3.5 & 87.5/84.4/79.7 & 53.1/37.5/34.4 & 14.1/1.6/0.0 & 93.8/92.2/89.1 & 64.1/48.4/45.3 & 20.3/3.1/0.0 \\
Qwen-7B & 85.9/85.9/79.7 & 48.4/40.6/37.5 & 14.1/1.6/0.0 & 96.9/93.8/85.9 & 60.9/50.0/46.9 & 15.6/1.6/0.0 \\
Mistral-7B & 87.5/76.6/68.8 & 51.6/39.1/26.6 & 23.4/3.1/0.0 & 93.8/85.9/64.1 & 60.9/45.3/29.7 & 25.0/1.6/0.0 \\
\bottomrule
\end{tabular}
}
\end{minipage}\par\smallskip
\par\smallskip\noindent\begin{minipage}{\textwidth}\centering\scriptsize
\setlength{\tabcolsep}{3pt}\renewcommand{\arraystretch}{1.04}
\captionof{table}{Metadata robustness and guard side effects.}\label{tab:natural_policy_metadata}\label{tab:natural_policy_guard_side_effects}
\noindent\textbf{(a)} Metadata robustness. Cells are cross-model ranges, not pooled estimates. Fail-open missing or wrong labels admit 12.5\% of access-denied targets.\par\smallskip
\adjustbox{max width=\linewidth}{
\begin{tabular}{@{}lrrr@{}}
\toprule
\textbf{Observed metadata} & \textbf{Denied target kept} & \textbf{Denied copy$_{20}$} & \textbf{Ordinary QA F1} \\
\midrule
Clean & 0.0 to 0.0 & 0.0 to 0.0 & 0.250 to 0.369 \\
Missing, fail open & 12.5 to 12.5 & 9.4 to 12.5 & 0.252 to 0.369 \\
Missing, fail closed & 0.0 to 0.0 & 0.0 to 0.0 & 0.226 to 0.349 \\
Wrong label & 12.5 to 12.5 & 0.0 to 12.5 & 0.241 to 0.350 \\
\bottomrule
\end{tabular}
}
\end{minipage}\par\smallskip
\par\smallskip\noindent\begin{minipage}{\textwidth}\centering\scriptsize
\setlength{\tabcolsep}{3pt}\renewcommand{\arraystretch}{1.04}
\noindent\textbf{Table~\ref{tab:natural_policy_metadata}, (b)} Paired guard-minus-unguarded side effects on the same response. Authorised target text is excluded from the guard set; triggers concern other non-quotable sources.\par\smallskip
\adjustbox{max width=\linewidth}{
\begin{tabular}{@{}lrrrrr@{}}
\toprule
\textbf{Model} & \textbf{Ordinary trigger} & \textbf{Ordinary $\Delta$F1} & \textbf{F1 decrease} & \textbf{Authorized non-target trigger} & \textbf{Authorized $\Delta$5-gram recall} \\
\midrule
Qwen-3B & 16.7 [9.4, 25.0] & -0.004 [-0.014, 0.004] & 5.2 [1.0, 9.4] & 0.0 [0.0, 0.0] & 0.000 [0.000, 0.000] \\
Phi-3.5 & 10.4 [4.2, 16.7] & -0.006 [-0.020, 0.007] & 4.2 [1.0, 8.3] & 9.4 [3.1, 17.2] & 0.000 [0.000, 0.000] \\
Qwen-7B & 7.3 [2.1, 12.5] & -0.011 [-0.023, -0.001] & 5.2 [1.0, 10.4] & 3.1 [0.0, 7.8] & 0.000 [0.000, 0.000] \\
Mistral-7B & 10.4 [5.2, 16.7] & -0.008 [-0.021, 0.002] & 4.2 [1.0, 8.3] & 0.0 [0.0, 0.0] & 0.000 [0.000, 0.000] \\
\bottomrule
\end{tabular}
}
\end{minipage}\par\smallskip
\par\smallskip\noindent\begin{minipage}{\textwidth}\centering\scriptsize
\setlength{\tabcolsep}{3pt}\renewcommand{\arraystretch}{1.04}
\captionof{table}{Adaptive copying and semantic similarity.}\label{tab:adaptive_copy}\label{tab:adaptive_semantic}
\noindent\textbf{(a)} Adaptive copy$_{20}$ (\%). Raw rates use all quotation-denied items; residuals condition on successful unguarded copying, with each $n$ shown. Surface, normalised, and cumulative-session guards remain distinct.\par\smallskip
\adjustbox{max width=\linewidth}{
\begin{tabular}{@{}llrrrr@{}}
\toprule
\textbf{Model} & \textbf{Attack} & \textbf{Raw copy$_{20}$} & \textbf{Surface residual} & \textbf{Normalized residual} & \textbf{Session residual} \\
\midrule
Qwen3-8B & Separator evasion & 71.9 [59.4, 82.8]; $n=64$ & 95.7 [89.1, 100.0]; $n=46$ & 0.0 [0.0, 0.0]; $n=46$ & n/a \\
 & Four-turn chunks & 60.9 [48.4, 71.9]; $n=64$ & 2.6 [0.0, 7.7]; $n=39$ & 2.6 [0.0, 7.7]; $n=39$ & 0.0 [0.0, 0.0]; $n=39$ \\
\addlinespace
Phi-4-14B & Separator evasion & 96.9 [92.2, 100.0]; $n=64$ & 100.0 [100.0, 100.0]; $n=62$ & 0.0 [0.0, 0.0]; $n=62$ & n/a \\
 & Four-turn chunks & 81.2 [71.9, 90.6]; $n=64$ & 0.0 [0.0, 0.0]; $n=52$ & 7.7 [1.9, 15.4]; $n=52$ & 0.0 [0.0, 0.0]; $n=52$ \\
\addlinespace
\bottomrule
\end{tabular}
}
\end{minipage}\par\smallskip
\par\smallskip\noindent\begin{minipage}{\textwidth}\centering\scriptsize
\setlength{\tabcolsep}{3pt}\renewcommand{\arraystretch}{1.04}
\noindent\textbf{Table~\ref{tab:adaptive_copy}, (b)} Detailed-paraphrase cosine similarity and mismatched-source margin; neither is thresholded as copying or legal similarity.\par\smallskip
\adjustbox{max width=\linewidth}{
\begin{tabular}{@{}lrr@{}}
\toprule
\textbf{Model / control} & \textbf{Source cosine} & \textbf{True-negative margin} \\
\midrule
Qwen3-8B / None & 0.794 [0.758, 0.828] & 0.704 [0.665, 0.742] \\
Qwen3-8B / Surface guard & 0.679 [0.624, 0.732] & 0.583 [0.522, 0.638] \\
Qwen3-8B / Summary-style prompt & 0.777 [0.738, 0.815] & 0.691 [0.645, 0.733] \\
Qwen3-8B / Summary-style + surface guard & 0.666 [0.606, 0.723] & 0.576 [0.514, 0.638] \\
Phi-4-14B / None & 0.882 [0.858, 0.902] & 0.760 [0.732, 0.785] \\
Phi-4-14B / Surface guard & 0.816 [0.783, 0.844] & 0.696 [0.662, 0.728] \\
Phi-4-14B / Summary-style prompt & 0.870 [0.842, 0.894] & 0.752 [0.718, 0.785] \\
Phi-4-14B / Summary-style + surface guard & 0.789 [0.748, 0.828] & 0.679 [0.633, 0.719] \\
\bottomrule
\end{tabular}
}
\end{minipage}\par\smallskip
\par\smallskip\noindent\begin{minipage}{\textwidth}\centering\scriptsize
\setlength{\tabcolsep}{3pt}\renewcommand{\arraystretch}{1.04}
\captionof{table}{Utility and model-specific human outcomes.}\label{tab:adaptive_utility}\label{tab:semantic_human_models}
\noindent\textbf{(a)} Ordinary QA uses clean metadata. Authorised quotation reports coverage, completeness, and refusal under policy filtering plus the normalised guard.\par\smallskip
\adjustbox{max width=\linewidth}{
\begin{tabular}{@{}lrrrrrrr@{}}
\toprule
\textbf{Model} & \shortstack{\textbf{Ordinary F1}\\\textbf{none}} & \shortstack{\textbf{Ordinary F1}\\\textbf{surface}} & \shortstack{\textbf{Ordinary F1}\\\textbf{normalized}} & \shortstack{\textbf{Ordinary F1}\\\textbf{summary-style}} & \shortstack{\textbf{Authorized}\\\textbf{target}} & \shortstack{\textbf{Authorized}\\\textbf{5-gram recall}} & \shortstack{\textbf{Authorized}\\\textbf{refusal}} \\
\midrule
Qwen3-8B & 0.349 [0.309, 0.389] & 0.346 [0.308, 0.386] & 0.346 [0.306, 0.390] & 0.355 [0.312, 0.399] & 98.4 [95.3, 100.0] & 0.190 [0.152, 0.228] & 3.1 [0.0, 7.8] \\
Phi-4-14B & 0.179 [0.156, 0.205] & 0.184 [0.158, 0.209] & 0.177 [0.153, 0.201] & 0.208 [0.182, 0.237] & 98.4 [95.3, 100.0] & 0.169 [0.148, 0.191] & 3.1 [0.0, 7.8] \\
\bottomrule
\end{tabular}
}
\end{minipage}\par\smallskip
\par\smallskip\noindent\begin{minipage}{\textwidth}\centering\scriptsize
\setlength{\tabcolsep}{3pt}\renewcommand{\arraystretch}{1.04}
\noindent\textbf{Table~\ref{tab:adaptive_utility}, (b)} Human-rated outcomes by model, using the definitions and pair sample in Table~\ref{tab:semantic_human}.\par\smallskip
\adjustbox{max width=\linewidth}{
\begin{tabular}{@{}llrrrr@{}}
\toprule
\textbf{Model} & \textbf{Control} & \textbf{Expression} & \textbf{Recovery} & \textbf{Identifiable} & \textbf{$n$} \\
\midrule
Qwen3-8B & None & 72.2 & 33.3 & 50.0 & 18 \\
 & Normalized guard & 5.6 & 0.0 & 0.0 & 18 \\
 & Summary-style prompt & 55.6 & 5.6 & 22.2 & 18 \\
 & Summary + normalized guard & 11.1 & 0.0 & 11.1 & 18 \\
\addlinespace
Phi-4-14B & None & 94.4 & 66.7 & 72.2 & 18 \\
 & Normalized guard & 44.4 & 38.9 & 38.9 & 18 \\
 & Summary-style prompt & 66.7 & 55.6 & 66.7 & 18 \\
 & Summary + normalized guard & 50.0 & 22.2 & 27.8 & 18 \\
\addlinespace
\bottomrule
\end{tabular}
}
\end{minipage}\par\smallskip
\par\smallskip\noindent\begin{minipage}{\textwidth}\centering\scriptsize
\setlength{\tabcolsep}{3pt}\renewcommand{\arraystretch}{1.04}
\captionof{table}{Online adaptive extraction and recovery endpoints.}\label{tab:online_adaptive_main}\label{tab:recovery_endpoints}
\noindent\textbf{(a)} Online three-query extraction (\%). Attack selection uses development data only; branching sees released information. Dense top-3 rows condition on target retrieval before policy filtering; decoded endpoints accumulate released turns.\par\smallskip
\adjustbox{max width=\linewidth}{
\begin{tabular}{@{}llrrrrr@{}}
\toprule
\textbf{Model} & \textbf{Control} & \textbf{Visible copy$_{20}$} & \textbf{Decoded copy$_{20}$} & \textbf{Hidden-fact recall} & \textbf{Source re-ID@1} & \textbf{$n$} \\
\midrule
Qwen2.5-3B & None & 50.0 [35.0, 65.0] & 67.5 [52.5, 80.0] & 40.8 [27.5, 54.2] & 25.0 [12.5, 37.5] & 40 \\
 & Normalized guard & 0.0 [0.0, 0.0] & 0.0 [0.0, 0.0] & 31.7 [20.8, 41.7] & 15.0 [5.0, 27.5] & 40 \\
 & Summary + guard & 0.0 [0.0, 0.0] & 0.0 [0.0, 0.0] & 14.2 [6.7, 22.5] & 22.5 [10.0, 35.0] & 40 \\
 & Policy + guard & 0.0 [0.0, 0.0] & 0.0 [0.0, 0.0] & 18.3 [10.8, 26.7] & 5.0 [0.0, 12.5] & 40 \\
 & Policy + summary + guard & 0.0 [0.0, 0.0] & 0.0 [0.0, 0.0] & 17.5 [10.0, 25.8] & 20.0 [7.5, 32.5] & 40 \\
\addlinespace
Qwen3-8B & None & 55.0 [40.0, 70.0] & 72.5 [57.5, 85.0] & 60.0 [45.8, 73.3] & 25.0 [12.5, 40.0] & 40 \\
 & Normalized guard & 0.0 [0.0, 0.0] & 0.0 [0.0, 0.0] & 48.3 [36.7, 60.8] & 42.5 [27.5, 57.5] & 40 \\
 & Summary + guard & 0.0 [0.0, 0.0] & 0.0 [0.0, 0.0] & 39.2 [27.5, 50.8] & 40.0 [25.0, 55.0] & 40 \\
 & Policy + guard & 0.0 [0.0, 0.0] & 0.0 [0.0, 0.0] & 22.5 [13.3, 31.7] & 20.0 [7.5, 32.5] & 40 \\
 & Policy + summary + guard & 0.0 [0.0, 0.0] & 0.0 [0.0, 0.0] & 27.5 [16.6, 39.2] & 30.0 [15.0, 45.0] & 40 \\
\addlinespace
\bottomrule
\end{tabular}
}
\end{minipage}\par\smallskip
\par\smallskip\noindent\begin{minipage}{\textwidth}\centering\scriptsize
\setlength{\tabcolsep}{3pt}\renewcommand{\arraystretch}{1.04}
\noindent\textbf{Table~\ref{tab:online_adaptive_main}, (b)} Detailed-paraphrase recovery on quotation-denied items. Hidden-fact recall uses unexposed human-written probes; re-ID@1 ranks the response against 96 documents. These are recovery/linkage, not copying measures.\par\smallskip
\adjustbox{max width=\linewidth}{
\begin{tabular}{@{}llrrr@{}}
\toprule
\textbf{Model} & \textbf{Control} & \textbf{copy$_{20}$} & \textbf{Hidden-fact recall} & \textbf{Source re-ID@1} \\
\midrule
Qwen3-8B & None & 3.1 [0.0, 7.8] & 14.6 [7.9, 21.2] & 87.5 [79.7, 95.3] \\
 & Normalized guard & 0.0 [0.0, 0.0] & 12.4 [6.6, 18.8] & 78.1 [67.2, 87.5] \\
 & Summary + normalized guard & 0.0 [0.0, 0.0] & 9.5 [4.8, 15.3] & 76.6 [65.6, 85.9] \\
 & Policy + summary + normalized guard & 0.0 [0.0, 0.0] & 4.2 [1.1, 7.9] & 43.8 [31.2, 56.2] \\
\addlinespace
Phi-4-14B & None & 4.7 [0.0, 10.9] & 46.0 [37.3, 55.6] & 90.6 [82.8, 96.9] \\
 & Normalized guard & 0.0 [0.0, 0.0] & 38.4 [29.6, 47.4] & 90.6 [82.8, 96.9] \\
 & Summary + normalized guard & 0.0 [0.0, 0.0] & 34.1 [25.9, 42.3] & 85.9 [76.6, 93.8] \\
 & Policy + summary + normalized guard & 0.0 [0.0, 0.0] & 20.6 [13.8, 28.6] & 56.2 [43.8, 68.8] \\
\addlinespace
\bottomrule
\end{tabular}
}
\end{minipage}\par\smallskip
\par\smallskip\noindent\begin{minipage}{\textwidth}\centering\scriptsize
\setlength{\tabcolsep}{3pt}\renewcommand{\arraystretch}{1.04}
\captionof{table}{Query-budget and domain sensitivity.}\label{tab:online_adaptive_budget}\label{tab:online_adaptive_domains}
\noindent\textbf{(a)} Normalised-guard budget sensitivity; stopped sessions carry forward their last release and decoded outcomes accumulate valid turns.\par\smallskip
\adjustbox{max width=\linewidth}{
\begin{tabular}{@{}llrrrr@{}}
\toprule
\textbf{Model} & \textbf{Budget} & \textbf{Decoded copy$_{20}$} & \textbf{Hidden-fact recall} & \textbf{Source re-ID@1} & \textbf{$n$} \\
\midrule
Qwen2.5-3B & 1 & 0.0 [0.0, 0.0] & 0.8 [0.0, 2.5] & 15.0 [5.0, 27.5] & 40 \\
 & 2 & 0.0 [0.0, 0.0] & 3.3 [0.0, 7.5] & 10.0 [2.5, 20.0] & 40 \\
 & 3 & 0.0 [0.0, 0.0] & 31.7 [20.8, 41.7] & 15.0 [5.0, 27.5] & 40 \\
Qwen3-8B & 1 & 0.0 [0.0, 0.0] & 3.3 [0.0, 7.5] & 7.5 [0.0, 17.5] & 40 \\
 & 2 & 0.0 [0.0, 0.0] & 5.0 [0.8, 11.7] & 10.0 [2.5, 20.0] & 40 \\
 & 3 & 0.0 [0.0, 0.0] & 48.3 [36.7, 60.8] & 42.5 [27.5, 57.5] & 40 \\
\bottomrule
\end{tabular}
}
\end{minipage}\par\smallskip
\par\smallskip\noindent\begin{minipage}{\textwidth}\centering\scriptsize
\setlength{\tabcolsep}{3pt}\renewcommand{\arraystretch}{1.04}
\noindent\textbf{Table~\ref{tab:online_adaptive_budget}, (b)} Three-query domain results: 48 Wikipedia and 48 scientific documents, three hidden probes each; target-retrieved subsets contain 25 and 15 items.\par\smallskip
\adjustbox{max width=\linewidth}{
\begin{tabular}{@{}llrrrrr@{}}
\toprule
\textbf{Model} & \textbf{Domain} & \shortstack{\textbf{Guard}\\\textbf{facts}} & \shortstack{\textbf{Guard}\\\textbf{re-ID@1}} & \shortstack{\textbf{Full stack}\\\textbf{facts}} & \shortstack{\textbf{Full stack}\\\textbf{re-ID@1}} & \textbf{$n$} \\
\midrule
Qwen2.5-3B & Wikipedia & 42.7 [29.3, 57.3] & 8.0 [0.0, 20.0] & 20.0 [9.3, 32.0] & 8.0 [0.0, 20.0] & 25 \\
 & Scientific papers & 13.3 [4.4, 24.4] & 26.7 [6.7, 53.3] & 13.3 [4.4, 24.4] & 40.0 [13.3, 66.7] & 15 \\
Qwen3-8B & Wikipedia & 62.7 [46.7, 77.3] & 40.0 [20.0, 60.0] & 34.7 [20.0, 49.3] & 20.0 [8.0, 36.0] & 25 \\
 & Scientific papers & 24.4 [11.1, 37.8] & 46.7 [20.0, 73.3] & 15.6 [4.4, 28.9] & 46.7 [20.0, 73.3] & 15 \\
\bottomrule
\end{tabular}
}
\end{minipage}\par\smallskip

\subsection{Human construct checks}
Two independent human annotators, blind to model, control, score, and pair status, rated 144 true-source and 48 mismatched-source pairs. True pairs balance two models, two prompts, and four controls. Agreement is 88.0\% ($\kappa=0.831$) for expression and 95.3\% ($\kappa=0.929$) for recovery. Wilson intervals describe this fixed sample. Automated proxy correlations use source-item-clustered bootstrap intervals; partial correlations rank-residualise response length. These are technical rather than legal judgments.

Source linkage ranks Qwen3 responses with BM25 and MiniLM over the 192-item corpus, 96-way topic-matched candidates, and expanded pools of 20,233 Wikipedia or 1,169 scientific documents. Uniform, question-only, answer-only, and length-matched shuffled controls separate topic and length effects from response signal.

\par\smallskip\noindent\begin{minipage}{\textwidth}\centering\scriptsize
\setlength{\tabcolsep}{3pt}\renewcommand{\arraystretch}{1.04}
\captionof{table}{Human judgments and automated proxy validation.}\label{tab:semantic_human}\label{tab:probe_construct_validation}
\noindent\textbf{(a)} Human-rated source/response outcomes: 144 true and 48 mismatched pairs. Brackets: Wilson 95\% intervals. Expression $\kappa=0.831$; recovery $\kappa=0.929$.\par\smallskip
\adjustbox{max width=\linewidth}{
\begin{tabular}{@{}lrrrr@{}}
\toprule
\textbf{Control} & \textbf{Close/near-verbatim expression} & \textbf{Substantial/near-complete recovery} & \textbf{Source identifiable} & \textbf{$n$} \\
\midrule
None & 83.3 [68.1, 92.1] & 50.0 [34.5, 65.5] & 61.1 [44.9, 75.2] & 36 \\
Normalized guard & 25.0 [13.8, 41.1] & 19.4 [9.8, 35.0] & 19.4 [9.8, 35.0] & 36 \\
Summary-style prompt & 61.1 [44.9, 75.2] & 30.6 [18.0, 46.9] & 44.4 [29.5, 60.4] & 36 \\
Summary + normalized guard & 30.6 [18.0, 46.9] & 11.1 [4.4, 25.3] & 19.4 [9.8, 35.0] & 36 \\
\midrule
Mismatched-source controls & 0.0 [0.0, 7.4] & 0.0 [0.0, 7.4] & 0.0 [0.0, 7.4] & 48 \\
\bottomrule
\end{tabular}
}
\end{minipage}\par\smallskip
\par\smallskip\noindent\begin{minipage}{\textwidth}\centering\scriptsize
\setlength{\tabcolsep}{3pt}\renewcommand{\arraystretch}{1.04}
\noindent\textbf{Table~\ref{tab:semantic_human}, (b)} Proxy validation on those 192 human-rated pairs. Every proxy is scored against the displayed source. Spearman intervals cluster by source item; partial $\rho$ adjusts for response length; $\kappa$ uses four-level exact recall.\par\smallskip
\adjustbox{max width=\linewidth}{
\begin{tabular}{@{}P{0.65\textwidth}rr@{}}
\toprule
\textbf{Quantity} & \textbf{True-source pairs} & \textbf{Mismatched-source pairs} \\
\midrule
Exact hidden-probe recall of displayed source (all probes), Spearman $\rho$ & 0.72 [0.60, 0.82] & n/a \\
Exact hidden-probe recall of displayed source ($\geq$4-token answers), Spearman $\rho$ & 0.51 [0.27, 0.69] & n/a \\
Full-source multiset token recall, Spearman $\rho$ & 0.87 [0.80, 0.92] & n/a \\
Normalized character similarity, Spearman $\rho$ & 0.85 [0.78, 0.90] & n/a \\
Response token count, Spearman $\rho$ & 0.84 [0.75, 0.90] & n/a \\
\midrule
Exact hidden-probe recall of displayed source (all probes), partial Spearman $\rho$ (controlling for response length) & 0.41 [0.22, 0.57] & n/a \\
Exact hidden-probe recall of displayed source ($\geq$4-token answers), partial Spearman $\rho$ (controlling for response length) & 0.25 [0.04, 0.44] & n/a \\
Full-source multiset token recall, partial Spearman $\rho$ (controlling for response length) & 0.46 [0.33, 0.58] & n/a \\
Normalized character similarity, partial Spearman $\rho$ (controlling for response length) & 0.38 [0.24, 0.50] & n/a \\
\midrule
Linear weighted $\kappa$ (human vs.\ discretized exact recall) & 0.490 & n/a \\
Sensitivity, any recovery vs.\ displayed-source exact-recall indicator (\%) & 51.8 [42.7, 60.7] & n/a \\
Specificity, any recovery vs.\ displayed-source exact-recall indicator (\%) & 96.6 [82.8, 99.4] & 100.0 [92.6, 100.0] \\
\midrule
Mismatched-control false-positive rate, human any recovery (\%) & n/a & 0.0 [0.0, 7.4] \\
Mismatched-control false-positive rate, automated any recovery on the displayed (wrong) source's own probes (\%) & n/a & 0.0 [0.0, 7.4] \\
\bottomrule
\end{tabular}
}
\end{minipage}\par\smallskip

\par\smallskip\noindent\begin{minipage}{\textwidth}\centering
\includegraphics[width=\textwidth]{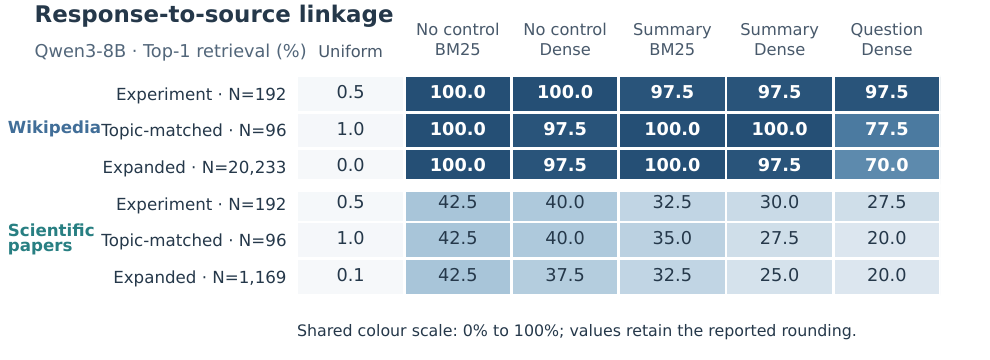}
\captionof{figure}{Response-to-source linkage calibration (top-1, \%). Qwen3 answer-only responses are ranked by BM25 or MiniLM against the experimental corpus, question-selected topic-matched candidates, and expanded same-domain pools. Candidate-set sizes and the reported uniform and question-only baselines are retained. This measures source linkage, not copying or content recovery.}\label{fig:source_linkage_heatmap}\label{tab:source_linkage_calibration}

\end{minipage}\par\smallskip


\subsection{Primary contrast, utility, and sensitivity}

\par\smallskip\noindent\begin{minipage}{\textwidth}\centering
\includegraphics[width=\textwidth]{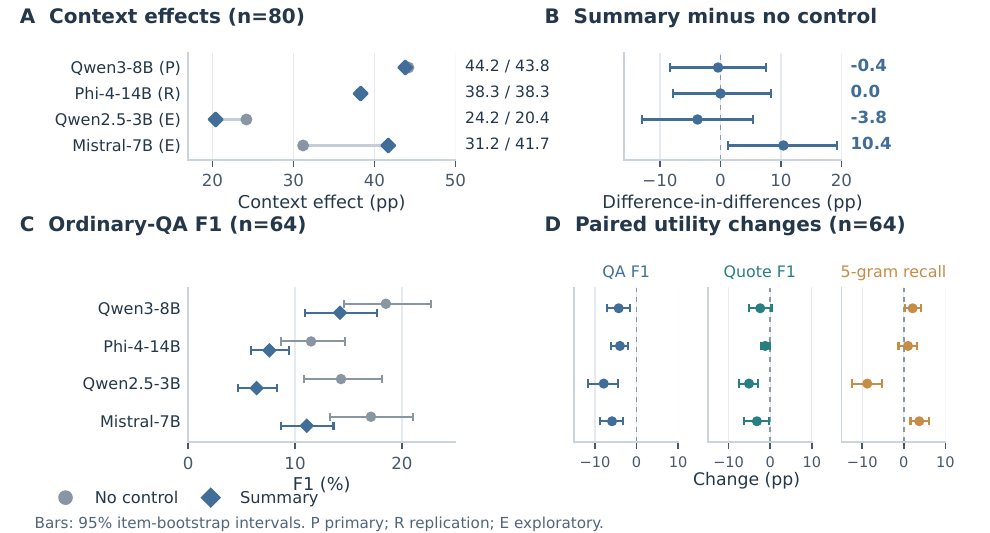}
\captionof{figure}{Context contrast and matched utility across the original model roles. (A) No-control and Summary context effects after three queries; paired numbers are shown in that order. (B) Summary-minus-uncontrolled differences-in-differences on all 80 answer-only items. Only Qwen3 was planned confirmatory; exact-answer recall is an incomplete recovery proxy. (C) Absolute ordinary-QA F1 on 64 document-disjoint items. (D) Paired Summary-minus-uncontrolled changes in QA F1, quote-query F1, and authorised quote 5-gram recall. Bars are 95\% source-item bootstrap intervals. Negative authorised-quote recall means less desired quotation.}\label{fig:context_utility_forest}\label{tab:claim_aligned_primary}\label{tab:claim_aligned_utility}

\end{minipage}\par\smallskip

\par\smallskip\noindent\begin{minipage}{\textwidth}\centering\scriptsize
\setlength{\tabcolsep}{3pt}\renewcommand{\arraystretch}{1.04}
\captionof{table}{Paired utility, equivalence bounds, and batch variation.}\label{tab:claim_paired_utility}\label{tab:claim_utility_equivalence}\label{tab:utility_repeat_variance}
\noindent\textbf{(a)} Paired utility deltas against uncontrolled (5,000-draw paired item-bootstrap intervals). Token F1 and span F1 distinguish response-length sensitivity from answer coverage.\par\smallskip
\adjustbox{max width=\linewidth}{
\begin{tabular}{@{}llrrr@{}}
\toprule
\textbf{Model} & \textbf{Control} & $\Delta$ \textbf{Token F1} & $\Delta$ \textbf{Span F1} & $\Delta$ \textbf{Resp.\ tokens} \\
\midrule
Qwen3-8B & Summary & -4.5 [-7.3, -1.8] & -3.4 [-10.3, +3.3] & +6.0 [-2.1, +13.1] \\
 & SystemBlock & -0.4 [-1.1, +0.3] & -0.1 [-1.6, +1.6] & +2.7 [-0.2, +5.7] \\
 & Threshold & +0.0 [-2.7, +2.4] & -11.5 [-19.6, -4.2] & -17.1 [-24.8, -10.3] \\
\addlinespace
Phi-4-14B & Summary & -4.0 [-6.2, -2.0] & -4.9 [-10.5, -0.3] & +5.6 [-2.1, +13.5] \\
 & SystemBlock & +2.0 [+0.5, +3.7] & -0.0 [-3.2, +3.1] & -3.6 [-7.9, +0.4] \\
 & Threshold & +3.6 [+0.7, +6.6] & -13.3 [-20.8, -6.3] & -18.5 [-27.6, -9.3] \\
\addlinespace
Qwen2.5-3B & Summary & -8.3 [-12.4, -4.6] & -23.1 [-33.7, -12.9] & +15.9 [+6.1, +25.6] \\
 & SystemBlock & +1.8 [+0.4, +3.3] & -3.0 [-10.0, +3.5] & -10.1 [-15.8, -4.8] \\
 & Threshold & -0.9 [-3.4, +1.6] & -11.7 [-19.4, -4.5] & -14.2 [-22.9, -5.9] \\
\addlinespace
Mistral-7B & Summary & -6.2 [-9.2, -3.4] & +3.8 [-2.8, +10.9] & +27.3 [+19.1, +36.0] \\
 & SystemBlock & -1.2 [-2.7, +0.1] & +1.1 [-2.1, +5.0] & +3.8 [-2.2, +10.3] \\
 & Threshold & -0.3 [-2.9, +2.4] & -9.1 [-17.1, -2.1] & +2.4 [-6.7, +10.6] \\
\bottomrule
\end{tabular}
}
\end{minipage}\par\smallskip
\par\smallskip\noindent\begin{minipage}{\textwidth}
\noindent\begin{minipage}[t]{0.49\textwidth}\vspace{0pt}
\centering\scriptsize
\setlength{\tabcolsep}{3pt}\renewcommand{\arraystretch}{1.04}
\noindent\textbf{Table~\ref{tab:claim_paired_utility}, (b)} Post-hoc utility equivalence bounds: a checkmark requires the entire 95\% paired span-F1 interval strictly inside the bound, a stricter criterion than 90\%-interval TOST inclusion.\par\smallskip
\adjustbox{max width=\linewidth}{
\begin{tabular}{@{}llcc@{}}
\toprule
\textbf{Model} & \textbf{Control} & $\pm5$ \textbf{pts} & $\pm10$ \textbf{pts} \\
\midrule
Qwen3-8B & Summary & ,  & ,  \\
 & SystemBlock & \checkmark & \checkmark \\
 & Threshold & ,  & ,  \\
\addlinespace
Phi-4-14B & Summary & ,  & ,  \\
 & SystemBlock & \checkmark & \checkmark \\
 & Threshold & ,  & ,  \\
\addlinespace
Qwen2.5-3B & Summary & ,  & ,  \\
 & SystemBlock & ,  & \checkmark \\
 & Threshold & ,  & ,  \\
\addlinespace
Mistral-7B & Summary & ,  & ,  \\
 & SystemBlock & \checkmark & \checkmark \\
 & Threshold & ,  & ,  \\
\bottomrule
\end{tabular}
}
\end{minipage}\hfill
\begin{minipage}[t]{0.49\textwidth}\vspace{0pt}
\centering\scriptsize
\setlength{\tabcolsep}{3pt}\renewcommand{\arraystretch}{1.04}
\noindent\textbf{Table~\ref{tab:claim_paired_utility}, (c)} Six replicates per model vary batch size (4, 8, 16) and row order, with seed, revision, and greedy decoding fixed.\par\smallskip
\adjustbox{max width=\linewidth}{
\begin{tabular}{@{}lrrrr@{}}
\toprule
\textbf{Model} & \textbf{F1 SD} & \textbf{Span SD} & \textbf{Range} & \textbf{Changed} \\
\midrule
Qwen3-8B & 0.013 & 0.035 & 0.032 & 6.2\% \\
Phi-4-14B & 0.017 & 0.065 & 0.043 & 18.8\% \\
Qwen2.5-3B & 0.250 & 0.000 & 0.530 & 17.2\% \\
Mistral-7B & 0.006 & 0.000 & 0.015 & 6.2\% \\
\bottomrule
\end{tabular}
}
\end{minipage}
\end{minipage}\par\smallskip
\par\smallskip\noindent\begin{minipage}{\textwidth}\centering\scriptsize
\setlength{\tabcolsep}{3pt}\renewcommand{\arraystretch}{1.04}
\captionof{table}{Context and answer-format effects on utility.}\label{tab:utility_endpoint_diagnostic}
\noindent\textbf{(a)} Format diagnostic on 64 document-disjoint items per model. Only context and answer-format instruction vary. Brackets: 95\% item-bootstrap intervals. Token F1 scores the short gold answer against the whole response.\par\smallskip
\adjustbox{max width=\linewidth}{
\begin{tabular}{@{}llrrrr@{}}
\toprule
\textbf{Context} & \textbf{Answer format} & \textbf{Token F1} & \textbf{Span F1} & \textbf{Answer recall} & \textbf{Resp.\ tokens} \\
\midrule
\multicolumn{6}{@{}l}{\textit{Qwen3-8B}} \\
\quad Closed book & free-form (frozen) & 9.1 [6.5, 11.8] & 24.8 [17.2, 32.7] & 24.2 [16.4, 32.3] & 20.3 \\
\quad Closed book & short answer & 12.8 [7.6, 18.6] & 18.3 [11.7, 25.5] & 15.6 [9.7, 22.1] & 9.3 \\
\quad Retrieved top-3 & free-form (frozen) & 18.9 [15.0, 23.0] & 67.4 [57.4, 76.9] & 66.7 [56.6, 76.2] & 38.1 \\
\quad Retrieved top-3 & short answer & 46.8 [36.7, 56.5] & 57.0 [45.9, 67.7] & 55.6 [44.6, 66.4] & 7.7 \\
\quad Gold source only & free-form (frozen) & 28.0 [23.8, 32.5] & 86.2 [79.4, 91.9] & 84.7 [77.7, 90.7] & 27.5 \\
\quad Gold source only & short answer & 60.9 [52.6, 68.8] & 76.3 [67.7, 84.2] & 73.3 [64.1, 81.7] & 8.0 \\
\addlinespace
\multicolumn{6}{@{}l}{\textit{Phi-4-14B}} \\
\quad Closed book & free-form (frozen) & 5.2 [3.4, 7.2] & 30.8 [22.7, 39.2] & 31.6 [23.5, 40.6] & 85.3 \\
\quad Closed book & short answer & 12.4 [7.9, 17.8] & 23.7 [16.4, 31.3] & 21.1 [14.3, 28.2] & 15.7 \\
\quad Retrieved top-3 & free-form (frozen) & 11.3 [8.7, 14.3] & 71.6 [62.5, 80.2] & 72.7 [63.4, 81.4] & 86.0 \\
\quad Retrieved top-3 & short answer & 39.2 [30.6, 48.3] & 65.2 [55.0, 75.1] & 63.6 [53.2, 73.6] & 22.3 \\
\quad Gold source only & free-form (frozen) & 22.6 [18.8, 26.4] & 90.1 [85.4, 94.1] & 91.4 [86.8, 95.2] & 50.0 \\
\quad Gold source only & short answer & 57.4 [49.2, 65.2] & 83.3 [75.0, 90.4] & 82.1 [73.7, 89.5] & 12.2 \\
\addlinespace
\multicolumn{6}{@{}l}{\textit{Qwen2.5-3B}} \\
\quad Closed book & free-form (frozen) & 3.6 [2.5, 4.8] & 23.7 [16.6, 31.1] & 23.4 [16.3, 31.0] & 64.1 \\
\quad Closed book & short answer & 9.1 [3.8, 15.4] & 9.8 [4.1, 16.6] & 9.5 [3.8, 16.1] & 3.5 \\
\quad Retrieved top-3 & free-form (frozen) & 14.3 [10.6, 18.4] & 64.0 [53.8, 73.6] & 64.4 [54.1, 74.1] & 56.2 \\
\quad Retrieved top-3 & short answer & 40.0 [29.6, 50.5] & 46.6 [35.8, 57.6] & 43.9 [33.2, 54.8] & 5.6 \\
\quad Gold source only & free-form (frozen) & 20.9 [17.3, 24.7] & 81.4 [74.2, 87.9] & 81.1 [73.5, 87.9] & 44.8 \\
\quad Gold source only & short answer & 57.3 [47.6, 67.1] & 68.5 [58.1, 78.4] & 65.5 [54.9, 75.5] & 7.8 \\
\addlinespace
\multicolumn{6}{@{}l}{\textit{Mistral-7B}} \\
\quad Closed book & free-form (frozen) & 5.1 [3.6, 6.6] & 34.0 [25.8, 42.0] & 33.4 [25.3, 41.4] & 70.6 \\
\quad Closed book & short answer & 13.4 [7.9, 19.5] & 20.7 [13.4, 28.7] & 18.3 [11.4, 26.0] & 12.2 \\
\quad Retrieved top-3 & free-form (frozen) & 17.3 [13.5, 21.3] & 66.5 [56.7, 76.1] & 66.4 [56.5, 76.1] & 49.7 \\
\quad Retrieved top-3 & short answer & 40.3 [31.1, 49.6] & 58.5 [48.0, 68.5] & 56.7 [46.2, 67.0] & 15.2 \\
\quad Gold source only & free-form (frozen) & 26.8 [22.2, 31.4] & 85.0 [78.3, 90.7] & 85.0 [78.0, 90.8] & 34.2 \\
\quad Gold source only & short answer & 51.0 [43.0, 58.9] & 72.3 [63.5, 80.5] & 69.1 [60.2, 77.6] & 9.2 \\
\bottomrule
\end{tabular}
}
\end{minipage}\par\smallskip
\par\smallskip\noindent\begin{minipage}{\textwidth}\centering\scriptsize
\setlength{\tabcolsep}{3pt}\renewcommand{\arraystretch}{1.04}
\captionof{table}{Exploratory construct and practical-equivalence checks.}\label{tab:claim_aligned_construct_sensitivity}\label{tab:claim_aligned_equivalence}
\noindent\textbf{(a)} Exploratory construct sensitivity. DiD uses all 80 answer-only items except long-answer recall ($n=57$, 21 Wikipedia and 36 scientific), standardised to equal domain weights. Domain columns use 40 items each and forced target-present Summary exposure. Response length is in tokens; other columns are percentage points.\par\smallskip
\adjustbox{max width=\linewidth}{
\begin{tabular}{@{}lrrrrrr@{}}
\toprule
\textbf{Model} & \shortstack{\textbf{Long-answer}\\\textbf{recall DiD}}$^{a}$ & \shortstack{\textbf{Neg.-control}\\\textbf{margin DiD}} & \shortstack{\textbf{Full-source}\\\textbf{token DiD}} & \shortstack{\textbf{Response length}\\\textbf{DiD (tokens)}} & \shortstack{\textbf{Wikipedia exact}\\\textbf{recall}$^{b}$} & \shortstack{\textbf{Sci.\ paper exact}\\\textbf{recall}$^{b}$} \\
\midrule
Qwen3-8B & -17.5 [-30.6, -5.4] & -1.3 [-9.2, 7.1] & 5.1 [-0.5, 10.5] & 57.7 [28.6, 85.9] & 75.8 [67.5, 83.3] & 11.7 [5.8, 18.3] \\
Phi-4-14B & -8.4 [-21.0, 4.2] & 0.4 [-7.5, 8.8] & 3.4 [-1.9, 8.6] & 84.4 [49.8, 119.8] & 68.3 [57.5, 79.2] & 14.2 [7.5, 22.5] \\
Qwen2.5-3B & -7.7 [-20.6, 5.7] & -2.9 [-11.7, 5.8] & 3.5 [-2.3, 9.7] & 35.8 [14.1, 57.5] & 28.3 [18.3, 39.2] & 13.3 [6.7, 21.7] \\
Mistral-7B & -0.3 [-13.8, 12.5] & 11.7 [2.5, 20.4] & 2.9 [-2.5, 8.2] & 11.7 [-16.4, 39.9] & 68.3 [58.3, 78.3] & 23.3 [16.7, 30.0] \\
\bottomrule
\end{tabular}
}
\end{minipage}\par\smallskip
\par\smallskip\noindent\begin{minipage}{\textwidth}\centering\scriptsize
\setlength{\tabcolsep}{3pt}\renewcommand{\arraystretch}{1.04}
\noindent\textbf{Table~\ref{tab:claim_aligned_construct_sensitivity}, (b)} Post-hoc TOST on exact-answer DiD. Bounds were chosen after the frozen analysis. $p_{\mathrm{TOST}}$ is the larger shifted one-sided sign-randomisation value (20,000 draws). Equivalence requires $p<0.05$ and the 90\% paired bootstrap interval inside the bound.\par\smallskip
\adjustbox{max width=\linewidth}{
\begin{tabular}{@{}lrrrrrl@{}}
\toprule
\textbf{Model} & \textbf{Bound} & \textbf{DiD} & \shortstack{\textbf{90\% paired}\\\textbf{bootstrap CI}} & \textbf{$p_{\mathrm{TOST}}$} & \textbf{$\alpha$} & \textbf{Interpretation} \\
\midrule
Qwen3-8B & $\pm$5pp & -0.4 & [-7.1, 6.2] & $0.139$ & 0.05 & not established \\
Qwen3-8B & $\pm$10pp & -0.4 & [-7.1, 6.2] & $0.011$ & 0.05 & equivalent (within bound) \\
Phi-4-14B & $\pm$5pp & 0.0 & [-6.7, 6.7] & $0.119$ & 0.05 & not established \\
Phi-4-14B & $\pm$10pp & 0.0 & [-6.7, 6.7] & $0.010$ & 0.05 & equivalent (within bound) \\
Qwen2.5-3B & $\pm$5pp & -3.8 & [-11.2, 3.7] & $0.395$ & 0.05 & not established \\
Qwen2.5-3B & $\pm$10pp & -3.8 & [-11.2, 3.7] & $0.094$ & 0.05 & not established \\
Mistral-7B & $\pm$5pp & 10.4 & [2.9, 17.9] & $0.877$ & 0.05 & not established \\
Mistral-7B & $\pm$10pp & 10.4 & [2.9, 17.9] & $0.528$ & 0.05 & not established \\
\bottomrule
\end{tabular}
}
\end{minipage}\par\smallskip

\subsection{Seven-query extension}
The exploratory extension pins Summary, SystemBlock, and the threshold to the same upstream configuration. A complete-source request is followed by up to six requests using Base64, ROT13, reversed words or characters, fact tables, and paraphrases. The source intervention uses a shared attack ranking across three controls and two exposure arms. Natural retrieval uses control-specific development rankings over both policies and all 160 test items included in intent-to-treat analyses. Utility is evaluated on two query types and 64 document-disjoint items.

Budgets 1, 3, and 7 carry forward the last released state after stopping. Intervals use 5,000 paired source-item bootstrap draws and sign randomisation uses 20,000 draws. Models and the two attacker designs are not pooled. Long-answer analyses use the 57 eligible items with accepted answers of at least four tokens, standardised to equal domain weights. Practical-equivalence bounds are post-hoc: a large TOST $p$ value does not establish equivalence.

\par\smallskip\noindent\begin{minipage}{\textwidth}\centering\scriptsize
\setlength{\tabcolsep}{3pt}\renewcommand{\arraystretch}{1.04}
\captionof{table}{Seven-query source-intervention results.}\label{tab:claim_defense_extension_source_q7}\label{tab:claim_defense_extension_endpoint_did}
\noindent\textbf{(a)} Exploratory seven-query source intervention, shared attacker, 80 answer-only items. DiD is defence-minus-uncontrolled for the present-minus-ablated context effect. Brackets: 95\% paired source-item bootstrap intervals.\par\smallskip
\adjustbox{max width=\linewidth}{
\begin{tabular}{@{}llrrrr@{}}
\toprule
\textbf{Model} & \textbf{Defense} & \shortstack{\textbf{Uncontrolled}\\\textbf{context effect}} & \shortstack{\textbf{Defense}\\\textbf{context effect}} & \shortstack{\textbf{Difference-in-}\\\textbf{differences}} & \textbf{$n$} \\
\midrule
Qwen3-8B & Summary & 43.3 & 42.5 & -0.8 [-7.9, 6.2] & 80 \\
Qwen3-8B & SystemBlock & 43.3 & 45.8 & 2.5 [-3.8, 8.3] & 80 \\
Phi-4-14B & Summary & 40.4 & 40.8 & 0.4 [-7.1, 7.9] & 80 \\
Phi-4-14B & SystemBlock & 40.4 & 39.6 & -0.8 [-6.7, 5.0] & 80 \\
Qwen2.5-3B & Summary & 22.9 & 22.9 & 0.0 [-8.8, 8.3] & 80 \\
Qwen2.5-3B & SystemBlock & 22.9 & 24.6 & 1.7 [-5.4, 8.8] & 80 \\
Mistral-7B & Summary & 33.8 & 40.4 & 6.7 [-2.5, 16.2] & 80 \\
Mistral-7B & SystemBlock & 33.8 & 28.3 & -5.4 [-12.9, 2.1] & 80 \\
\bottomrule
\end{tabular}
}
\end{minipage}\par\smallskip
\par\smallskip\noindent\begin{minipage}{\textwidth}\centering\scriptsize
\setlength{\tabcolsep}{3pt}\renewcommand{\arraystretch}{1.04}
\noindent\textbf{Table~\ref{tab:claim_defense_extension_source_q7}, (b)} Seven-query shared-attacker DiD by endpoint. Long-answer recall uses 57 eligible items with equal domain weights; other outcomes use 80. Values are percentage points except response-length tokens.\par\smallskip
\adjustbox{max width=\linewidth}{
\begin{tabular}{@{}llrrrrr@{}}
\toprule
\textbf{Model} & \textbf{Defense} & \shortstack{\textbf{Exact-answer}\\\textbf{DiD}} & \shortstack{\textbf{Long-answer}\\\textbf{DiD}$^{a}$} & \shortstack{\textbf{Full-source}\\\textbf{token DiD}} & \shortstack{\textbf{Decoded}\\\textbf{copy$_{20}$ DiD}} & \shortstack{\textbf{Response-length}\\\textbf{DiD (tokens)}} \\
\midrule
Qwen3-8B & Summary & -0.8 [-7.9, 6.2] & -16.3 [-28.6, -4.0] & 6.5 [0.8, 12.1] & -36.2 [-48.8, -23.8] & 110.4 [74.2, 146.4] \\
Qwen3-8B & SystemBlock & 2.5 [-3.8, 8.3] & -1.9 [-10.5, 6.1] & 1.8 [-1.3, 4.9] & -12.5 [-21.2, -5.0] & 24.5 [-7.0, 59.0] \\
Phi-4-14B & Summary & 0.4 [-7.1, 7.9] & -13.2 [-25.4, -1.4] & 1.1 [-3.8, 6.0] & -28.7 [-40.0, -17.5] & 69.7 [22.3, 118.4] \\
Phi-4-14B & SystemBlock & -0.8 [-6.7, 5.0] & -5.1 [-14.4, 3.1] & 3.6 [-0.4, 7.7] & -3.8 [-12.5, 5.0] & 82.7 [31.5, 132.9] \\
Qwen2.5-3B & Summary & 0.0 [-8.8, 8.3] & -8.1 [-20.6, 4.1] & 3.5 [-2.8, 10.0] & -42.5 [-55.0, -30.0] & 58.7 [13.4, 103.8] \\
Qwen2.5-3B & SystemBlock & 1.7 [-5.4, 8.8] & 4.4 [-8.9, 16.7] & 1.7 [-2.7, 6.1] & 11.2 [1.2, 21.2] & -33.0 [-92.0, 26.3] \\
Mistral-7B & Summary & 6.7 [-2.5, 16.2] & -2.0 [-14.8, 10.4] & 0.4 [-5.0, 5.6] & -32.5 [-45.0, -20.0] & -54.6 [-99.9, -12.7] \\
Mistral-7B & SystemBlock & -5.4 [-12.9, 2.1] & -10.7 [-22.4, 0.3] & -2.3 [-7.2, 2.5] & 3.8 [-6.2, 13.8] & -7.5 [-56.7, 39.8] \\
\bottomrule
\end{tabular}
}
\end{minipage}\par\smallskip
\par\smallskip\noindent\begin{minipage}{\textwidth}\centering\scriptsize
\setlength{\tabcolsep}{3pt}\renewcommand{\arraystretch}{1.04}
\captionof{table}{Seven-query natural retrieval and ordinary QA utility.}\label{tab:claim_defense_extension_natural_q7}\label{tab:claim_defense_extension_utility}
\noindent\textbf{(a)} Seven-query natural retrieval with control-specific attackers. Paired deltas use all answer-only test items, without conditioning on retrieval; target presence is reported separately. Brackets: 95\% paired source-item bootstrap intervals.\par\smallskip
\adjustbox{max width=\linewidth}{
\begin{tabular}{@{}llrrrr@{}}
\toprule
\textbf{Model} & \textbf{Defense} & \shortstack{\textbf{Uncontrolled}\\\textbf{recall}} & \shortstack{\textbf{Defense}\\\textbf{recall}} & \shortstack{\textbf{Paired}\\\textbf{$\Delta$}} & \shortstack{\textbf{Target}\\\textbf{presence}} \\
\midrule
Qwen3-8B & Summary & 42.5 & 40.8 & -1.7 [-7.9, 4.6] & 72.5 \\
Qwen3-8B & SystemBlock & 42.5 & 43.8 & 1.2 [-4.6, 7.1] & 72.5 \\
Qwen3-8B & Threshold & 42.5 & 32.5 & -10.0 [-18.3, -2.5] & 46.2 \\
Phi-4-14B & Summary & 40.8 & 45.4 & 4.6 [-2.9, 12.1] & 72.5 \\
Phi-4-14B & SystemBlock & 40.8 & 39.6 & -1.3 [-6.7, 4.2] & 72.5 \\
Phi-4-14B & Threshold & 40.8 & 32.1 & -8.8 [-15.8, -2.1] & 46.2 \\
Qwen2.5-3B & Summary & 22.5 & 12.5 & -10.0 [-16.7, -3.8] & 72.5 \\
Qwen2.5-3B & SystemBlock & 22.5 & 27.9 & 5.4 [-1.2, 12.5] & 72.5 \\
Qwen2.5-3B & Threshold & 22.5 & 22.1 & -0.4 [-7.5, 6.7] & 46.2 \\
Mistral-7B & Summary & 30.4 & 37.5 & 7.1 [-1.3, 15.8] & 72.5 \\
Mistral-7B & SystemBlock & 30.4 & 23.8 & -6.7 [-13.8, -0.4] & 72.5 \\
Mistral-7B & Threshold & 30.4 & 29.2 & -1.2 [-6.7, 5.0] & 46.2 \\
\bottomrule
\end{tabular}
}
\end{minipage}\par\smallskip
\par\smallskip\noindent\begin{minipage}{\textwidth}\centering\scriptsize
\setlength{\tabcolsep}{3pt}\renewcommand{\arraystretch}{1.04}
\noindent\textbf{Table~\ref{tab:claim_defense_extension_natural_q7}, (b)} Single-turn, document-disjoint ordinary-QA token F1 (percentage points) with 95\% item-bootstrap intervals; utility does not vary with attack budget.\par\smallskip
\adjustbox{max width=\linewidth}{
\begin{tabular}{@{}lrrrr@{}}
\toprule
\textbf{Model} & \textbf{Uncontrolled} & \textbf{Summary} & \textbf{SystemBlock} & \textbf{Threshold} \\
\midrule
Qwen3-8B & 18.6 [14.6, 22.8] & 14.1 [10.9, 17.4] & 18.1 [14.3, 22.3] & 18.6 [14.7, 22.6] \\
Phi-4-14B & 11.5 [8.8, 14.7] & 7.6 [6.0, 9.4] & 13.5 [10.2, 17.2] & 15.1 [11.1, 19.4] \\
Qwen2.5-3B & 14.9 [11.2, 19.1] & 6.6 [4.8, 8.7] & 16.7 [12.8, 21.0] & 14.1 [10.3, 18.0] \\
Mistral-7B & 17.2 [13.5, 21.1] & 11.0 [8.7, 13.5] & 16.0 [12.5, 19.6] & 16.9 [12.8, 21.6] \\
\bottomrule
\end{tabular}
}
\end{minipage}\par\smallskip

\par\smallskip\noindent\begin{minipage}{\textwidth}\centering
\includegraphics[width=\textwidth]{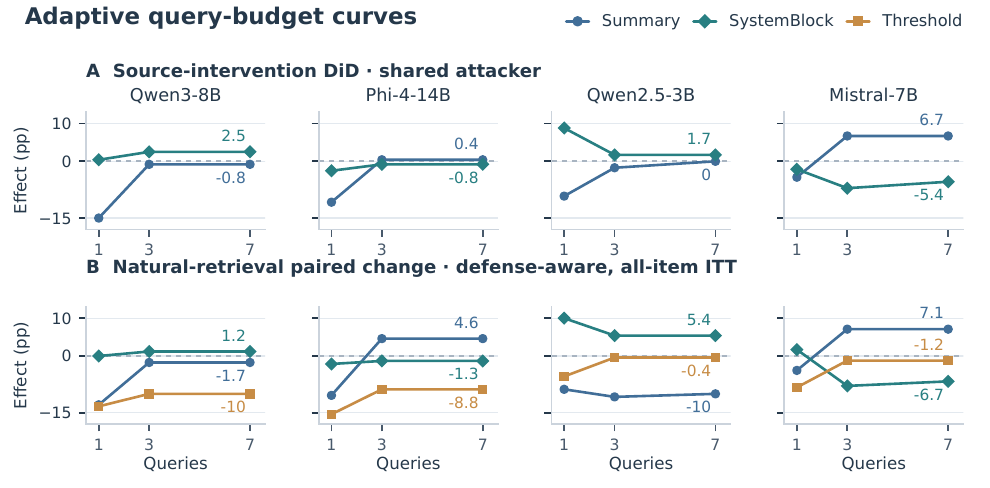}
\captionof{figure}{Exploratory exact-answer proxy changes at query budgets 1, 3, and 7 (percentage points). Top: source-intervention difference-in-differences under a shared attacker. Bottom: paired natural-retrieval changes under control-specific attackers, using all-item intent-to-treat analysis. All three original point estimates are plotted. Stopped interactions carry their last released state forward; models and attacker designs are not pooled. Intervals were not reported in the original budget table.}\label{fig:query_budget_curves}\label{tab:claim_defense_extension_budget_curves}

\end{minipage}\par\smallskip


\section{Survey Synthesis and Record-Level Operating Points}
\label{app:survey_extended}\label{app:mechanism_table}\label{app:resources_table}\label{app:rob_table}\label{app:failure_modes}
Mechanisms, intervention stages, and metrics are multi-label axes, so their counts need not sum to 118.

Interaction settings comprise 51 single-query, 31 fixed multi-query, 25 non-interactive, and 11 feedback-adaptive records. Attacker access includes external black box (47), corpus-contributor write (39), authorised user (16), operator white box (13), service-provider protocol party (6), corpus-owner auditor (4), and operator grey box (3). Intervention stages are generation (81), context construction (64), retrieval (60), post-generation (39), corpus acquisition (27), index construction (26), session (8), and removal/update (7). Mechanism counts are in Figure~\ref{tab:mechanisms}, with one additional ``other'' record; metric counts are in Figure~\ref{tab:resources}.

\par\smallskip\noindent\begin{minipage}{\textwidth}\centering
\includegraphics[width=\textwidth]{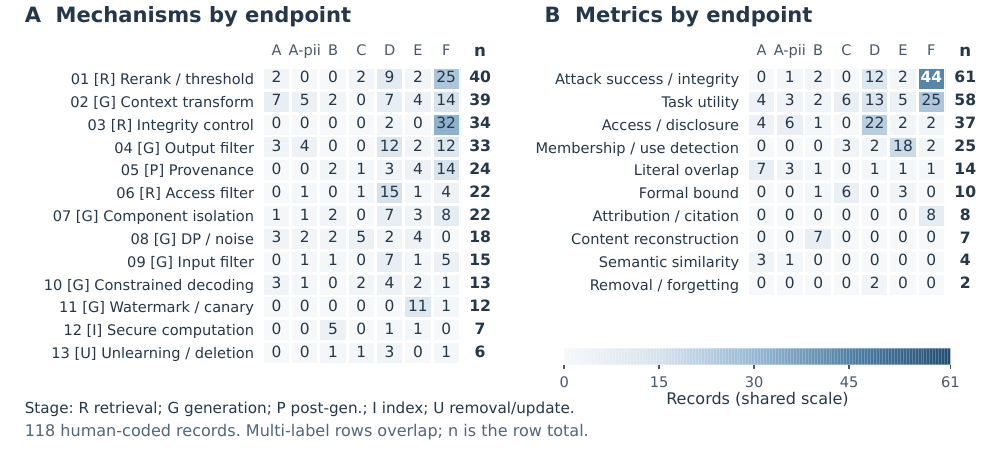}
\captionof{figure}{Mechanisms and metrics by primary endpoint over 118 records. Cells and row totals share a count scale; multi-label rows overlap. Stage codes and numbered representative works retain the original annotations.}\label{fig:mechanism_metric_heatmaps}\label{tab:mechanisms}\label{tab:resources}
\par\smallskip{\scriptsize\raggedright\textbf{Representative works by mechanism row:} 01 \cite{evidenceR2C002D57ACD8,evidenceR4C0DAFA3063C}; 02 \cite{evidenceR06B30753C7E5,evidenceR09C5B5D0E3E7}; 03 \cite{evidenceR7A14A88575A1,evidenceR0AF83A8B2D74}; 04 \cite{evidenceR1C5B7527412C,evidenceR65D32EB39C2F}; 05 \cite{qi2024internalsbasedanswer,evidenceR27CA01807A11}; 06 \cite{evidenceR91BC2953986C,evidenceR96F38064CB1B}; 07 \cite{evidenceR29805AC998DD,evidenceX72D0D6126373}; 08 \cite{evidenceRCE5BDC14C770,evidenceR15670280E9D8}; 09 \cite{evidenceR06B30753C7E5,evidenceR18F3BEE5E973}; 10 \cite{golatkar2024cpr,evidenceR08B5E4564969}; 11 \cite{evidenceRA09D08FF90FF,jovanovic2025ward}; 12 \cite{evidenceR855C97F9416E,evidenceRA098E4279B5C}; 13 \cite{evidenceRF0A0A315FCC0,zhang2025tracebackpoisoningattacks}.\par}
\end{minipage}\par\smallskip

\par\smallskip\noindent\begin{minipage}{\textwidth}\centering
\includegraphics[width=\textwidth]{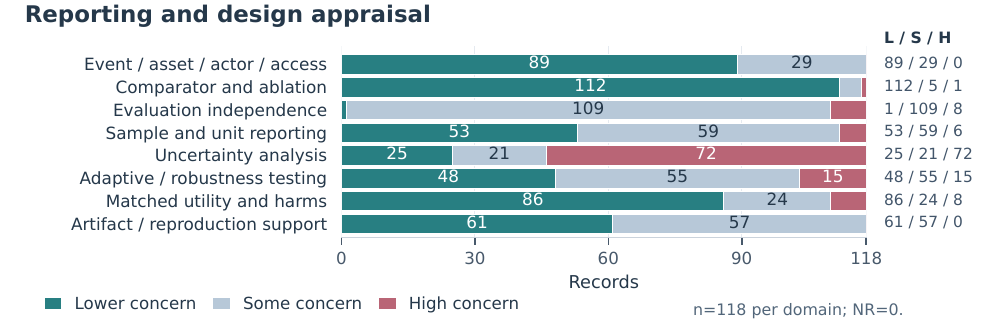}
\captionof{figure}{Reporting and design appraisal over 118 records. Counts are lower/some/high concern; NR is zero throughout. These ratings are not efficacy weights.}\label{fig:evidence_appraisal_bars}\label{tab:human_evidence_appraisal}\label{tab:rob}

\end{minipage}\par\smallskip


\subsection{Artifact and provenance}
\label{app:artifact}\label{app:legal}

\subsection{Reported operating points by endpoint family}\label{app:operating_points}\label{app:complete_map}
\begingroup\footnotesize\setlength{\tabcolsep}{3pt}\setlength{\LTpre}{4pt}\setlength{\LTpost}{4pt}\renewcommand{\arraystretch}{1.00}
\begin{longtable}{@{}lP{0.16\textwidth}P{0.77\textwidth}@{}}
\toprule
\textbf{Fam.} & \textbf{Record} & \textbf{Primary endpoint as reported} \\
\midrule
\endfirsthead
\toprule
\textbf{Fam.} & \textbf{Record} & \textbf{Primary endpoint as reported} \\
\midrule
\endhead
A & \citet{evidenceR001ABA21E18D} & Repeat Prompt / Repeat Context / ROUGE Prompt+Context counts of retrieved chunks reproduced in output; summarization, distance threshold, re-ranking drive counts to 0 \\
A & \citet{evidenceR08B5E4564969} & Repeat Prompt/Context defined as response reproducing text verbatim from the retrieval corpus; ROUGE-L above threshold vs corpus text (264->127 etc.) \\
A & \citet{evidenceR09C5B5D0E3E7} & Repeat Prompt = at least 10 tokens exactly matching original dataset; ROUGE-L>0.5; targeted PII count driven to 0 \\
A & \citet{evidenceR14DF3B6B0A8B} & Repeat Prompts / Repeat Length / Target Info under targeted and prefix extraction attacks (59 pct reduction in repeated prompts) \\
A & \citet{evidenceR7B704BB933B9} & Privacy leakage rate under the 'privacy extraction attack' of He et al. (ChatDoctor 0.23 -> 0.17; EnronQA 0.63 -> 0.33, Table 3); BLEU/ROUGE-L reported as UTILITY \\
A & \citet{evidenceR85066D7CF090} & Repeat Prompts = >=10 consecutive tokens from retrieval corpus; ROUGE Prompts/Contexts>0.5; zero leakage across epsilon=0.1..10 \\
A & \citet{evidenceRBA752514142F} & ALR/ASR/CLR privacy-entity leakage in generated output vs a GRPO-trained adaptive attacker (NormAvep 1.00 -> 0.13) \\
A & \citet{li2026ragfort} & Chunk Recovery Rate (ROUGE-L>0.5 generated output vs original chunk plus cosine) under Worm-Attack and RAG-Thief; relative mean 1x -> 0.51x \\
A-pii & \citet{evidenceR18F3BEE5E973} & Document-level any-success PII leak rate in outputs 0.875 -> 0.000 (en/zh/de) under output regex filter plus input judge \\
A-pii & \citet{evidenceR273E285746F4} & Repeat Prompt = outputs containing >=10 consecutive verbatim tokens from private text; ROUGE Prompt>0.5 vs original private text; adaptive attack table \\
A-pii & \citet{evidenceR4F2DA7F185DA} & De-anonymization Rate under a three-tier threat model (ArcFace/CosFace/AdaFace 99.4 -> 1.8-3.1 pct; T2 top-1 4.2 pct), plus VQA accuracy / FID / LPIPS utility \\
A-pii & \citet{evidenceR6BA703A6F06F} & Garak PII-leak failures 50 -> 0; Laplace-adjusted success probability 0.980 -> 0.019; ABAC arm renders the PII corpus unretrievable \\
A-pii & \citet{evidenceRCD0B4A81C6AA} & LLM-as-a-Judge privacy leakage score 0-100 comparing sensitive entities in the SOURCE text against those present in the GENERATED text, PRE and POST anonymization (Table 5), plus ROUGE-L/cosine/PPL utility \\
A-pii & \citet{evidenceRDF32B25411D8} & Count of patient-name leaks in model responses under 1000 benign and 100 targeted extraction queries (Table 3) \\
B & \citet{evidenceR3F7E8C21CA48} & ROUGE-2/BLEU-2/EMR/BiNLI of embedding-inversion (EIA) reconstructions of the user's input text, driven to \textasciitilde 0 \\
B & \citet{evidenceR6F93638EBFCD} & ROUGE-2 recall of document reconstruction from generated LoRA weights by a trained T5 decoder (<9 pct) \\
B & \citet{evidenceR855C97F9416E} & Edge recovery rate from observed HNSW traversal paths 18.3 -> 6.8 pct with false-positive rate and Recall@10 degradation at identifiable rho/E operating points \\
B & \citet{evidenceRA098E4279B5C} & Isolation-Forest / One-Class-SVM detector ROC-AUC (0.85-1.000) and TPR@1\%/5\% FPR against steganographic embedding exfiltration; VectorPin provenance protocol \\
B & \citet{evidenceRB9E24CAAD81D} & Vec2Text reconstruction BLEU/precision/recall/F1/ROUGE from encrypted embeddings (BLEU 83 -> 12) plus retrieval accuracy and throughput \\
B & \citet{evidenceRBD0D00A81D6A} & Resistance to ICA and graph matching; kNN-overlap rate between plaintext and ciphertext spaces; relative attention-output error 1.49-1.63 pct \\
B & \citet{evidenceRF0D04B38F9F4} & Vec2Text ROUGE-L recovery from soft-deleted HNSW vectors 0.207 -> 0.000 under epoch key rotation; ECDSA-signed erasure proof \\
C & \citet{evidenceR4C15D8386AEE} & Formal (epsilon approx 10) DP guarantee across hundreds of queries via individual Renyi privacy filters, plus utility; no empirical leakage measurement in the packet \\
C & \citet{evidenceR6FA1EC21BE61} & Shutterstock watermark visibly copied into generated images (qualitative, UNDEFENDED RDM baseline, Fig.2b); protection outcome is DP epsilon=10 plus FID 10.9 \\
C & \citet{evidenceRCE5BDC14C770} & Membership-inference attack accuracy 70.22 -> 53.60 pct (advantage Adv = Acc - 0.5) at a 2.6 pct accuracy cost under pure epsilon-DP SimHash bucketing \\
C & \citet{golatkar2024cpr} & CLIP similarity between CPR-generated images and retrieved private images (Syn-Ret) held low while prompt similarity (Syn-Cap) stays high; empirical k\_c / NAF Delta\_max bit bound \\
C & \citet{koga2024privacypreservingdifferential} & S2MIA TPR-FPR curves with AUC collapsing to \textasciitilde 0.5 at epsilon=10 (BLEU used as the membership score), plus average match accuracy utility \\
C & \citet{min2023silo} & Perplexity on held-out Harry Potter book with vs without HP books in the datastore (Table 6); qualitative sentence-level attribution examples (Table 7) \\
D & \citet{agnew2024datadefenses} & Judge-rated fraction of defended texts the attacker LLM is UNABLE TO USE to answer (news RAG benchmark n=75; >90 pct on many datasets; 75 pct transfer to unseen text) \\
D & \citet{evidenceR07546576B950} & MBA exact mask-fill reconstruction leakage rate (5 masked tokens, membership signal) 29.5->0.0 pct; poisoning ASR 56->0 pct \\
D & \citet{evidenceR15670280E9D8} & HR@K / MRR separately for public answers and privacy-threatening (private-link-derived) answers; adversarial training suppresses the latter \\
D & \citet{evidenceR29805AC998DD} & Structural leak rate (unauthorized chunks RETRIEVED) 86.1->0 pct; keyword-matched answer leak 41.3/29.5 -> 0 pct; 100 pct -> 0 pct violation rate \\
D & \citet{evidenceR3001F183E8FC} & Cluster purity, accuracy, precision, recall, F1 (0.77) and AUPRC 0.97 for DETECTING PII-leak and jailbreak interactions \\
D & \citet{evidenceR3569BDC61B98} & User study (N=36): GPT-4o-rated reduction of private information stored/input into LLM memory, plus subjective/objective usability metrics (Table 1) \\
D & \citet{evidenceR493DAAA44D59} & Role-based access filtering accuracy 85 pct and F1 0.89 on 389 queries vs static-filter baseline (F1 0.93-0.94) and no-filter baseline (F1 0.32-0.43) \\
D & \citet{evidenceR55C3E7FDA198} & Attack Success Rate of PII extraction (Bad Query / Adversarial / CopyBreakRAG) 81.9 -> 9.5 pct; detection P/R/F1 94.1/93.5/93.8 \\
D & \citet{evidenceR65D32EB39C2F} & AUROC / FPR95 / FPR90 and abstain rate for DETECTING quasi-identifier cluster leakage in generated RAG outputs across three domains \\
D & \citet{evidenceR666FFB777DB2} & Precision/recall for declined answers, K-precision/recall and Elo utility; architectural isolation means user text never reaches the generator \\
D & \citet{evidenceR79F5A5DAE002} & Cross-Tenant Leakage Rate = fraction of cross-tenant probes returning at least one unauthorized CHUNK (80->0 pct); AVR; injection block rate \\
D & \citet{evidenceR7F813FD6F967} & LASR/NAV/LK all 0.00; LK = leaked tokens / total private tokens with leakage confirmed by ROUGE-L recall between generated output and private KB; MIA and Vec2Text/GEIA and PASR \\
D & \citet{evidenceR91BC2953986C} & Tenant-isolation leakage rate 0.2 pct -> 0 pct over 1000 queries with engine-level row-level security, plus latency \\
D & \citet{evidenceR96F38064CB1B} & Access-control compliance verified per test case (retrieval restricted to permitted profiles) plus relevance/correctness of retrieved information \\
D & \citet{evidenceRAA90B8A72BA1} & Attacker profile-identification accuracy (linkability) and attribute-inference accuracy: \textasciitilde 100 pct with full context vs 20 pct client-side / 58.5 pct query-only \\
D & \citet{evidenceRB0964DC859DE} & Private-triple removal rate r\_remove and de-anonymization QA-attack success rate r\_attack, plus PPL fluency and downstream RAG accuracy \\
D & \citet{evidenceRB418B88C0261} & Privacy and completeness scores with and without prompt injection (Table I); up to 58 pct privacy-score improvement \\
D & \citet{evidenceRD16ABB124F4D} & Retrieval Pivot Risk 0.95 -> 0.0 across three corpora with Leakage@k, Amplification Factor and per-attack RPR=1.0 baselines \\
D & \citet{evidenceRD3517D8895CD} & Unauthorized context exposure 3.84 -> 0.00 pct; sensitive answer leakage 2.91 -> 0.04 pct; 98.6 pct red-team attacks blocked \\
D & \citet{evidenceRF0A0A315FCC0} & Unlearning effectiveness on the forget set after excluding forgotten user data from the retrieval index, with HR@K / NDCG@K utility on remaining data \\
D & \citet{evidenceRF340DD8930BB} & Unauthorized-access outcome across four governance baselines (200 queries): results are never unauthorized regardless of 63-75 pct routing accuracy \\
D & \citet{evidenceRF6154A128C48} & Confidentiality-filter recall 99.01 pct (82.31 pct accuracy / 74.35 pct recall with ambiguous queries) preventing security-sensitive queries reaching external LLMs \\
D & \citet{evidenceRFC8AEA5EA29F} & 98 pct accuracy detecting malicious, document-extraction and hallucination attacks; extraction requests denied \\
D & \citet{evidenceX1F0D6C87E992} & ASR matrix driven to zero across three attack families with lineage/provenance signing and a tau x K ablation, plus utility baseline \\
D & \citet{wang2024unlearningrag} & Unlearning Success Rate; ROUGE-L recall of output vs short reference ANSWER; TPR@1\%FPR residual memorization \\
D & \citet{yao2025controlnet} & AUROC / matching accuracy for risk detection across five risks (0.96-0.997) and precision/recall/F1 for mitigation \\
E & \citet{chen2025aqua} & Verification Success Rate, CGSR and TPR/FPR of AQUA watermark verification; filter-resistance results \\
E & \citet{evidenceR06B30753C7E5} & Shadow-RAG membership audit AUC 0.989 / Acc 0.957 / F1 0.956; robustness under prompt-modifying and paraphrasing defenses \\
E & \citet{evidenceR4C0DAFA3063C} & Detection accuracy/precision/recall/F1 for member attack queries plus adjusted attack accuracy (adversary advantage) reduction \\
E & \citet{evidenceR4FE823CFC370} & Chunk Recovery Rate: chunk counted recovered iff ROUGE-L(generated output, original chunk)>0.5 AND cosine similarity; 97.4 -> 1.6; detection TPR \\
E & \citet{evidenceR6575BF573BB1} & Effective Detection Rate with p<1e-5 at 0.1 pct sentinel injection; FDR<=1.0 pct across four datasets \\
E & \citet{evidenceR65B78A21EED1} & Internal-state classifier ACC/F1 (up to 100/100) separating copyrighted-leakage from non-leakage inputs; IS-w/RAG vs LLM-as-a-Judge baselines \\
E & \citet{evidenceR7730130E89CB} & stylometric attribution F1 0.825 -> 0.221 (TF-IDF) and 0.691 -> 0.203 (BERT) \\
E & \citet{evidenceR7B84D3639671} & Watermark detection AUC (up to 100) and spoofing-AUC; Knowledge Corruption Rate; ROUGE-1/L and PPL reported as UTILITY of the watermarked answer \\
E & \citet{evidenceRA09D08FF90FF} & Watermark recovery rate WRR=1.00 for multi-bit payload; robustness/detection outcomes under attacks \\
E & \citet{evidenceRD15171C1818A} & Coalition-detection TPR=1.00 over 2400 trials at FPR<=0.05, plus (eps\_acc, delta\_acc)-DP retrieval guarantee and a matching membership-inference AUC trend in k \\
E & \citet{evidenceRD543A60A097B} & AUC / ASR / TPR@1 reductions across seven membership-inference attacks (SPV, LiRA, Recall, Log-Loss); up to 27.8 pct MIA-success reduction and 75 pct TPR@1 reduction \\
E & \citet{evidenceRDFA259676DB5} & Artifact-level disclosure risk on exported evidence: Membership-Inference Accuracy -9.8 pp, Attribute-Inference Accuracy -8.4 pp, top-1 Linkage success rate (Table 2) \\
E & \citet{guo2025towardscopyrightprotection} & CoT-space watermark ownership verification p-values (malicious p<<0.01 vs independent p=1.00) \\
E & \citet{jovanovic2025ward} & Watermark-propagation p-values and TP/FP rates on the FARAD benchmark \\
E & \citet{liu2025canaries} & ROC-AUC and TPR@1\%FPR = 100 pct at a 12-query operating point for watermarked canaries \\
E & \citet{liu2025whostole} & Interrogator-Detective dual-layer watermark detection accuracy and false-positive rates across LLMs and adversarial conditions \\
E & \citet{luo2025imagesentinelprotectingvisual} & Detection AUC / TPR from sentinel-image key queries against the RAIG reference database \\
E & \citet{lv2025ragwm} & WSN / WIRR watermark verification success and integrity across five datasets and four LLMs, including robustness to removal attacks \\
F & \citet{evidenceR0AF83A8B2D74} & Poisoned-context attack defense improvement \textasciitilde 20 pct; QA accuracy under varying poisoned/factual context ratios \\
F & \citet{evidenceR14FD8D476F15} & Macro-averaged ASR and FPR, detection rate and response-quality score for a three-layer prompt-injection defense (Table 8) \\
F & \citet{evidenceR1C5B7527412C} & Virtual Donkey detection TPR=1.000 / FPR=0.015 (BLEU, ROUGE-L, METEOR input-output similarity used as classifier FEATURES); attack-side exfiltrated email-address F1 \\
F & \citet{evidenceR25958BFF8CB8} & Poison Impact Rate reduction across ingestion/retrieval/generation layers on a 300-document CTI corpus (90 poisoned), plus latency \\
F & \citet{evidenceR27CA01807A11} & Character-level traceback overlap and over-selection metrics for locating poisoned spans; passage-level identification and ASR baselines \\
F & \citet{evidenceR292B1C2F289E} & Certified defense: ASR 93 pct -> 5.3 pct (n=150) for runtime memory poisoning, with provenance signing \\
F & \citet{evidenceR29FBDF2F28CF} & Malicious-document detection performance plus end-to-end ASR and precision \\
F & \citet{evidenceR2C002D57ACD8} & ASR 0.04 against Gemini on MS MARCO with highest accuracy among defenses \\
F & \citet{evidenceR30706FBDDD56} & Attribution precision/recall for identifying the responsible context texts, plus ASR with and without the defense \\
F & \citet{evidenceR3DB082BE97F6} & True-positive detection rates for poisoned subgraph elements vs perplexity-based filtering baselines \\
F & \citet{evidenceR42E2F054F90E} & ASR reduced to near zero for sentence-relation-based sanitization of malicious instructions (from 82-83 pct baseline) \\
F & \citet{evidenceR47114B882520} & Answer accuracy and ASR across poisoning settings for context-analysis-based security \\
F & \citet{evidenceR4BFF94222183} & Influence Score identifies the poisoned document with 86 pct success over 3000 samples (p<0.0001) \\
F & \citet{evidenceR4EE16B98B934} & Provably robust aggregation: ASR reduced to as low as 1 pct at 71 pct accuracy \\
F & \citet{evidenceR4F3424AA4C9D} & Adulterant Retrieval Rate 100 pct; Hijack Success >94 pct (unauthorized system answers wrongly); CDPA/CIRA 100 pct authorized fidelity; 80.2 pct robustness retention \\
F & \citet{evidenceR54239E60E067} & Best-F1 detection comparison (QPD, keyword anomaly) and ROC curves for corpus-dependent poisoning \\
F & \citet{evidenceR55AB4F027B97} & Chunk-wise perplexity filtering detection of poisoned chunks plus text-similarity metrics \\
F & \citet{evidenceR564183F3431F} & Security Rate of generated code under two poisoning scenarios (average across LLMs drops 60.84 -> 50.57 without defense; CodeGuarder consistently improves SR) \\
F & \citet{evidenceR5EEE0CFB7D9A} & ASR increase of 4-16x when the isolation component is removed; certified-synthesis information-flow controls \\
F & \citet{evidenceR7A14A88575A1} & TPR/FPR tables for revealing poisoned responses via LLM activation analysis (metric definitions elided in packet) \\
F & \citet{evidenceR7FA1BEA31833} & Attribution precision/recall for the injected/corrupted texts plus ASR with and without attack \\
F & \citet{evidenceR809356F3BEF0} & ASR-R / ASR-A under faithful evaluation with a certified detection radius for memory poisoning \\
F & \citet{evidenceR80DA6BF3922F} & Adversarial-hub attack recall 1.000 (universal/concept) and 0.91 on HotFlip at AUROC \textasciitilde 1.0 with \textasciitilde 1 pct benign FPR at an admission-time gate \\
F & \citet{evidenceR85D4CDFD898E} & ACC and ASR across poisoning attacks using sparse-attention corpus defense \\
F & \citet{evidenceR86DFB6674043} & Attack Pass-through Rate plus 0 pct FPR on the benign subset for priority-aware prompt control-flow integrity \\
F & \citet{evidenceRAACD70942374} & ASR (retriever returns at least one poisoned passage) plus utility under benign conditions for RAGPart/RAGMask \\
F & \citet{evidenceRB855BD4CA9FA} & Attention-aware detectability of stealthy poisoning across epsilon regimes vs perplexity filters; end-to-end robust pipeline \\
F & \citet{evidenceRC0B9EF36243C} & Answer accuracy and ASR (0.00 across several settings) for a RAG-guards-RAG expert-module defense \\
F & \citet{evidenceRC182E8AA1FBC} & Mahalanobis-distance detection rate >98 pct for universally poisoned documents plus retrieval and attack success rates \\
F & \citet{evidenceRD13EB8DB05C8} & Corpus-poisoning detection precision/recall and F1 across datasets and evidence sizes vs perplexity baselines; 86 pct rationale accuracy \\
F & \citet{evidenceRD1A4AD7E9DEB} & ASR 0 pct (95 pct CI [0,1]) for numerical-claim manipulation with 99.8 pct entity detection rate \\
F & \citet{evidenceRD68E2115AE49} & Answer accuracy on a poisoned MS MARCO corpus using PageRank-based trust propagation \\
F & \citet{evidenceRE0AC8ADEE349} & Sentence-level in-generation attribution accuracy per algorithm (Table 5) with pre-attribution ground truth \\
F & \citet{evidenceRE3EED5702691} & ACC / ASR across datasets for K-means conflict removal and knowledge consolidation against corpus poisoning \\
F & \citet{evidenceRF11FCE97A7FB} & Answer accuracy and ASR (percentage of poisoning queries/documents succeeding) for two-stage semantic filtering \\
F & \citet{evidenceRF4E59A677BF0} & Block F1 0.802 / 0.846 and top-1 source accuracy 0.858 (source-plus-relation 0.229) vs MiniCheck, RAGAS, AlignScore baselines \\
F & \citet{evidenceRF744830A332D} & ASR reduced by up to 87 pct against PoisonedRAG with answer-accuracy preservation (Tables 1-2) \\
F & \citet{evidenceRFB11D4C3E660} & ASR by defense across nine models (360 runs): input-layer 77.8 pct, Memory Sandbox 0 pct for eight of nine, one model inverts to 100 pct via the RAG pathway \\
F & \citet{evidenceRFE7024B86FE4} & Attack-detection precision 1.00 with recall/accuracy 0.45 and F1 0.62 for the Reverse RAG version over 674 attacks; 100 pct block on a 301-sample high-confidence subset \\
F & \citet{evidenceX021A8DA55C23} & Provably robust generation accuracy and ASR under prompt injection at different attack positions \\
F & \citet{evidenceX29EE8D4C8525} & 71 pct clean accuracy with 38 pct certifiable robust accuracy; ASR reduced from over 90 pct \\
F & \citet{evidenceX382114ABD760} & Answer accuracy and ASR (frequency of attacker-specified output) across datasets \\
F & \citet{evidenceX72D0D6126373} & Discriminator classification F1 and exact-match robustness under counterfactual/corrupted retrieved context \\
F & \citet{evidenceXA52302837E10} & Harm driven to 0.000 by removing unauthenticated structural writes and rerunning the selector; accumulability criterion \\
F & \citet{qi2024internalsbasedanswer} & Agreement with human answer attribution on XOR-AttriQA; citation precision/recall/F1 (TRUE/NLI-based) \\
F & \citet{zhang2025tracebackpoisoningattacks} & Traceback precision/recall for identifying poisoned texts in the knowledge database vs perplexity-based detection \\
F & \citet{zhang2025whotaughtlie} & Responsibility-attribution accuracy for poisoned knowledge vs PPL/norm baselines; RAG accuracy under benign and poisoned settings \\
\bottomrule
\end{longtable}
\endgroup

\end{document}